\documentclass[aps,twocolumn,amsmath,amssymb,preprintnumbers]{revtex4}
\usepackage{amsmath} \usepackage{amsfonts} \usepackage{amssymb}
\usepackage{bbm}
\usepackage{epsfig}
\usepackage{graphics}
\usepackage{graphicx}
\newcommand{\ba}{\begin{array}}
\newcommand{\ea}{\end{array}}
\newcommand{\be}{\begin{equation}}
\newcommand{\ee}{\end{equation}}
\newcommand{\bear}{\begin{eqnarray}}
\newcommand{\ear}{\end{eqnarray}}
\newcommand{\n}{{\nonumber}}
\begin{document}

\title[ ]{Functional renormalization for quantum phase transitions with non-relativistic bosons}

\author{C. Wetterich}
\affiliation{Institut  f\"ur Theoretische Physik\\
Universit\"at Heidelberg\\
Philosophenweg 16, D-69120 Heidelberg}

\begin{abstract}
Functional renormalization yields a simple unified description of bosons at zero temperature, in arbitrary space dimension $d$ and for $M$ complex fields. We concentrate on nonrelativistic bosons and an action with a linear time derivative. The ordered phase can be associated with a nonzero density of (quasi) particles $n$. The behavior of observables and correlation functions in the ordered phase depends crucially on the momentum $k_{ph}$, which is characteristic for a given experiment. For the dilute regime $k_{ph}\gtrsim n^{1/d}$ the quantum phase transition is simple, with the same ``mean field'' critical exponents for all $d$ and $M$. On the other hand, the dense regime  $k_{ph}\ll n^{1/d}$  reveals a rather rich spectrum of features, depending on $d$ and $M$. In this regime one observes for $d\leq 3$ a crossover to a relativistic action with second time derivatives. This admits order for $d>1$, whereas $d=1$ shows a behavior similar to the low temperature phase of the classical two-dimensional  $O(2M)$-models. 
\end{abstract}

\maketitle

\section{Introduction}
\label{intro}
The ground state of many quantum systems can undergo a second order phase transition if the density, concentration, external fields or some effective coupling constants are varied. An example is the transition from para- to ferro-magnetism for bosonic atoms with spin. Bosonic quasiparticles also describe the quantum phase transition for strongly correlated electrons in case of anisotropic antiferromagnetic order. We address here only systems where the low energy excitations are bosons - they can be relevant also for correlated electrons, if the fermionic excitations are gapped. 

An effective description accounts for a transition from an ``ordered phase'' with a nonzero continuously varying density of bosonic excitations to a ``disordered phase'' where this density vanishes. At zero temperature the boson density can often be characterized by a condensate, which is described by the nonvanishing expectation value of a (complex) field. In a broad sense such a condensate can be associated with order, while for the disordered phase the expectation value vanishes. Such transitions between a disordered or ``symmetric'' phase and an ordered phase with spontaneous symmetry breaking are therefore described by Bose-Einstein condensation in a gas of interacting bosons. In this picture the parameter driving the phase transition can be associated with an effective chemical potential $\sigma$. Universality of the critical behavior near second order phase transitions implies that many key features of quantum phase transitions are independent of the particular ``microscopic'' physical systems. With ultracold bosonic atoms new ways of experimental investigation of such systems open up. 

For dilute systems, many features of quantum phase transitions for bosons are well understood, and many details of the critical behavior in various dimensions are known. The basic aspects are visible in a mean field theory and perturbation theory \cite{1D}. The use of several methods, including mapping to fermionic systems and bosonization in one dimension, together with strong universality arguments based on the renormalization group, allows for the computation of critical exponents and correlation functions \cite{SB}. One finds a rather simple picture with mean field critical exponents. This picture is valid, however, only as long as the system is sufficiently dilute. In this paper we extend the discussion to dense systems. For one- and two-dimensional systems we will find qualitative changes. They are induced by the fluctuations of the Goldstone boson, which is characteristic for the spontaneous breaking of a continuous symmetry. In three dimensions, these effects are logarithmic - still sufficient for a cure of the infrared problems in many previous treatments \cite{1A}. This cure is similar to other renormalization group approaches \cite{1B,1C}.

In order to define the notion of dense and dilute, one should compare a typical physical length scale, $l$, with the average distance between particles, $D\sim n^{-1/d}$. For dense systems one has $l\gg D$, whereas dilute systems obey $l\ll D$. As a first attempt one could try to use the correlation length $\xi$ as physical length scale. This works well in the disordered phase where $\xi$ is finite away from the phase transition. In the ordered phase, however, the spontaneous symmetry breaking of an abelian continuous symmetry leads to superfluidity. For a nonzero condensate the system has always a gapless (``massless'') mode - the Goldstone boson. The correlation length is infinite such that the system would appear ``dense'' for any nonvanishing $n$. 

In practice, a given experiment will always involve an effective momentum scale $k_{ph}$, for example the inverse of the wave length used to probe the system. Technically, the ``physical momentum scale'' $k_{ph}$ may correspond to the momentum in some relevant Green's functions and act as an (additional) infrared cutoff for the fluctuations. The smallest possible value of $k_{ph}$ is given by the inverse size of the experimental probe. Instead of $\xi^{-1}$ we may therefore compare the physical momentum scale $k_{ph}$ with the scale $k_F\sim n^{1/d}$. For $k_F\ll k_{ph}$ the particle density only induces small corrections and the Bose gas is dilute. In contrast, the dense regime for $k_{ph}\ll k_F$ corresponds to a situation where a characteristic inter-particle distance $D\sim k^{-1}_F$ is small as compared to a typical experimental length scale $l\sim k^{-1}_{ph}$. We use the concepts ``dilute'' and ``dense'' here in a rather general sense since we do not specify $l$ a priori. 

In the ordered phase the quantum fluctuations with low momenta, $\vec q^2=k^2~,~k\ll k_F$, are dominated by the Goldstone boson. We will call this the ``Goldstone regime''. In three dimensions the Goldstone fluctuations play a quantitative role, but do not change the qualitative behavior, except for the extreme infrared. (Due to a logarithmic running of dimensionless couplings the qualitative changes may only occur for exponentially small momentum scales.) Typically, fluctuations on length scales larger than the scattering length $a$ give only small corrections. For $k_{ph}\ll a^{-1}$ the precise value of $k_{ph}$ becomes unimportant - the most important effective infrared cutoff is set by $a^{-1}$. In this case we may consider a system with $ak_F\ll 1$ as dilute, independent of $k_{ph}$. In contrast, for $d=1,2$ the Goldstone fluctuations always play an important role for $k_{ph}\ll k_F$. In this case the value of $k_{ph}$ matters and needs to be considered as a separate physical scale. 

The simple critical behavior of the dilute regime always applies for the disordered phase since $n=0$. In contrast, the understanding of the ordered phase is more subtle, in particular for lower dimensional systems, $d=1$ or $d=2$. The dilute regime $k_{ph}\gg k_F$ remains simple, with similar properties as for the disordered phase. As $k_{ph}$ becomes smaller than $k_F$ we have to deal with a dense Bose gas where $n$ sets a new scale. Following the scale dependence of ``running'' renormalized couplings one observes a crossover to the ``Goldstone regime'', with new qualitative properties. We argue that for $d=1$ and $d=2$ the Goldstone regime is effectively described by a relativistic action with two time derivatives. It therefore shares common features with the classical $O(2M)$-models in dimension $d+1$. In particular, for $d=1$ and $M=1$ this implies the characteristic behavior of the low temperature phase in the Kosterliz-Thouless \cite{KT} phase transition.

We propose here a simple unified picture for the properties of the quantum phase transition which is valid on all scales. It is based on the functional renormalization group \cite{WW} for the average action \cite{CWAV}, \cite{CWFE}, \cite{BTW}. Within a simple $\phi^4$-model it describes the quantum phase transition for an arbitrary number of space dimensions $d$ and an arbitrary number of components $M$. Within the same model we can explore the flow in the disordered and the ordered phase. For low dimensions $d=1$ or $d=2$ we find several interesting crossover phenomena, indeed associated to the nontrivial physics of Goldstone bosons in low dimensions. This crossover persists for $d=3$, but wide scale separations occur due to logarithmic running. The case $d=3$ can be considered as the boundary dimension for the relevance of the Goldstone regime. 

All of the relevant physics is non-perturbative (with a few exceptions) and involves long range excitations. We do not limit our investigation to small interaction strength. We therefore rely heavily on the capability of modern approaches to functional renormalization where the variation of an effective infrared cutoff enables the exploration of systems with massless excitations (infinite correlation length) in a nonperturbative context for arbitrary $d$ \cite{CWAV,CWFE,BTW,CWR}. For the regime $k_{ph}\gg k_F$ the merits of our approach lie, for the time being, more in the simplicity of the unified picture rather than in new quantitative results. In contrast, the flow for the dense systems, $k_{ph}\ll k_F$, reveals features that have attracted less attention so far.

Our approach is based on a functional integral formulation where the bosonic excitations are associated to a complex field $\chi$. It is formulated in a $d+1$ dimensional euclidean space with $d$ space dimensions and an euclidean time $\tau$. (For nonzero temperature $T$ euclidean time parameterizes a torus with circumference $T^{-1}$.) The transition from the Hamiltonian formulation with operators to the functional integral (or Lagrange formulation) with fields is sometimes subtle \cite{ZJ}, \cite{GW}, \cite{SB}. Two classes of systems can be distinguished, according to the presence of a linear $\tau$-derivative or not. In a rather general approach we may consider a microscopic or ``classical'' action
\begin{equation}\label{A1}
{\cal S}=\int\limits_x\chi^*
\left(\bar S\partial_\tau-\bar V\partial^2_\tau-\frac{\Delta}{2M_B}-\sigma\right)\chi+{\cal S}_{int}
\end{equation}
where $\int\limits_x=\int d\tau\int d^d\vec x$. We will assume that $S_{int}$ describes a local interaction, involving powers of $\chi^*\chi$ without derivatives. For arbitrary $\bar S$ and $\bar V$ the action (\ref{A1}) is invariant under euclidean time reversal $\tau\leftrightarrow -\tau~,~\chi\leftrightarrow\chi^*$. 

The case $\bar S=0$ is special, however. The system possesses now an enhanced rotation symmetry $SO(d+1)$, mixing space coordinates $\vec x$ and the time coordinate $\tau$. Indeed, a simple multiplicative rescaling of time or space coordinates brings the derivatives to the form $(\partial^2_\tau+\Delta)$. The relativistic excitation spectrum can be directly seen by analytic continuation to ``real time'', $\tau=it$. After suitable rescalings we may set $\bar V =1,~2M_B=1$, such that eq. (\ref{A1}) reduces to the classical $O(2M)$-model in dimension $d+1$, if $\chi$ has $M$ complex components. (For $M>1$ suitable sums over components are implied in eq. (\ref{A1})). Functional renormalization has already provided a unified picture for the phase transition in classical $O(N)$-models for arbitrary $d$ \cite{CWFE,BTW,CWR}, including the Kosterlitz-Thouless phase transition for $d+1=2$ and $M=1$ \cite{GKT}. Due to the enhanced symmetry the vanishing of the coefficient linear in $\partial_\tau~(\bar S=0)$ is stable under the renormalization flow.

In this paper we search for a similar unified picture for the ``nonrelativistic bosons'' with $\bar S\neq 0$. We will concentrate on the simplest case $\bar V=0$ where by a suitable rescaling we may choose $\bar S=1$. In Minkowski space the microscopic action is now invariant under Galilei-transformations. We emphasize, however, that $\bar V=0$ is not protected by a symmetry and second $\tau$-derivatives will be generated by the functional renormalization flow. For $T=0$ this comes together with the higher order gradient terms requested by Galilei invariance.

Even though we concentrate in this paper on the Galilei-invariant setting with $\bar V=0$, the discussion in sect. \ref{crossover}  will also cover the more general case of a microscopic action \eqref{A1} with $\bar S\neq 0~,~\bar V\neq 0$. We first specialize to one component and extend the discussion to $M$-components in sect. \ref{mcomponentmodels}. An overview over the different  regimes for arbitrary $d,M$ and  $\bar S/\sqrt{\bar V}$, together  with our main results, can be found in the conclusions. 

Our paper is organized as follows. In order to specify our model and to fix notations we recall the functional integral formulation and functional renormalization in sect. \ref{functional-integration}. In sect. \ref{floweq} we derive the flow equations for the renormalized couplings for arbitrary $d~(M=1)$. Sect. \ref{disordered} gives a brief description of running couplings in the disordered phase. We discuss the fixed points and the associated scaling behavior relevant for the quantum phase transition in sect. \ref{scaling}. Sects.  \ref{disordered} and \ref{scaling} reproduce the known results for non-relativistic quantum phase transitions \cite{SB} within the framework of functional renormalization. They may be skipped by the reader familiar with the subject. In sect. \ref{orderd} we derive the flow equations for the ordered phase, first within a truncation where second $\tau$-derivatives are neglected. Sect. \ref{linearandgoldstone} distinguishes the ``linear regime'' relevant for dilute systems from the Goldstone regime which is important for dense systems. For $d<2$ our simplest truncation yields an attractive fixed point with nonzero order parameter and density. For $d=1$ this fixed point will persist for extended truncations, while it turns out to be an artefact of the truncation for $d>1$. 

For $d\leq 3$ the large time behavior within the ordered phase is governed by a term quadratic in the $\tau$-derivatives $\sim V$. We show in sect. \ref{crossover} how this term is generated by the flow, even if it vanishes for the microscopic action. For dense systems and $d<3$, the ``relativistic dynamic term'' $\sim V$ will always dominate over the term linear in $\partial_\tau$. The running coupling $S(k)$ vanishes with a power of $k$. In contrast, for the boundary dimension $d=3$ the vanishing of $S(k)$ is only logarithmic. The corresponding change in the propagator induces qualitative changes for the renormalization flow and phase structure for one- and two-dimensional systems. They constitute the main result of the present paper. We find that the long distance behavior of correlation functions is similar as for the classical $O(2)$-models in $d+1$ dimensions. In particular, for $d=1$ the ordered phase of the non-relativistic model behaves similar to the low temperature phase of the two-dimensional Kosterlitz-Thouless phase transition. We associate the Tomonaga-Luttinger liquid \cite{TL} to this phase. In sect. \ref{goldstoneregime} we discuss the Goldstone regime in terms of non-linear $\sigma$-models. Our approach permits a unified view of the linear $\varphi^4$-models and non-linear $\sigma$-models. Sect. \ref{mcomponentmodels} is devoted to an extension of our discussion to $M$ complex bosonic fields. We present conclusions and outlook in sect. \ref{conclusions}.

\section{Functional integral and functional renormalization}
\label{functional-integration}
We start with the partition for a nonrelativistic bosonic particle for $M=1$ 
\begin{eqnarray}\label{1}
Z&=&\int {\cal D}\chi\exp(-{\cal S}[\chi]),
\end{eqnarray}
with action $S[\chi]$ given by eq. \eqref{A1}, with $\bar S=1,\bar V=0$. 
The complex field $\chi$ may be expressed by its Fourier modes
\begin{equation}\label{2}
\chi(x)=\chi(\tau,\vec x)=\int\limits_{\vec q}e^{i\vec q\vec x}\chi(\tau,\vec q)=\int_qe^{iqx}\chi(q),
\end{equation}
with
\begin{eqnarray} 
q =(q_0,\vec q)~,~\int_q&=&\int_{\vec q}\int_{q_0}~,~\int_{q_0}=\frac{1}{2\pi}
\int\limits^\infty_{-\infty}dq_0,\nonumber\\
\int_{\vec q}&=&(2\pi)^{-d}\int d^d\vec q~.
\end{eqnarray}
For nonzero temperature $T$ the Euclidean time $\tau$ parameterizes a circle with circumference $\Omega_\tau=T^{-1}$ and the Matsubara frequencies $q_0=2\pi nT~,~n\in{\mathbbm Z}$, are discrete, with $\int_{q_0}=T\sum\limits_n$. 

In this paper we are interested in quantum phase transitions for $T=0$. This phase transition occurs as the parameter $\sigma$ is varied from positive to negative values. We regularize the theory by a momentum cutoff $\vec q\ ^2<\Lambda^2$ and take $\Lambda\to\infty$ when appropriate. Furthermore, we assume the invariance of the classical action (\ref{1}) under a global abelian symmetry of phase rotations  $\chi\to e^{i\varphi}\chi$, corresponding to a conserved total particle number
\begin{equation}\label{3}
N=\int_{\vec x}n(\vec x)=\Omega_d\int_{\vec q}n(\vec q),
\end{equation}
with $\Omega_d$ the volume of $d$-dimensional space $(\Omega_{d+1}=\Omega_d\Omega_\tau)$. Following the Noether construction we can express $n(\vec q)$ by the two point correlation function
\begin{equation}\label{4}
n(\vec{q})=\frac{1}{\Omega_{d+1}}\int_{q_0}\langle\chi^*(q_0,\vec q)\chi(q_0,\vec q)\rangle-\frac12.
\end{equation}
We may associate $\sigma =\bar \sigma -\Delta \sigma$ with a chemical potential. Here our normalization of the additive shift $\Delta\sigma$ \cite{GW} and of $n(\vec q)$ is such that $N=0$ for $\sigma<0$ and $N\neq 0$ for $\sigma>0$. If we interprete $n=N/\Omega_d$ as the number density of some bosonic quasi particle, the quantum phase transition is a transition from a state with no particles to a state with nonzero particle density. 

By a suitable rescaling of units of $x,\tau$ and $\chi$ we can replace $2M_B\to 1~,~\sigma \to \tilde\sigma=2M_B\sigma/\hat k^2$. Here we may use some arbitrary momentum unit $\hat k$ in order to make all quantities dimensionless, or we may retain dimensionful parameters by employing $\hat k=1$.  (The parameters in $S_{int}$ have to be rescaled accordingly, see Ref. \cite{DW}.) In the following we will work with a basis of real fields $\chi_1,\chi_2$ defined by 
$\chi(x)=\frac{1}{\sqrt{2}}\big(\chi_1(x)+i\chi_2(x)\big)$ such that $\chi_a(-q)=\chi^*_a(q)$. The connected part of the two point function describes the propagator ${\cal G}$
\begin{equation}\label{4B}
\langle\chi^*_a(q)\chi_b(q')\rangle={\cal G}_{ab}(q,q')+
\langle\chi^*_a(q)\rangle
\langle\chi_b(q')\rangle.
\end{equation}
For a translation invariant setting, ${\cal G}$ is diagonal in momentum space
\begin{equation}\label{4C}
{\cal G}_{ab}(q,q')=\bar G_{ab}(q)\delta(q-q'),
\end{equation}
with $\delta(q-q')=(2\pi)^{d+1}\delta(q_0-q'_0)\delta^d(\vec q-\vec q\ ')$. Also, translation invariance implies for a possible order parameter $\langle\chi_a(q)\rangle=\sqrt{2}\bar\phi_0\delta(q)\delta_{a1}$ with real $\bar\phi_0$. Here we have chosen the expectation value in the one-direction without loss of generality.

We assume a repulsive two particle interaction $(\bar\lambda>0)$
\be\label{8A}
{\cal S}_{int}=\frac{\bar\lambda}{2}\int_x\big(\bar\chi^*(x)\chi(x)\big)^2=
\frac{\bar\lambda}{8}\int_x\big(\chi_a(x)\chi_a(x)\big)^2.
\ee
After the rescaling the mass dimensions are 
$x\sim\mu^{-1}~,~\tau\sim\mu^{-2}~,~\vec  q\sim\mu~,~q_0\sim\mu^2~,~\chi\sim\mu^{\frac{d}{2}}~,~\tilde\sigma\sim\mu^2~,~n\sim\mu^d$, such that $\bar\lambda\sim\mu^{2-d}$. Already at this point one sees the crucial role of the dimension $d$. For $d=3$ the coupling $\bar\lambda$ has the dimension of a length. After a suitable renormalization it corresponds to the scattering length $a\sim\lambda$. In the vacuum $(T=0~,~n=0~,~\tilde\sigma=0)$ the renormalized interaction strength $\lambda$ sets the only scale, besides the ultraviolet cutoff $\Lambda$. As a consequence, those correlation functions that are independent of $\Lambda$ can only depend on dimensionless combinations, as $\lambda\vec q$ and $\lambda^2 q_0$. For example, the two point function takes the form $\bar G=\bar A(\vec q^2+iSq_0+Vq^2_0)^{-1}$ with real functions $\bar A,S$, and $V/\lambda^2$ depending on these dimensionless combinations. A nonzero density introduces an additional scale $k_F$. The long distance physics will now depend on the dimensionless concentration $c=ak_F\sim\lambda n^{1/3}$. 

In one dimension $(d=1)$ the interaction strength scales $\bar\lambda\sim\mu$. Now a length scale is set by $\lambda^{-1}$. For nonzero density the macroscopic physics can depend on dimensionless combinations, as $n/\lambda$. Indeed, for $\bar\lambda\to\infty$ the repulsion becomes infinite such that a particle can never pass another particle. (This permits the mapping to a non-interacting Fermi gas \cite{SB}.) The combination $n/\lambda$ is a measure of how many interparticle distances a particle can travel before being repulsed. It therefore defines an effective volume where it can move.

The case $d=2$ is special because $\bar\lambda$ is dimensionless. In the vacuum no length scale except the cutoff is present. The running coupling vanishes logarithmically for large distances (see sect. \ref{disordered}), such that the long distance physics is described by a free theory. In the two point correlation $\bar G=\bar A(\vec q^2+iSq_0+Vq^2_0)^{-1}$ the functions $S,\bar A$ can only depend on $q_0/\vec q^2$ and $\lambda$, besides cutoff effects involving $\vec q^2/\Lambda^2$. No coupling $V$ is allowed. A nonzero density sets again a further scale $k_F\sim n^{1/2}$. For momentum scales below $k_F$ the running of $\lambda$ gets modified and the macrophysics is no longer a free theory. In particular, the dimensionless combination $Vn$ will now play an important role.

We will conveniently work with the effective action $\Gamma[\bar{\phi}]$ which generates the $1 PI$ correlation functions. It obtains by introducing local linear sources $j(x)$ for $\chi(x)$ and performing a Legendre transform of $\ln Z[j]$, with $\bar{\phi}(x)=\langle\chi(x)\rangle_{|j}$ in the presence of sources
\begin{eqnarray}\label{19A}
\Gamma[\bar\phi]&=&-\ln Z[j]\\
&&+\int_q\big(\bar\phi^*(q)j(q)+j^*(q)\bar{\phi}(q)\big).\nonumber
\end{eqnarray}

The difference between $S$ and $\Gamma$ results from quantum fluctuations. We include these fluctuation effects stepwise by introducing first an infrared cutoff which suppresses the fluctuations with momenta $\vec{q}\ ^2<k^2$. This is done by adding to the action (\ref{1}) an infrared cutoff term \cite{CWAV}
\begin{equation}\label{24}
\Delta_k {\cal S}=\int_qR_k(\vec{q})\chi^*(q)\chi(q).
\end{equation}
In turn, the effective action is now replaced by the average action $\Gamma_k$ which depends on $k$ \cite{CWFE}, \cite{BTW}. With $R_k(\vec{q})$ diverging for $k\to\infty$ all fluctuations are suppressed in this limit and one finds $\Gamma_{k\to\infty}=S$. On the other hand $R_k(\vec{q})=0$ for $k\to 0$ implies $\Gamma_{k\to 0}=\Gamma$. The average action therefore interpolates smoothly between the classical action for $k\to \infty$ and the effective action for $k\to 0$. Its dependence on $k$ obeys an exact flow equation \cite{CWFE}
\begin{eqnarray}\label{25}
\partial_k\Gamma_k[\bar\phi]=\frac12 {\rm Tr} \{\partial_k {\cal R}
(\Gamma^{(2)}_k[\bar\phi]+{\cal R})^{-1}\}
\end{eqnarray}
with ${\cal R}(q,q')=R_k(\vec{q})\delta(q-q')$. The second functional derivative $\Gamma^{(2)}_k$ is given by the full inverse propagator in the presence of ``background fields'' $\bar{\phi}$. For a homogeneous background field one has $\Gamma^{(2)}_k(q,q')=\bar P(q)\delta(q-q')$ with $\bar P$ a matrix in the space of fields $(\bar{\phi}_1,\bar{\phi}_2)$. The trace involves a momentum integration and a trace over internal indices. Taking functional derivatives of eq. (\ref{25}) yields the flow of all $1$PI-vertices or associated Green's functions. Eq. (\ref{25}) therefore describes infinitely many running couplings. 
For homogeneous background fields $\bar{\phi}$ eq. (\ref{25}) takes the explicit form (with $tr$ the internal trace)
\begin{equation}\label{26}
\partial_k\Gamma_k=\frac{\Omega_{d+1}}{2}tr\int_q\partial_kR_k(\vec{q})
\big(\bar P(q)+R_k(\vec{q})\big)^{-1}.
\end{equation}
The precise shape of the cutoff function $R_k$ is, in principle, arbitrary.

For $\partial_kR_k$ decaying sufficiently fast for large $\vec q^2$ the $\vec q$-integration on the r.h.s. of the flow equation \eqref{25} or \eqref{26} is ultraviolet finite. Instead of an explicite ultraviolet cutoff for the momentum integration, we can therefore define our model by specifying the form of $\Gamma_k$ at some cutoff scale $k=\Lambda$. The short distance physics is now given by the ``initial value'' $\Gamma_\Lambda$. For example, the bare coupling $\bar\lambda$ in the action can be replaced by a coupling $\lambda_\Lambda$, given by the fourth derivative of $\Gamma_\Lambda$. This definition has the advantage that momentum integrals can always be performed over an infinite range. The relation between the action $S$ (with momentum cutoff) and the microscopic effective action $\Gamma_\Lambda$ (without momentum cutoff) can be established by a one loop calculation. (In particular, this absorbs the shift $\Delta\sigma$ in the chemical potential which is generated by the transition from a Hamiltonian formalism to the functional integral \cite{GW}.)

Our task will be to follow the flow of $\Gamma_k$ from an initial value given at $k=\Lambda$ towards $k=0$. From $\Gamma_{k=0}=\Gamma$ the $1$PI-correlation functions of the quantum theory can be extracted by simple functional  differentiation. Despite its conceptually simple one loop form, the exact flow equation (\ref{25}) remains a complicated functional differential equation. For approximate solutions we truncate the most general form of $\Gamma_k$. In the present investigation we will use very simple truncations, involving only a small number of couplings. The minimal set involves only three $k$-dependent renormalized couplings $S,m^2$ and $\lambda$, according to the truncation
\begin{equation}\label{A2}
\Gamma_k=\int_x\left\{\phi^*(S\partial_\tau-\Delta+m^2)\phi+\frac\lambda2(\phi^*\phi)^2\right\}.
\end{equation}
Nevertheless, many characteristic properties of the quantum phase transition in arbitrary dimension $d$ will be accounted for by this truncation. This also holds away from the phase transition for the disordered phase and for the dilute regime of the ordered phase. For the dense regime of the ordered phase in $d=1,2$ we should add a term containing a second $\tau$ derivative as in eq. (\ref{A1}), such that the minimal set consists of four running coupling $m^2,\lambda,S$ and $V$. The coupling $V$ is also needed for $d=3$ if one attempts quantitative accuracy or a correct description of the long distance asymptotics.

In this paper we mainly concentrate on non-relativistic bosons with a linear $\tau$-derivative in the action. The initial value $\Gamma_\Lambda$ is then given by eq. \eqref{A2}, with $S_\Lambda=1~,~m^2_\Lambda=-\tilde\sigma$ and $\lambda=\lambda_\Lambda$. (We will only briefly comment on the more general case where a second $\tau$-derivative $\sim V_\Lambda$ is added to the microscopic action.) The microscopic average action $\Gamma_\Lambda$ defines the model, which has only two parameters in our case, namely the rescaled chemical potential $\tilde\sigma$ and the microscopic interaction strength $\lambda_\Lambda$. This should be distinguished from the truncation of $\Gamma_k$ for $k<\Lambda$. In principle, all couplings allowed by the symmetries will be generated by the flow. This holds even though $\Gamma_\Lambda$ has only two parameters. Restricting $\Gamma_k$ to a finite number of couplings defines the approximation scheme. 

In the formal setting the physical $n$-point functions are only recovered for $k\to 0$. Nevertheless, the properties of $\Gamma_k$ for $k>0$ also admit a physical interpretation. A typical experimental situation has neither infinite volume nor observation devices working at infinite wavelength. This induces a characteristic experimental or ``physical'' momentum scale $k_{ph}$, as  mentioned in the introduction. Formally, this scale appears in the form of nonvanishing ``external'' momenta for the Green's functions which are relevant for a given observation. Often $k_{ph}$ acts as an effective infrared cutoff such that the evolution of these Green's functions (with finite momenta) stops once $k$ becomes smaller than $k_{ph}$. On the other hand, for $k\gg k_{ph}$ the external momenta are not relevant so that one may investigate the Green's functions or appropriate derivatives at zero momentum. In a simplified approach we may therefore associate the Greens-functions derived from $\Gamma_{k=0}$ at finite physical momentum $|\vec q|\approx k_{ph}$ with the Greens function extracted from $\Gamma_{k_{ph}}$ at zero external momentum. In this picture we simply should stop the flow of $\Gamma_k$ at the physical scale $k_{ph}$ rather than considering the limit $k\to 0$. The experimentally relevant Green's functions can then be extracted from $\Gamma_{k_{ph}}$. Of course, such a procedure gives only a rough idea. In general, the Greens-functions will depend on several momenta. Even if only one momentum $\vec q$ is involved, the precise way how the flow is stopped by a physical infrared cutoff involves ``threshold effects'' \cite{BTW}. As a consequence, the proportionality coefficient between $|\vec q|$ and $k_{ph}$ will depend on the particular definition of the $n$-point function. 

The average action $\Gamma_k$ has the same symmetries as the microscopic action, provided one chooses a cutoff $R_k$ consistent with the symmetries. We sketch in appendix G the consequences of Galilei symmetry and local $U(1)$ symmetry for the general form of $\Gamma_k$ at $T=0$ - more details can be found in Ref. \cite{FW}.

\section{Flow equations for pointlike interactions}
\label{floweq}
We first truncate the average action in the pointlike approximation and keep only the lowest time and space derivatives
\begin{eqnarray}\label{aa1}
\Gamma_k&=&\int_x\big\{Z_\phi\bar{\phi}^*\partial_\tau\bar{\phi}-\bar A
\bar{\phi}^*\Delta\bar{\phi}+u(\bar A\bar{\phi}^*\bar \phi)\big\}\nonumber\\
&=&\int_x\big\{S\phi^*\partial_\tau\phi-\phi^*\Delta\phi+u(\phi^*\phi)\big\}.
\end{eqnarray}
Here we have introduced $S(k)=Z_\phi(k)/\bar A(k)$ and the renormalized field
\begin{equation}\label{aa2}
\phi=\bar A^{1/2}\bar{\phi}.
\end{equation}
We use notations where quantities with a bar denote the couplings of the ``unrenormalized field'' $\bar\phi$, whereas the renormalized couplings of the field $\phi$ have no bar. At the scale $k=\Lambda$ one has $\bar A(\Lambda)=\bar A_\Lambda=1$ so that $\bar\phi$ and $\phi$ coincide. In general, the couplings $Z_\phi$ and $\bar A$ are evaluated at a nonzero value of the renormalized field $\phi_0$. As a consequence, they need not to be equal in the ordered phase, even in presence of Galilei symmetry for $T=0$ (cf. app. G). 

We choose the infrared cutoff function \cite{Litim} 
\begin{equation}\label{aa3}
R_k=\bar A(k^2-\vec q\ ^2)\theta(k^2-\vec q\ ^2).
\end{equation}
This cutoff violates Galilei symmetry, but our truncation will neglect counterterms associated to anomalous Ward identities - they vanish for $k\to 0$. 
The initial values of $\Gamma_\Lambda$ will be taken as
\begin{eqnarray}
&&Z_{\phi,\Lambda}=\bar A_\Lambda=1~,~u_\Lambda= m^2_\Lambda \phi^*\phi+\frac12\lambda_\Lambda(\phi^*\phi)^2,\nonumber\\
&& m^2_\Lambda=-\tilde\sigma.
\end{eqnarray}
Besides the rescaled chemical potential $\tilde\sigma$ our model depends on the strength of the repulsive interaction, $\lambda_\Lambda>0$. By a rescaling of the momentum unit $\hat k\to\hat k/\alpha$ the parameters and fields scale as $m^2_\Lambda\to \alpha^2 m^2_\Lambda~,~\lambda_\Lambda\to \alpha^{2-d}\lambda_\Lambda~,~\bar\phi\to\alpha^{\frac d2}\bar\phi$. Physical results for dimensionless quantities can therefore only depend on scaling invariant combinations as $\lambda_\Lambda(m^2_\Lambda)^{\frac{d-2}{2}}$, $\vec q\ ^2/m^2_\Lambda, q_0/m^2_\Lambda, \vec q\ ^2/\Lambda^2$. We notice again the special role of $d=2$ where $\lambda_\Lambda$ is dimensionless.

The phase is determined by the properties of the effective potential $u$ for $k\to 0$ (or $k\to 1/L$ with $L$ the macroscopic size of the experimental probe). In the ordered phase the minimum of $u$ occurs for $\bar\phi_0(k)\neq0$ and one observes spontaneous symmetry breaking (SSB) of the global $U(1)$-symmetry. In contrast, the disordered or symmetric phase (SYM) has $\bar\phi_0=0$. For $d=1$ we will encounter the boundary case where $\bar\phi_0(k)\neq 0$ for arbitrarily small $k$, while $\bar \phi_0(k=0)=0$. (Typically $\bar\phi_0(k)$ vanishes with some power of $k$.) Since many properties of this phase are analogous to the SSB phase for $d>1$ we will use the name ``ordered phase'' also for this case, even though long range order does not exist in a strict sense for the infinite volume limit.

The flow of the average potential $u$ follows by evaluating eq. (\ref{25}) for space- and time-independent $\bar\phi$, i.e. eq. (\ref{26}). We use the fact that the potential depends only on the invariant $\rho=\phi^*\phi$ and $u_k(\rho)=\Gamma_k(\bar\phi)/\Omega_{d+1}$. In our truncation the flow equation reads, using $t=\ln(k/\Lambda)$,
\begin{equation}\label{aa4}
\partial_t u_{|\bar\phi}=\frac12\int_q tr\{\partial_t R_k\bar{G}\}.
\end{equation}
Here the propagator is a $2\times 2$ matrix $\bar{G}=\bar A^{-1}G$,
\begin{equation}\label{aa5}
G^{-1}=\left(
\begin{array}{ccc}
\tilde q^2+u'+2\rho u''&,&-Sq_0\\
Sq_0&,&\tilde q^2+u'\end{array}\right),
\end{equation}
with $\tilde q^2=\vec q\ ^2$ for $\vec q\ ^2>k^2$ and  $\tilde q^2=k^2$ for $\vec q\ ^2<k^2$. Primes denote derivatives with respect to $\rho$. Introducing the anomalous dimension
\begin{equation}\label{aa6}
\eta=-\partial_t\ln\bar A
\end{equation}
we compute in app. A the flow equation for the average potential (at fixed $\phi$ instead of fixed $\bar\phi$) as
\begin{eqnarray}\label{aa9}
\partial_tu&=&\eta\rho u'+\frac{4 v_d}{d S}k^{d+2}\left(1-\frac{\eta}{d+2}\right)\nonumber\\
&&\frac{k^2+u'+\rho u''}{\sqrt{k^2+u'}\sqrt{k^2+u'+2\rho u''}},
\end{eqnarray}
where
\begin{equation}\label{A.7A}
v^{-1}_d=2^{d+1}\pi^{\frac d2}\Gamma\left(\frac d2\right).
\end{equation}
Eq. (\ref{aa9}) is a nonlinear differential equation for a function of two variables $u(\rho,k)$, if $\eta(k)$ and $S(k)$ are known. One may solve equations of this type numerically \cite{AdTetr}. 

We will choose here an even more drastic truncation and use a polynomial expansion around the minimum of $u$. In the symmetric regime the minimum of $u$ is at $\rho=0$ and we approximate
\begin{equation}\label{aa12}
u=m^2\rho+\frac12\lambda\rho^2.
\end{equation}
The corresponding flow equations for $m^2=u'(0)~,~\lambda=u''(0)$ read
\begin{eqnarray}\label{aa13}
\partial_t m^2&=&\eta m^2,\\
\partial_t\lambda&=&2\eta\lambda+
\frac{4v_d}{dS}
\left(1-\frac{\eta}{d+2}\right)
\frac{k^{d+2}}{(k^2+m^2)^2}\lambda^2.\nonumber
\end{eqnarray}
Inspection of eqs. (\ref{aa10}), (\ref{aa11}) shows that the system (\ref{aa13}) is closed and does not involve higher derivatives of the potential as $u^{(3)}$ and $u^{(4)}$. For the SSB regime, with minimum of $u(\rho)$ at $\rho_0\neq 0$, one expands
\begin{equation}\label{aa14}
u=\frac\lambda2(\rho-\rho_0)^2.
\end{equation}
In this case the flow equations for $\rho_0$ and $\lambda$ also involve $u^{(3)}$ and $u^{(4)}$. Neglecting these higher order couplings in our simplest truncation one finds, from $u'(\rho_0)=0~,~u''(\rho_0)=\lambda$, the flow of the minimum
\begin{eqnarray}\label{aa15}
\partial_t\rho_0&=&-\frac1\lambda\partial_t u'(\rho_0)\\
&=&-\eta\rho_0+\frac{2v_d}{dS}
\left(1-\frac{\eta}{d+2}\right)\lambda\rho_0\nonumber\\
&&\frac{k^{d+1}}{\sqrt{k^2+2\lambda \rho_0}}
\left(\frac{1}{k^2}-\frac{3}{k^2+2\lambda\rho_0}\right).\nonumber
\end{eqnarray}
The flow of the quartic coupling obeys now
\begin{eqnarray}\label{aa16}
&&\partial_t\lambda=2\eta\lambda-\frac{2v_d}{dS}\left(1-\frac{\eta}{d+2}\right)\lambda^2
\frac{k^{d+1}}{\sqrt{k^2+2\lambda\rho_0}}\\
&&\left\{\frac{1}{k^2}-\frac{3}{k^2+2\lambda\rho_0}
-\frac32\lambda\rho_0\left(\frac{1}{k^4}-\frac{9}{(k^2+2\lambda\rho_0)^2}\right)\right\}.\nonumber
\end{eqnarray}

In the symmetric regime we find a (partial) fixed point for $m^2=0$, while in the SSB one has a fixed point for $\rho_0=0$. These points coincide, with a quartic potential $u=\frac12\lambda\rho^2$. In turn, the flow for the quartic coupling
\begin{equation}\label{aa17}
\partial_t\lambda=2\eta\lambda+\frac{4v_d}{dS}
\left(1-\frac{\eta}{d+2}\right)k^{d-2}\lambda^2
\end{equation}
has a fixed point for $\lambda=0$, corresponding to a free theory. In order to understand the flow pattern we will need, however, the flow of appropriately rescaled dimensionless quantities and the behavior of $S$ and $\eta$.

For a computation of $\eta$ and $S$ we need the flow of the inverse propagator
\begin{equation}\label{x1}
\left(\Gamma^{(2)}_k\right)_{ab}(q',q'')=\frac{\delta^2\Gamma_k}{\delta\phi^*_a(q')\delta\phi_b(q'')}=
\bar P_{ab}(q')\delta(q'-q'').
\end{equation}
The flow of $\bar P_{ab}$ obtains by the second functional derivative of the exact flow equation (\ref{25})
\begin{eqnarray}\label{x2}
\partial_t\bar P_{ab}(q)&=&\frac12\bar\phi^2\int_{q'}\partial_t R_k(q')(\bar G^2)_{cd}(q')\\
&&\{\gamma_{ade}\bar\gamma_{bfc}\bar G_{ef}(q'+q)+\bar\gamma_{bde}\gamma_{afc}\bar G_{ef}(q'-q)\}\nonumber
\end{eqnarray}
with
\begin{equation}\label{x3}
\bar G(q')=\big(\bar P(q')+R_k(\vec q\ ')\big)^{-1}.
\end{equation}
We have omitted here a term $\sim\Gamma^{(4)}$ which does not contribute to $\eta$ or $S$ in our truncation of momentum independent vertices. The cubic couplings $\bar\phi\gamma$ are specified by
\begin{eqnarray}\label{46}
\frac{\delta\Gamma^{(2)}_{cd}(p',p'')}{\delta\bar\phi^*_a(q')}&=&\gamma_{acd}\bar\phi\delta(p'-p''+q'),
\nonumber\\
\frac{\delta\Gamma^{(2)}_{cd}(p',p'')}{\delta\phi_b(q'')}&=&\bar\gamma_{bcd}\bar\phi\delta(p'-p''-q'')
\end{eqnarray}
and read in our truncation
\begin{eqnarray}\label{b11}
\gamma_{acd}&=&\bar\gamma_{acd}=\sqrt{2}\bar A^2\Big\{u''(\delta_{a1}\delta_{cd}+\delta_{c1}\delta_{bd}+\delta_{d1}\delta_{ac})\nonumber\\
&&+2\rho u^{(3)}\delta_{a1}\delta_{c1}\delta_{d1}\big\}.
\end{eqnarray}
The anomalous dimension $\eta$ and the flow of $S$ are defined by
\begin{equation}\label{x5}
\eta=-\frac{1}{\bar A}\frac{\partial}{(\partial\vec{q}\ ^2)}\partial_t\bar P_{22}(q)_{|q=0}
\end{equation}
and
\begin{equation}\label{x6}
\eta_S=-\partial_t\ln S=-\eta-
\frac{1}{S\bar A}\frac{\partial}{\partial q_0}\partial_t\bar P_{21}(q)_{|q=0}.
\end{equation}

Many qualitative features for arbitrary dimension $d$ can already be seen in the extremely simple truncation of this section. Nevertheless, an important missing ingredient for the dense regime is the  second order $\tau$-derivative $\sim V\phi^*\partial^2_\tau\phi$ discussed in sect. \ref{crossover} and appendix C. We will see that this plays a central role for the infrared behavior, and appropriate  corrections should be included in the flow equation \eqref{aa9}. Beyond this, the extension of the truncation is more a matter of quantitative improvement. The most general pointlike interactions are accounted for by eq. (\ref{aa1}). For example, including in $u$ a term $\sim \rho^3$ describes pointlike six-point vertices, as discussed in app. E.  The leading order in a systematic derivative expansion needs, in addition to $V$, a term $-\frac 14 \bar Y\bar\rho\Delta \bar\rho-\frac14\bar Y_t\bar\rho\partial^2_\tau\bar\rho$ with $\bar\rho=\bar\phi^*\bar\phi$. This contains momentum dependent interactions. The next to leading order in the derivative expansion has $\bar A,\bar S,\bar V$ and $\bar Y,\bar Y_t$ depending on $\bar\rho$. All these approximations have been successfully implemented for ``relativistic'' models with second order $\tau$-derivatives and have led to a precise picture for $O(N)$-models in arbitrary $d$ \cite{PN}, \cite{GKT}. 

\section{Disordered Phase}
\label{disordered}
In the next four sections we discuss the simplest truncation. In the symmetric regime $(\bar{\phi}_0=0)$ the cubic couplings $\sim\bar\phi\gamma$ vanish. From eq. (\ref{x2}) we find in our truncation of momentum independent vertices
\begin{equation}\label{a12a}
\eta=0~,~\eta_S=0.
\end{equation}
In terms of the dimensionless mass term and quartic coupling
\begin{equation}\label{a12b}
w=m^2/k^2~,~\tilde\lambda=\frac{\lambda k^{d-2}}{S}
\end{equation}
we obtain
\begin{eqnarray}\label{a12c}
\partial_t m^2&=&0\quad,\quad\partial_t w=-2 w,\nonumber\\
\partial_t\tilde\lambda&=&(d-2)\tilde\lambda+
\frac{4v_d}{d}(1+w)^{-2}\tilde\lambda^2.
\end{eqnarray}
Since $m^2,\bar A$ and $S$ do not depend on $k$, the quantum propagator $(\Gamma^{(2)})^{-1}$ is given by the classical propagator (for real frequencies $\omega=-iq_0$)
\begin{equation}\label{39A}
G=(-\omega+\vec q\ ^2+m^2_\Lambda)^{-1}.
\end{equation}
The non-renormalization property of $G$ for $T=0,m^2_\Lambda>0$ is believed to be exact since the situation describes the vacuum with zero particle number \cite{SB}. This is also the reason for the closed form of eq. (\ref{a12c}) which does not involve higher order $n$-point functions.

As long as $k^2\gg m^2$ (or $w\ll 1)$ the quartic coupling $\lambda$ runs while for $k^2\ll m^2(w\gg1)$ the running effectively stops. For $d<2$ and $w=0$ the combination $\tilde\lambda$ is attracted towards an infrared fixed point at 
\begin{equation}\label{FFA}
\tilde\lambda_*=\frac{(2-d)d}{4v_d}.
\end{equation}
In the vicinity of this fixed point $\lambda$ decreases with $k$
\begin{equation}\label{a12d}
\lambda\sim\tilde\lambda_* k^{2-d}
\end{equation}
and the repulsive interaction tends to be shielded by the fluctuation effects. For $d>2$ there is no fixed point for $\tilde\lambda\neq 0$. Again $\tilde\lambda$ decreases for $k\to 0$. Now the running of $\lambda$ stops in the infrared even for $w=0$. 

The explicite solution of the flow equation for $\lambda$ in the range $w\ll 1$ reads for $d\neq 2$
\begin{equation}\label{a12e}
\lambda(k)=\lambda_\Lambda\left[1+\frac{4v_d\lambda_\Lambda}{d(d-2)S}
(\Lambda^{d-2}-k^{d-2})\right]^{-1}.
\end{equation}
We note the different behavior for $d>2$ and $d<2$. For $d>2$ the fluctuation effects on $\lambda$ are dominated by the short distance physics, i.e. momenta of the order $\Lambda$ (ultraviolet domination). One expects the precise value of the effective quartic coupling to depend sensitively on the microscopic details. In contrast, for $d<2$ the long-distance physics dominates (infrared domination). For systems with a characteristic physical infrared cutoff $k_{ph}$ the value of the effective coupling is given by $\lambda(k_{ph})$. For $d<2$  the corrections are dominated by the fluctuation effects with infrared momenta $\vec q^2\approx k^2_{ph}$. If the microscopic coupling $\lambda_\Lambda$ is large enough, $\lambda_\Lambda\gg\lambda_{c}(k_{ph})$,
\begin{equation}\label{B11a}
\lambda_{c}(k_{ph})=\frac{(2-d)dS}{4v_d}k^{2-d}_{ph},
\end{equation}
the value of $\lambda_\Lambda$ becomes unimportant
\begin{equation}\label{B11b}
\lambda(k_{ph})\approx\lambda_{c}(k_{ph}).
\end{equation}
The system has lost memory of the microscopic details except for the value of $m^2_\Lambda$. For $w=0$ the value $\lambda_{c}(k_{ph})$ is actually an upper bound for the allowed values of $\lambda(k_{ph})$. For $k_{ph}\to 0$ the model becomes non-interacting, $\lambda(k_{ph})\to 0$. This ``triviality property'' is analogous to the relativistic model, as relevant for the upper bound on the Higgs mass in the standard model of particle physics. For $m^2>0$ one effectively replaces $k^2_{ph}\to c m^2$ with $c$ a proportionality constant of order one. 

The boundary between the qualitatively different role of fluctuations occurs at the ``upper critical dimension'' $d_c=2$. For $d>d_c$ the critical behavior is well approximated by mean field theory, with mean field theory critical exponents. For $d<d_c$ the fixed point behavior (\ref{FFA}) influences the critical physics as far as the interaction strength is concerned. At the critical dimension $d=2$ the running of $\lambda$ for $w\ll 1$ becomes logarithmic
\begin{equation}\label{a12f}
\lambda(k)=\lambda_\Lambda\left[1+\frac{\lambda_\Lambda}{4\pi S}\ln\frac{\Lambda}{k}\right]^{-1}.
\end{equation}

\section{Scaling solutions and quantum phase transition}
\label{scaling}
It is instructive to discuss the critical behavior in terms of the scaling solutions. Possible scaling solutions correspond to the fixed points for $w$ and $\tilde\lambda$, i.e. to values where both $\partial_t w$ and $\partial_t\tilde\lambda$ vanish. For all $d$ one has the trivial fixed point
\begin{equation}\label{A12a}
(A):\quad w_*=0~,~\lambda_*=0.
\end{equation}
Small deviations from this fixed point, with $w>0$, grow for $k\to0$. The fixed point $(A)$ is unstable in the $w$-direction, thus $w$ (or $m^2$) is a relevant parameter. For $d>2$ fixed point $(A)$ is infrared stable in the $\tilde\lambda$-direction, i.e. $\tilde\lambda$ is an irrelevant coupling. However, for $d<2$ $\tilde\lambda$ the coupling  becomes a relevant parameter, too. Fixed point $(A)$ has two IR-unstable directions for $d<2$. Indeed, the flow of $\tilde\lambda$ is attracted towards a second fixed point
\begin{equation}\label{A12b}
(B):\quad \tilde w=0~,~\tilde\lambda=\tilde\lambda_*,
\end{equation}
with $\tilde\lambda_*$ given by eq. (\ref{FFA}). Fixed point $(B)$ has only one relevant parameter $w$ whereas $\tilde\lambda$ becomes irrelevant. The critical behavior is dominated by fixed point $(B)$, except for very small $\lambda_\Lambda$ where one observes a ``crossover'' of the flow from the vicinity of $(A)$ towards $(B)$. Both fixed points $(A)$ and $(B)$ are located exactly on the phase transition $m^2=0$. 

The value of $\tilde\lambda$ does not affect the flow of $w$ or the anomalous dimension $\eta$ or $\eta_S$. We therefore find for the symmetric phase a mean field critical behavior for $m^2_\Lambda\to 0$. This equally applies for both fixed points $(A)$ and $(B)$ which are distinguished only by the value of $\lambda$. There is no running of $m^2$ and the anomalous dimension $\eta$ as well as $\eta_S$ vanish. The correlation length $\xi=m^{-1}(k\to 0)$ simply obeys  
\begin{equation}\label{a12g}
\xi=\frac{1}{m(k\to 0)}
=\frac{1}{m_\Lambda}=|\tilde \sigma|^{-1/2}=|\tilde\sigma|^{-\nu}
\end{equation}
and the correlation time (for $m^2_\Lambda>0)$ is given by
\begin{equation}\label{140A}
\tau_c=\frac{1}{m^2_\Lambda}=|\tilde\sigma|^{-1}=\xi^2=\xi^z.
\end{equation}
The time averaged correlation function for $m^2_\Lambda=0$ decays according to the canonical dimension $(d>2)$
\begin{equation}\label{a12h}
\langle\bar\phi^*(\vec r)\bar\phi(0)\rangle\sim |\vec r|^{2-d},
\end{equation}
as given by the $d$-dimensional Fourier-transform of eq. (\ref{39A}) for $m^2_\Lambda=0, \omega=0$.

The corresponding critical exponents are the mean field exponents \cite{UZ}, \cite{SB}
\begin{equation}\label{140B}
\nu=\frac12~,~\eta=0~,~z=2.
\end{equation}
In the present case, the critical exponents follow from naive scaling arguments. More generally, the critical exponent $\eta$ corresponds to the anomalous dimension for the scaling solution. Indeed, if we evaluate the propagator for $\vec q\ ^2>0$, the external momentum acts like an infrared cutoff $(|\vec q|\sim k_{ph})$, such that $\bar A\sim k^{-\eta}\to(\vec q\ ^2)^{-\eta/2}$. At the phase transition the static propagator $(q_0=0)$ behaves as  $\bar G=G/\bar A\sim(\vec q\ ^2)^{-1+\eta/2}$, which is precisely the definition of the critical exponent $\eta$. 

The value of $\eta_S$ for the scaling solution determines the dynamical critical exponent $z$,
\begin{equation}\label{FFA1}
z=2+\eta_S.
\end{equation}
The dynamical critical exponent $z$ relates the $\vec q^2$-dependence and the $q_0$-dependence of the renormalized inverse propagator away from the phase transition
\begin{eqnarray}\label{f4a}
&&G^{-1}(q_0=0,\vec q)=\vec q\ ^2+\xi^{-2},\\
&&G^{-1}(q_0,\vec q=0)=iS(q_0)q_0+\xi^{-2}=i\tilde cq_0^{2/z}+\xi^{-2}.\nonumber
\end{eqnarray}
If the zeros of $G^{-1}(q_0)$ occur for a value of $q_0$ with positive real part, $Re(q_0)=\tau^{-1}_c$,  the correlation function for real time $t$ decays exponentially with a typically dissipation time $\tau_c$, implying for $\vec q\ ^2\ll\xi^{-2}$ 
\begin{equation}\label{f4b}
\langle\varphi(t,\vec q)\varphi^*(0,\vec q)\rangle\sim \exp (-t/\tau_c).
\end{equation}
Assuming that for the zero of $G^{-1}$ one has $Re(q_0)\sim Im(q_0)$ one can relate the dissipation time $\tau_c$ to the correlation length $\xi$
\begin{equation}\label{f4c}
(\tau_c)^{2/z}\sim\xi^2~,~\tau_c\sim\xi^z.
\end{equation}

A nonzero external $q_0$ will replace the infrared cutoff in the propagator, $k^2\to S(q_0)q_0$, such that 
\begin{equation}\label{f4d}
S(q_0)q_0\sim k^{-\eta_S}q_0\to [S(q_0)q_0]^{-\eta_S/2}q_0.
\end{equation}
The scaling
\begin{equation}\label{f4e}
\big(S(q_0)q_0)^{\frac{2+\eta_S}{2}}\sim q_0~,~S(q_0)\sim q_0^{-\frac{\eta_S}{2+\eta_S}}
\sim q_0^{\frac 2z-1}
\end{equation}
yields the relation (\ref{FFA1}) between $z$ and $\eta_S$. A simpler argument compares the scaling of a characteristic $\hat q_0$ with $k~,~\hat q_0\sim k^z$, where $\hat q_0$ is determined such that the $q_0$-dependent part in $G^{-1}$ has the same size as the IR cutoff $k^2$
\begin{equation}\label{f4f}
S(k)\hat q_0\sim k^2\sim k^{-\eta_S}\hat q_0~,~\hat q_0\sim k^{2+\eta_S}\sim k^z.
\end{equation}
This yields, of course, the same relation (\ref{FFA1}). 

Obviously, the scaling arguments leading to eq. \eqref{FFA1} depend strongly on the absence of any other relevant cutoff. They will not be valid for dense systems where $\sqrt{2\lambda\rho_0}$ constitutes an important infrared cutoff for the radial fluctuations. For the dense systens we find for all $d$ a small momentum behavior $G^{-1}\sim(\vec q^2+q^2_0/v^2)$, such that the same type of scaling arguments yields $z=1$. 

For $d>2$ mean field theory is expected to be a valid approximation. For $d<2$, however, the strong dependence of $\lambda$ on $k$ will result in a momentum dependence of the effective vertex, with $k$ replaced by 
$\sqrt{\vec p\ ^2}$, and $\vec p$ a characteristic external momentum of the vertex. The approximation of a pointlike interaction becomes inaccurate and one may question the validity of a mean field tretament. Nevertheless, relations $\partial_tm^2=0,\eta=\eta_S=0$ continue to hold (cf. eq. (\ref{39A})), implying the mean field critical exponents (\ref{140B}) for all $d$. Also the equation for a momentum dependent quartic coupling will remain closed. Only the value of $\tilde\lambda_*$ and the precise evolution of the quartic coupling $\lambda$ will be modified for extended truncations. For $d=1$ and $\lambda_\Lambda\to\infty$ our model corresponds to ``hard core bosons''. In $d=1$ this is equivalent to a model of free spinless fermions and the universality class for  fixed point $(B)$ is therefore known \cite{SB}, confirming that  eq. (\ref{140B}) is exact for $d=1$. 

The quantum phase transition at $m^2_\Lambda=0$ is the only phase transition that we discuss explicitely in the present paper. Its scaling properties in the symmetric phase are quite simple. Our functional renormalization group equations account well for these scaling properties, establishing them as a reasonable starting point for $T>0$ in a straightforward generalization where  the $q_0$-integration in the appendices $A$ and $B$ is replaced by a Matubara sum. The simple features of the quantum phase transition discussed in this paper, namely the location exactly at $n=0$, the non-renormalization of $m^2$ and the vanishing $\eta$ and $\eta_S$, are all particular for the non-relativistic models with $\bar V=0$. (We will see in \ref{crossover} that $V=0$ is stable with respect to the flow in the symmetric phase.) Starting with $\bar V\neq 0$ will change all these properties - for example, the relativistic model with $\bar V\neq 0~,~\bar S=0$ shares none of them. We also do not investigate here the possibility that a very strong repulsion annihilates the order even for $\bar V=0~,~T=0~,~n\neq 0$. (This would lead to a different type of quantum phase transition at some critical value of $\lambda_\Lambda$, perhaps characterized by the critical behavior of the relativistic $O(2M)$- or $U(M)$-models.)

We close this section by a remark that a line of fixed points exist for all values $m^2_\Lambda>0$. The associated scaling solutions reflect, however, a different scaling behavior \cite{SB}. Indeed, for $k^2\ll m^2$ we may use the variables $m^2$ and 
\begin{equation}\label{51A}
\hat\lambda=\frac{k^{d+2}}{m^4 S}\lambda.
\end{equation}
From eq. (\ref{aa13}) and for $\eta=\eta_S=0$ we extract 
\begin{equation}\label{51B}
\partial_t\hat\lambda=(d+2)\hat\lambda +\frac{4v_d}{d}\hat\lambda^2
\end{equation}
and observe that an infrared stable fixed point $\hat\lambda=0$ exists for all $d>0$. As $k^2$ crosses the ``threshold'' $m^2$ the flow of $\lambda$ shows a crossover from fixed point $(B)$ (or $(A)$) to the fixed point of eq. (\ref{51B}). This is, of course, a fancy way of stating that $\lambda$ stops running. 

\section{Ordered Phase}
\label{orderd}
We next turn to the ordered phase.  This will be characterized by a richer spectrum of physical phenomena, since even for $T=0$ the particle density is nonvanishing. We will see that for $d<2$ the long distance physics is always characterized by an effective theory with strong interactions. The quantum phase transition to the disordered phase remains simple for $d=3$ and small coupling since fluctuation effects play a minor role. Such a simple description also applies for $d\leq 2$ in the dilute regime, as long as the momenta and energies of the process considered are larger than a characteristic momentum $k_F$ or a characteristic energy $\epsilon_F$ related to the density. (In our normalization $\epsilon_F=k^2_F$.) For smaller momenta and energies, however, the density $n$ sets a new scale
\begin{equation}\label{51C}
k_F=\left(\frac{dn}{8v_d}\right)^{1/d}.
\end{equation}
(The normalization of $k_F\sim n^{1/d}$ is somewhat arbitrary and we have chosen it here in analogy with a Fermi gas of particles with spin $1/2$.) One expects that this scale strongly influences the long distance behavior. For $d=1,2$ one finds qualitatively new phenomena whenever the ``physical momentum'' $k_{ph}$ is smaller than $k_F$ (dense regime). For $d=3$ and small couplings these effects may matter only on exponentially small scales, since the running of dimensionless couplings is logarithmic. 

The new physics for non zero density is directly related  to the possibility of a condensate. The flow equations will be influenced by the ``renormalized order parameter'' $\rho_0(k)>0$, which denotes the value of $\rho$ for which the average potential $U_k$ is minimal. For $T=0$ the symmetries relate the asymptotic value $\rho_0=\rho_0(k=0)$ to the density, cf. app. G,
\be\label{62A}
n=\rho_0.
\ee
(For dimensionless fields this reads $n=\rho_0\hat k^d$.) In terms of the original fields $\bar\phi$ the asymptotic order parameter $\bar\rho_0=\bar\rho_0(k=0)$ denotes the condensate density $n_c$
\be\label{62B}
n_c=\bar\rho_0,
\ee
such that the condensate fraction $\Omega_c$ reads
\be\label{62C}
\Omega_c=\frac{n_c}{n}=
\frac{\bar\rho_0}{\rho_0}=\bar A^{-1}.
\ee
We will encounter the notion of a ``local condensate'' $\bar\rho_0(k)$, even if no long range order exists, i.e. if $\bar\phi_0(k\to 0)=0$. In this perspective $k$ can be associated with the inverse size of a domain and $\bar\phi_0(k)$ measures the expectation value of the order parameter in such a domain. We refer to the ``SSB-{\em regime}'' of the flow whenever $\rho_0(k)\neq 0$. The ordered {\em phase} or the phase with spontaneous symmetry breaking (synonymous) are characterized by a nonzero $\rho_0$ at the end of the running, i.e. for $k=0$ or $L^{-1}$. Technically, the running of the couplings in the SSB-regime is  more involved due to the presence of the cubic couplings for $\bar{\phi}_0\neq 0$.

In the SSB-regime the two modes $\phi_1$ and $\phi_2$ show a different behavior. With an expectation value $\phi_0$ in the $1$-direction, $\phi_1$ denotes the radial mode (longitudinal mode) which is typically ``massive'' or ``gapped'', with ``mass term'' $2\lambda\rho_0$. In contrast, the ``Goldstone mode'' (transverse mode) $\phi_2$ is massless. In  the SSB-regime, the relative size of the contributions from the Goldstone and radial modes is governed by the dimensionless ratio
\begin{equation}\label{A11a}
w=\frac{2\lambda\rho_0}{k^2}.
\end{equation}
In eq. (\ref{aa15}), (\ref{aa16}) we note that for $w=2$ the flow of the unrenormalized parameters $\bar\rho_0=\rho/\bar A, ~\bar\lambda=\bar A^2\lambda$ vanishes. For $w<2$ (and $\eta<d+2)$ one finds for $k\to 0$ an increasing $\bar\rho_0$ and increasing $\bar\lambda$, whereas for $w>2$ both quantities decrease. 

Within the SSB-regime we distinguish between two limiting cases. The ``linear regime'' refers to $w\ll 1$ where both the radial and Goldstone mode are equally important. In contrast, in the ``Goldstone regime'' for $w\gg 1$ the radial mode plays a subleading role and the dominant physics is related to the behavior of the Goldstone modes. 

From eqs. \eqref{aa15}, \eqref{aa16} we extract the flow of $w$ 
\begin{eqnarray}\label{A11b}
\partial_tw&=&w\left\{-2+\eta+\frac{3v_d}{2d}\left(1-\frac{\eta}{d+2}\right)
\frac{\tilde\lambda w}{\sqrt{1+w}}\right.\nonumber\\
&&\left.\left(1-\frac{3}{1+w}\right)\left(1+\frac{3}{1+w}\right)\right\}.
\end{eqnarray}
In eq. (\ref{A11b}) we encounter again the dimensionless combination
$\tilde{\lambda}={\lambda k^{d-2}}/{S}$.
Its evolution obeys $(\eta=-\partial_t\ln \bar A,\eta_S=-\partial_t\ln S)$
\begin{eqnarray}\label{A112}
\partial_t\tilde{\lambda}&=&(d-2+2\eta+\eta_S)\tilde{\lambda}\nonumber\\
&&-\frac{2v_d}{d}\left(1-\frac{\eta}{d+2}\right)
\frac{1}{\sqrt{1+w}}\left(1-\frac{3}{1+w}\right)\nonumber\\
&&\frac{1-2w-\frac34w^2}{1+w}\tilde{\lambda}^2.
\end{eqnarray}

The anomalous dimension $\eta$ is computed in appendix B and we find in our truncation 
\begin{equation}\label{z10}
\eta=\frac{2v_d}{d}\tilde\lambda w(1+w)^{-3/2}.
\end{equation}
It vanishes both for $w\to0$ and $w\to\infty$. For $\eta_S$ we find (app. B) 
\begin{equation}\label{B14a}
\eta_S=-\eta+\frac{v_d}{2d}\left(1-\frac{\eta}{d+2}\right)
\tilde\lambda w\frac{8-4w-3w^2}{(1+w)^{5/2}}.
\end{equation}
We note that the leading term for large $w$
\begin{equation}\label{y3}
\eta_S =-\frac{3v_d}{2d}
\left(1-\frac{\eta}{d+2}\right)\tilde\lambda w^{1/2}
\end{equation}
cancels in the flow of $\tilde\lambda$ the term $\sim\tilde\lambda^2w^{1/2}$.

We note that $\eta_S$ can take large negative values. In this context we emphasize that the relation between $\eta_S$ and the dynamical critical exponent $z$ in eq. (\ref{FFA1}) holds only as long as the first order $\tau$-derivative dominates. For $V\neq 0$ and $\eta_S<-1$ the relativistic dynamic term will dominate, yielding simply $z=1$. Furthermore, in the ordered phase Goldstone bosons dominate the correlation function at large distances in space or time. We will see in sect. X that the decay of the correlation function is powerlike, 
$\bar G^{-1}\sim \big[\vec q\ ^2+(q_0/v)^2\big]^{1-\eta/2}$, leading effectively to $z=1$, independently of all other details. We observe an apparent clash with the critical exponent $z=2$ in eq. \eqref{140B}.

We may insert our results for $\eta$ and $\eta_S$ into eqs. (\ref{A11b}), (\ref{A112}) 
\begin{eqnarray}\label{Q1}
\partial_tw&=&w\left\{-2+\frac{v_d}{2d}\frac{\tilde\lambda w}{\sqrt{1+w}}
\left(3+\frac{4}{1+w}-\frac{27}{(1+w)^2}\right)\right.\nonumber\\
&&\left.-\frac{3v^2_d}{d^2(d+2)}\frac{\tilde\lambda^2 w^2}{(1+w)^2}
\left(1-\frac{9}{(1+w)^2}\right)\right\},\nonumber\\
\partial_t\tilde\lambda&=&\tilde\lambda\Big\{d-2+\frac{v_d}{d}\tilde{\lambda}(2-w)^2(1+w)^{-5/2}
\nonumber\\
&&-\frac{2v^2_d\tilde\lambda^2 w}{d^2(d+2)}
\frac{4-6w-w^2}{(1+w)^4}\Big\}.
\end{eqnarray}
These two coupled nonlinear differential equations for the two couplings $w$ and $\tilde\lambda$ already yield several characteristic features of the ordered phase for arbitrary $d$. However, the understanding of the dense regime requires an extension of the truncation by the ``relativistic dynamic term'' $\sim V$. This will be necessary in order to get the correct behavior for $w\to \infty$ (sect. IX). 

\section{Linear and Goldstone regimes}
\label{linearandgoldstone}
It is instructive to consider first the linear regime in the limit $w\to0$, where
\begin{eqnarray}\label{Q3}
\partial_tw&=&-2w\left[1+\frac{5v_d}{d}\tilde\lambda w-\frac{12 v^2_d}{d^2(d+2)}(\tilde\lambda w)^2\right]+\dots,\nonumber\\
\partial_t\tilde\lambda&=&\tilde\lambda\left[d-2+
\frac{4v_d}{d}\tilde\lambda-\frac{8v^2_d}{d^2(d+2)}\tilde\lambda^2w\right]+\dots
\end{eqnarray}
For $k\to 0$ one finds that $w$ increases. The dimensionless interaction strength $\tilde\lambda$ decreases for $d\geq 2$, while it increases for $d<2$ and small $\tilde\lambda$. For $w=0$ we recover the two fixed points $(A)$ and $(B)$ already found in the symmetric phase. As before, $(A)$ is IR-stable in the $\tilde\lambda$-direction for $d>2$ and unstable for $d<2$. Fixed point $(B)$ at 
$w_*=0~,~\tilde\lambda_*=(2-d)d/(4v_d)$ exists for $d<2$ and is IR-stable in the $\tilde\lambda$-direction. For both fixed points $w$ is a relevant parameter. For fixed point $(B)$ we find for $k\to 0$ that 
$2\lambda\rho_0=w k^2=W$ approaches a constant, as well as $S$ and $\bar A$. Since $\lambda\to S\tilde\lambda_* k^{2-d}$, we find that $\rho_0(k)$ increases for $d<2$ and $w\ll1$ according to 
\begin{equation}\label{A14a}
\rho_0(k)=\frac{W}{2S\tilde\lambda_*}k^{d-2}.
\end{equation}
This behavior stops once $w$ reaches a value of the order one. 

The quantum phase transition occurs for $w=0$. The critical behavior is characterized  by fixed point $(A)$ for $d>2$ and $(B)$ for $d<2$. Since the fixed points are the same for the ordered and disordered phases we also obtain the same scaling behavior. At this point everything may look very simple.

A closer look at the ordered phase reveals, however, that fixed point $(A)$ or $(B)$ cannot describe all aspects of the quantum phase transition. It is not clear how the exponents $\nu,\eta$ and $\eta_S$ should be defined in the ordered phase. The correlation length $\xi$ could be defined in the radial direction, $\xi_R=(2\lambda\rho_0)^{-1/2}$ where $\lambda\rho_0$ should be evaluated for $k=\xi^{-1}_R$. In the Goldstone direction, however, the correlation length is infinite. We emphasize that the correlation function for the complex field $\bar\phi$ is dominated by the propagator for Goldstone bosons $\bar G_{22}$ which does not exhibit a finite correlation length but rather shows a powerlike decay for large $|\vec x|$, 
\begin{eqnarray}\label{C6a}
&&\langle\bar \phi^*(x)\bar\phi(0)\rangle -\bar\phi^2_0\nonumber\\
&&=\frac12 \langle\bar\phi_1(x)\bar\phi_1(0)\rangle-\bar\phi^2_0+\frac12
\langle\bar\phi_2(x)\bar\phi_2(0)\rangle\nonumber\\
&&=\frac12\big(\bar G_{11}(x)+\bar G_{22}(x)\big).
\end{eqnarray}
Since $\bar G_{11}$ decays faster than $\bar G_{22}$ only the latter matters for $|\vec x|\to\infty$.

Similarly, Goldstone bosons dominate the occupation number (\ref{4}) for the small momentum modes
\begin{equation}\label{C6b}
n(\vec q)=\hat k^d\left[\bar\phi^2_0\delta(\vec q)+\frac12 \int_{q_0}\big(\bar G_{22}
(q_0,\vec q)+\bar G_{11}(q_0,\vec q)\big)\right]-\frac12.
\end{equation}
We will see that the shape of $\bar G_{22}$ for small $\vec q\ ^2$ and $q_0$ can become nontrivial and is no longer governed by the ``quantum critical fixed point'' (A) or (B). 

We may also study the critical behavior of the (bare) order parameter $\bar\rho_0(k\to 0)\sim \tilde\sigma^{\beta/2}$. The flow for $k\to 0$ will necessarily involve the flow in the region of large $w$ and one may wonder if this can be described by the fixed points $(A)$ or $(B)$ anymore. We will even find that for $d=1$ the order vanishes in a strict sense, $\bar\rho_0(k\to0)\to0$. The definition of $\beta$ seems not to be meaningful anymore. These particularities of the correlation length and the bare order parameter cannot be explained by simple extrapolations from  the fixed points $(A)$, $(B)$ for which  Goldstone bosons play no particular role.

What actually happens is a crossover phenomenon between the scaling associated to fixed points $(A)$ or $(B)$ for the quantum phase transition and the Goldstone regime where the gapless Goldstone mode dominates. This crossover depends on the scale $k_{ph}$ of characteristic momenta of an experiment. The crucial quantity is the ratio
\begin{equation}\label{D6a}
w(k_{ph})=\frac{2\lambda(k_{ph})\rho_0(k_{ph})}{k^2_{ph}}.
\end{equation}
Only for $w(k_{ph})\lesssim 1$ are the scaling laws of the quantum phase transition  given by the fixed points $(A)$ or $(B)$. In the opposite limit one has to explore the Goldstone regime $w\gg 1$. We discuss in appendix F that $\sqrt{2\lambda\rho_0}$ plays the role of the momentum scale $k_F\sim n^{1/d}$ associated to the density. The linear regime therefore applies for dilute systems, while the Goldstone regime is relevant for dense systems. We notice that for any small nonzero order parameter $\rho_0$ (corresponding to a situation near the phase transition) there is always a range of very small momenta $k_{ph}$ such that $w(k_{ph})\gg 1$. The extreme long range behavior is always dominated by the physics of Goldstone bosons for which the fixed points $(A)$, $(B)$ are not revelant. This also matters in practice since the macroscopic size of an experiment corresponds to very small $k_{ph}$. 

We show the different regimes which are relevant for the quantum phase transition in Fig. 1. The ``linear'' regime is governed by the fixed points $(A)$ or $(B)$. In the disordered phase this finds a simple extension to the ``massive regime'' where the flow simply stops due to the presence of an infrared cutoff $\sim m$, cf. eq. \eqref{51B}. In the ordered phase, however, the ``Goldstone regime'' is qualitatively different. The long distance physics shows new features which cannot be explained by the fixed points $(A)$, $(B)$. We note that for a fixed nonzero $k_{ph}$ the immediate vicinity of the phase transition for $\sigma\to 0$ is always governed by the quantum critical point, while for fixed $\sigma>0$ one always enters the Goldstone regime as $k_{ph}\to 0$.

\begin{figure}[htb]
\centering
\includegraphics[scale=1.5]{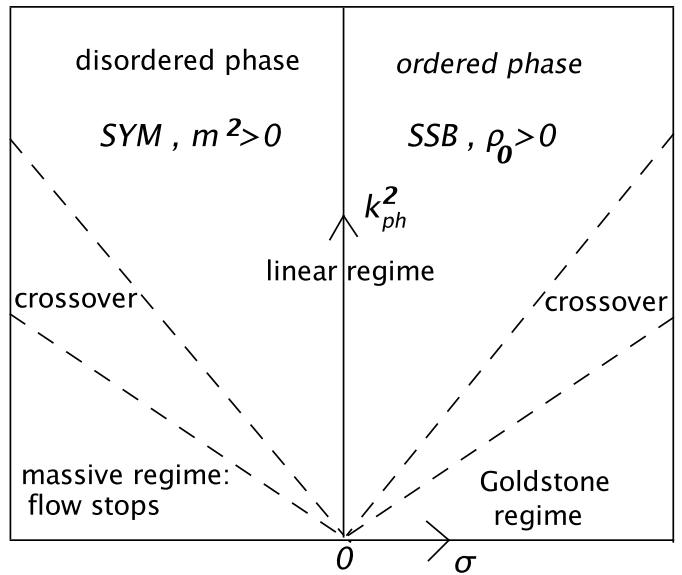}
\caption{Schematic view of different regimes for quantum phase transitions. Dense systems are described by the Goldstone regime.}
\label{functional-ren-fig1}
\end{figure}

We will next explore the Goldstone regime. This will bring us to the most important results of this paper. For the qualitative behavior of the flow away from the exact location of the phase transition $(\tilde\sigma\neq0)$ we observe an important difference between the disordered and ordered phase. In the disordered phase the running of the couplings $m^2, \lambda$ simply stops once $k\xi\ll 1$ and the consequences of scaling are immediate. In the ordered phase, however, we encounter the massless Goldstone fluctuations at all scales, including $k\xi_R\ll 1$. Correspondingly, the flow equations in the regime $w\gg 1$ will be nontrivial and we should explore their consequences. 

We first work with our simplest truncation and extend it subsequently in the following sections. Within the truncation (\ref{aa1}), (\ref{aa14}) we will find a new fixed point of eq. (\ref{Q1}) for $d<2$ and a nontrivial scaling behavior for $2<d<3$. As more couplings are included in extended truncations we find that the fixed point persists for $d=1$, while it turns out to be an artefact of the truncation for $d>1$. The precise properties of the fixed point are quite sensitive to the truncation, and the ``lowest order results'' of the simplest truncation have to be interpreted with care. Indeed, for $0<d<2$ the flow equations in the ordered phase \eqref{Q1} exhibit an additional fixed point for $w_*\neq 0$
\begin{equation}\label{A14b}
(C):\quad w_*\neq 0~,~\tilde\lambda_*\neq 0.
\end{equation}
The characteristic fixed point values obtained in this truncation by a numerical solution of eq. (\ref{Q1}) are shown in table I. In appendix E we have performed a similar computation including a local six point vertex in the truncation. Comparison of tables I and II reveals a very strong truncation dependence for $d$ near two, while the results are more robust for $d=1$. 

\medskip
\noindent
\begin{tabular}{|l|l|l|l|l|l|l|}
\hline
$d$&$w_*$&$\tilde\lambda_*$&$\eta$&$\eta_S$&$t_*$&$d+\eta_S+\eta$\\\hline
$1$&$2.257$&$28.3$&$1.73$&$-2.65$&$-8$&$0.08$\\
$1.9$&$7.33$&$23.8$&$0.335$&$-2.012$&$-80$&$0.23$\\
$1.99$&$127.3$&$5.85$&$0.021$&$-1.99$&$-500$&$0.021$\\\hline
\end{tabular}

\medskip
\noindent
table I: Fixed point values for $(C)$

\medskip
Fixed point $(C)$ is infrared attractive in all directions. Within the restricted space of couplings considered in our simple truncation this is an example of ``self-organized criticality''. For $0<d<2$ the flow for any initial value $m^2_\Lambda<0~,~\lambda_\Lambda>0$ will finally end in fixed point $(C)$. In table 1 we also indicate a characteristic value of $t=\ln(k/\Lambda)$ for which the fixed point is reached (for initial $w$ and $\tilde\lambda$ of the order one). As $d$ approaches two the fixed point behavior sets in only at extremely large distances.

Since for fixed point $(C)$ $w$ and $\tilde\lambda$ take constant values one finds in the simplest truncation
\begin{equation}\label{A14c}
\rho_0(k)=\frac{w_*}{2\tilde\lambda_*}\frac{k^d}{S(k)}~,~\lambda(k)=\tilde\lambda_*S(k)k^{2-d}.
\end{equation}
With 
\begin{equation}\label{A14d}
S=S_0\left(\frac{k}{k_0}\right)^{-\eta_S}~,~\bar A=A_0\left(\frac{k}{k_0}\right)^{-\eta}
\end{equation}
we infer
\begin{eqnarray}\label{A14e}
\rho_0\sim k^{d+\eta_S}~,~\bar\rho_0\sim k^{d+\eta_S+\eta},\nonumber\\
\lambda\sim k^{2-d-\eta_S}~,~\bar\lambda\sim k^{2-d-\eta_S-2\eta}.
\end{eqnarray}
For $\eta_S<-d$ the renormalized order parameter $\rho_0(k)$ increases with $k$, while for $d+\eta_S+\eta>0$ the bare order parameter $\bar \rho_0=\bar A^{-1}\rho_0$ vanishes for $k\to 0$. From the values of $d+\eta_S+\eta$ in table I one would infer that no long range order is present for $d<2$. (For the one-dimensional boson gas we find in this simple truncation that $\bar\rho_0$ vanishes $\sim k^{0.08}$.) Then there is no meaningful definition of the critical exponent $\beta$ for $d=1$. Also
\begin{equation}\label{A5a}
\xi_R(k)=\big(2\lambda(k)\rho_0(k)\big)^{-1/2}=\frac{1}{\sqrt{w_*}k}
\end{equation}
always diverges for $k\to 0$, due to the existence of the fixed point for $d<2$. For such a behavior there would be  no meaningful definition of a correlation length even for the radial mode, due to the strong impact of Goldstone fluctuations. 

Within our simplest truncation one would conclude that for $d=1$ a quantum phase transition exists, but the high density phase actually shows no long range order in a strict sense. It exhibits a powerlike decay of the correlation functions both for the radial and Goldstone modes. We may still call this phase an ``ordered phase'' in a somewhat weaker notion: The renormalized order parameter $\rho_0(k)$ does not vanish, implying the distinction between Goldstone and radial modes and several other features characteristic for an ordered phase. Also the order parameter $\bar\rho_0$ vanishes only asymptotically for $k\to0$. For a system with a characteristic infrared cutoff $k_{ph}\neq 0$ one can effectively observe order. A similar behavior has been found \cite{GKT} for classical phase transitions, e.g. the Kosterlitz-Thouless \cite{KT} phase transition.

For $d=1$ one expects for the ordered phase a behavior similar to a Tomonaga-Luttinger liquid \cite{TL} with dynamical exponent $z=1$ and a correlation function
\begin{eqnarray}\label{A5c}
&&\langle\bar\varphi^*(q_0,\vec q)\bar\varphi(q'_0,\vec q\ ')\rangle \sim
\big((q_0/v)^2+\vec q\ ^2\big)^{-\left(1-\frac\eta2\right)}\delta(q-q'),\nonumber\\
&&\langle\bar\varphi^*(\tau,\vec r)\bar\varphi(0,0\rangle\sim
(v^2\tau^2+\vec r^2)^{-\frac\eta2}.
\end{eqnarray}
The relativistic form of the propagator suggests that the ``relativistic kinetic term'' involving two $\partial_\tau$-derivatives should not be neglected for low dimensions. We therefore will enlarge our truncation and include the coupling $V$ in sect. \ref{crossover}. This modifies the qualitative characteristics for the flow in the Goldstone regime for $d=1,2$. For $d=2$ we will find that both $\rho_0$ and $\bar\rho_0$ settle to constant values as $k\to 0$. The fixed point $(C)$ disappears - it is an artefact of a too simple truncation. For $d=1$ we find indeed a relativistic correlation function (\ref{A5c}) with $z=1$. The flow shows again a (shifted) fixed point $(C)$, constant $w$ and $\rho_0$ and $\lambda\sim k^2~,~\bar\rho_0\sim k^\eta$. 

The qualitative new features induced by the coupling $V$ limit the direct use of fixed point $(C)$ in the simplest truncation (which neglects $V$). Nevertheless, the properties of the flow equation (\ref{Q1}) remain interesting in several aspects. One concerns the ``initial flow'' before a substantial relativistic kinetic term $\sim V$ has been generated. We discuss a few details of fixed point $(C)$ for the system (\ref{Q1}) in appendix D.

Let us finally briefly explore the behavior of eq. \eqref{Q1} for large $w$ - details can be found in appendix F. For $d > 3$ one finds that the flow of $\rho_0$ and $\lambda$ stops as $w\sim k^{-2}$ grows to large values for $k\to 0$. Also the anomalous dimensions $\eta$ and $\eta_S$ vanish. For $d<3$ the flow of the combination $\lambda w^{1/2}$ is attracted towards a partial fixed point. Again, the asymptotic behavior behavior for $k\to 0$ is characterized by constant $\rho_0$ and $\bar A~,~\eta\to 0$. However, one now finds asymptotically vanishing $\lambda\sim S\sim k^{-\eta_S}~,~\eta_S=2(d-3)$. For $d<2$ an initially very large value of $w$ decreases, consistent with the attractor property of fixed point $(C)$.

\section{Crossover to relativistic models for low dimensions}
\label{crossover}
For the Goldstone regime in $d=1$ and $d=2$ an important qualitative shortcoming of our simplest truncation becomes visible if we include the term with two time derivatives in an extended truncation
\begin{equation}\label{R1}
\Gamma_V=-V\int_x\phi^*\partial^2_\tau\phi.
\end{equation}
A nonvanishing coupling $V$ will always be generated by the flow of $\Gamma_k$ in the SSB regime, even if one starts with $V=0$ in the ``classical action'' at the microscopic scale $\Lambda$. This contrasts with the symmetric regime, relevant for the disordered phase, where an initially vanishing $V$ remains zero during the flow. For $d=3$ the additional coupling $V$ induces quantitative changes, but for small coupling the qualitative changes in ``overall thermodynamic quantities'', like density, pressure, order parameter and phase diagram, are moderate since the modifications of the infrared running only concern logarithms. Still, for more detailed features, like occupation numbers for small momenta, the coupling $V$ is dominant. For the ordered phase in $d=1,2$, however, the relativistic dynamic term'' (\ref{R1}) will dominate over the term linear in $\partial_\tau$ and radically modify basic aspects of the macroscopic properties. In the Goldstone regime the coupling $S$ vanishes for $k\to 0$ such that the flow of the effective action is attracted to a (partial) fixed point with enhanced ``relativistic'' $SO(d+1)$ symmetry. This approximate relativistic symmetry qualitatively changes the properties of fixed point $(C)$. For $d=1$ there will be a line of fixed points with different $\rho_0$, while the bare order parameter $\bar \rho_0$ vanishes $\sim k^\eta$. For $d=2$ the fixed point $(C)$ disappears. The flow for $k\to0$ will yield $w\to \infty$ and both $\rho_0$ and $\bar\rho_0$ settle at constant values, with $\eta=0$.

We emphasize that the enhanced $SO(d+1)$ symmetry concerns only the leading dynamic and gradient terms for the Goldstone mode. It is not expected to become a symmetry of the full effective action since the Lorentz symmetry is not compatible with Galilei symmetry for $T=0$. For example, an $SO(d+1)$ violating term with two time derivatives for the radial mode is possible, cf. app. G. 

For an initially vanishing or very small $V$ a nonzero value is generated by the flow equation $(\tilde\lambda=\lambda S^{-1}k^{d-2})$
\begin{eqnarray}\label{R2}
\partial_tV&=&-\alpha_V\frac{S^2}{k^2},\\
\alpha_V&=&\frac{5v_d}{d}\left(1-\frac{\eta}{d+2}\right)\tilde\lambda w\left(1+\frac w2\right)(1+w)^{-5/2}.\nonumber
\end{eqnarray}
(Details of the computation of the flow equation for $V$ can be found in appendix C.) The relative importance of the kinetic terms linear or quadratic in $\partial_\tau$ can be measured by the ratio
\begin{equation}\label{R3}
s=\frac{S}{k\sqrt{V}}.
\end{equation}
As long as $s$ remains larger than one one may guess that the effects of $S$ could remain important. Indeed, a naive scaling criterion  for equal importance of the terms $\sim S$ or $V$ is given by $Vq_0\approx S$ with $Sq_0\approx k^2$ such that $Vk^2\approx S^2$. We will argue, however, that for the Goldstone boson physics the relevant scale is $\sqrt{2\lambda\rho_0}$ rather than $k$. The effects of the coupling $V$ therefore become dominant for $V\gg S^2/(2\lambda\rho_0)$ or $S\ll \sqrt{w}$.

For $s\to 0$ the effective action shows an enhanced $SO(d+1)$ symmetry, where $\tau'=\tau/\sqrt{V}$ acts like an additional space coordinate. From eq. (\ref{R2}) it is clear that the evolution of $V$ essentially stops for $k\to0$ if $S$ decreases faster than $k$ (and $\alpha_V$ remains bounded). This will be the case for $\eta_S<-1$, but a weaker condition will be sufficient for an effective stop of the running of $V$. Indeed, from app. C we get the flow equation for $s$
\begin{equation}\label{R4}
\partial_ts=-(1+\eta_S)s+\frac12A_V(s,w,\tilde\lambda)s^3
\end{equation}
where 
\begin{equation}\label{R5}
\lim_{s\to\infty}A_V=\alpha_V~,~\lim_{s\to 0~,~w\to\infty}
A_Vs^2\sim\tilde\lambda w^{-2}.
\end{equation}
One concludes that $s$ is driven to zero if $\eta_S<-1$. This presumably happens for $d=1$ and $d=2$. In this case the trajectories corresponding to an enhanced $SO(d+1)$-symmetry are attractive - the long distance physics becomes effectively relativistic. For $\eta_S>-1$ large values of $s$ decrease and small values increase, suggesting a partial fixed point $s_*(\tilde\lambda,w)$. If this occurs for large $s$ we find $s_*\sim\tilde\lambda^{-1/2}w^{1/4}$. The relevant question for omitting the linear dynamic term $\sim S$ in the Goldstone regime is $s/\sqrt{w}\ll 1$. This condition is reached for $k\to 0$ if V and $S/\lambda$ go to constants, while $S$ goes to zero. Constant values of $V$ and $S/\lambda$ are suggested also on physical grounds since these quantities correspond to thermodynamic observables, see eq. \eqref{G.27} in app. G.

In this context we note that $S=0$ is always a (partial) fixed point, due to an enhanced discrete symmetry $\tau\to  -\tau$ (while keeping $\phi$ fixed). (This additional discrete symmetry is preserved by our cutoff $R_k(16)$, even  though this cutoff does not respect the $SO(d+1)$ symmetry - see app. C for a discussion on this issue.) For $\eta_S<-1$ the fixed point at $s=0$ is  IR-attractive, while for $\eta_S>-1$ it becomes repulsive. For $d=3$, where $\eta_S>-1$, the flow therefore ends  for $k\to0$ with nonzero $s$, corresponding to a violation of $SO(d+1)$ symmetry in the radial sector. For $d=3$ one expects that $V$  stops running for $k\to 0$ due to $w\to\infty$. For large $w$ one finds in eq. (\ref{R2}) $\alpha_V=(\lambda/2\rho_0)^{1/2}k^2/S$ and $\partial_tV\sim S\sim (\ln k_0/k)^{-1}$. For $\lambda/S\to$ const we therefore have logarithmic behavior
\be\label{85a}
\lambda\sim S\sim\frac{1}{\ln(k_0/k)}
\ee
and we note the difference as compared  to the simplest truncation \eqref{F.6A}, where $\lambda$ decreases with the square of the inverse logarithm. This implies that $s$ diverges $\sim k^{-1}\ln(k_0/k)$ such that for $d=3$ the large $s$ regime applies. 

Also for $d=1,2$ the running of $V$ stops, this time due to $SO(d+1)$ symmetry. The running of $\rho_0$ and $\lambda$ within the Goldstone regime in the relativistic models has been intensively studied by non-perturbative flow equations \cite{BTSE}, \cite{BTW}. For $w\to\infty$ the running of $\rho_0$ stops. On the other hand, the fluctuations of the Goldstone modes produce a fixed point for the dimensionless coupling $\lambda k^{d-3}$ for all $d<3$. One infers for the effective momentum dependence of the quartic coupling
\be\label{85A}
\lambda(\vec q^2)\sim(\vec q^2)^{\frac{3-d}{2}}.
\ee
Comparison with the simplest truncation \eqref{B6a} shows that $\eta_S$ has to be replaced by $d-3$ instead of $2(d-3)$. This underlines again the crucial importance of the relativistic kinetic term for the long distance physics in all dimension $d\leq 3$.

The summary of the situation in the Goldstone regime is rather simple. For all $d$ the asymptotic value for $V$ reaches a constant as $k\to 0$. For $d=3$ also $S$ becomes almost constant (it vanishes only logarithmically), whereas for $d=1$ and $d=2$ the flow rapidly approaches an enhanced $SO(d+1)$-symmetry due to $S$ vanishing with a power law $S\sim k^{-\eta_S}~,~S/k\to 0$. The value of the renormalized order parameter $\rho_0$ approaches a constant for $d\geq 1$. For $d>1$ the anomalous dimension $\eta$ vanishes for $w\to\infty$ and also $\bar\rho_0$ become constant. The renormalized quartic coupling shows a scaling behavior according to its canonical dimension in the relativistic model. 

For $d=1$ and $d=2$ the consequences of the ``relativistic asymptotics'' are immediate - the Goldstone regime is described by the classical $O(2)$-model in $d+1$ dimensions. With $\tau'=\tau/\sqrt{V}~,~q'_0=q_0\sqrt{V}$ the correlation function for large distances in space and time (or small momenta $\vec q,q_0$) obey $\big(G=(G_{11}+G_{22})/2)$ 
\begin{eqnarray}\label{XXBB}
G\sim (\vec q\ ^2+q'^2_0)^{-1}~&,&~\bar G\sim(\vec q\ ^2+q'^2_0)^{-1+\eta/2},\nonumber\\
G\sim (\vec r\ ^2+\tau'\ ^2)^{\frac{1-d}{2}}~&,&~\bar G\sim (\vec r\ ^2+\tau'^2)^{\frac{1-d-\eta}{2}}.
\end{eqnarray}
(We recall that $G$ is dominated by the Goldstone contribution.) One may generalize the concept of dynamical critical exponent also for situations without a finite correlation length. For $d=1,2$ the effective dynamical critical exponent takes the ``relativistic value'' $z=1$. For $d=2$ the Goldstone regime is described by the three-dimensional classical model. It is well known that $\rho_0$ and $\bar\rho_0$ settle to constants, with $\eta(k\to0)=0$. 

At this point we can already extend our discussion to an arbitrary number $M$ of complex fields. The potential $u(\rho)$, the gradient term and the relativistic dynamical term $\sim V$ all obey an extended $O(2M)$-symmetry. For our truncation, the asymptotic behavior for the flow equations in the SSB regime is therefore well known for $d=1$ and $d=2$. Since $S$ vanishes asymptotically, and $S$ is the only term in our truncation that violates the $O(2M)$ symmetry, the asymptotic behavior of the flow is given by the classical $O(2M)$-models in $d+1$ dimensions. (A more general discussion of $M$-component models will be given in sect. \ref{mcomponentmodels}.) In particular, for $d=2$ one finds a simple description of order for arbitrary $M$ in terms of the three-dimensional classical $O(2M)$ models.

For $d=1$ the two dimensional classical model applies. By virtue of the Mermin-Wagner theorem we know that no long range order exists with a spontaneously broken continuous symmetry. Since any $\bar\rho_0\neq 0$ would lead to spontaneous breaking of the $U(1)$-symmetry we can conclude $\bar\rho_0(k\to 0)=0$. The way how the Mermin-Wagner theorem is realized depends on the number of components $M$ \cite{{CWFE}}. For $M>1$ both $\rho_0(k)$ and $\bar\rho_0(k)$ reach zero at some positive value $k_{SR}$. For $k_{ph}<k_{SR}$ no order persists, while for $k_{ph}>k_{SR}$ the system behaves effectively as in the presence of order. Typically, ordered domains exist with size $L_d\lesssim k^{-1}_{SR}$. Since the running of $\rho_0$ is only logarithmic the scale $k_{SR}$ can be exponentially small. For an experimental probe with size $L$ one has $k_{ph}>L^{-1}$ so that for practical applications an ``ordered phase'' will persist. The typical size of ordered domains is then larger than the size of the system. (This issue has been discussed in detail for classical antiferromagnetism in two dimensions \cite{BBW}.) 

For $M=1$, in contrast, $\rho_0$ reaches a constant value for $k\to 0$. Only the bare order parameter vanishes due to a nonvanishing anomalous dimension, $\bar\rho_0\sim k^\eta$, such that order does not exist in a strict sense. In the corresponding classical model this situation describes the ``low temperature phase'' related to the Kosterlitz-Thouless phase transition. For practical purposes this phase behaves like an ordered phase, with powerlike decay of the correlation function $\bar G$ (\ref{XXBB}) due to the existence of a Goldstone boson. This is also the characteristic behavior of a Tomonaga-Luttinger liquid. It is well known from the classical $O(2)$ model in two dimensions that the low temperature phase is characterized by a line of fixed points which may be labelled by $\rho_0=\rho_0(k\to 0)$. The anomalous dimension depends on $\rho_0$ \cite{CWFE,GKT}
\begin{equation}\label{R6}
\eta=\frac{1}{4\pi\sqrt{V}\rho_0}.
\end{equation}
It seems plausible that $\rho_0$ depends on the effective chemical potential $\sigma$ such that we predict an anomalous dimension depending on $\sigma$. 

It is remarkable that the main qualitative features for $d=1$ and $k\to 0$, namely a nonzero $\rho_0$, vanishing $\bar\rho_0$, and a positive anomalous dimension $\eta>0$, are already visible from fixed point $(C)$ in the simple truncation of sect. \ref{orderd}. Not surprisingly, however, the quantitative accuracy for the anomalous dimensions is very poor if the coupling $V$ is omitted. We may indeed address the properties of the Goldstone regime in the perspective of the properties of fixed points in presence of the coupling $V$. For $d=2$ one has the well known Wilson-Fisher fixed point of the three dimensional classical model. It corresponds to $S=0$. The question of how close trajectories approach the Wilson-Fisher  fixed point depends on the microscopic parameters $\sigma$ and $\lambda_\Lambda$ as well as on a possible microscopic coupling $V(\Lambda)$. Quantum phase transitions with critical behavior different from eq. (\ref{140B}) can be associated with the Wilson-Fisher fixed point. In this case $z=1$ and the critical exponents $\nu$ and $\eta$ of the three-dimensional $O(2M)$ model apply. For $d=2$ this type of phase transition presumably becomes relevant for large enough microscopic couplings $V(\Lambda)$. For $V(\Lambda)=0$, as considered in this paper, the quantum critical fixed point discussed in sect. V is relevant. For this quantum critical fixed point a vanishing relativistic coupling $V=0$ is stable with respect to the flow. In our truncation we infer from eq. (\ref{R2}) that for $w=0$ one has $\alpha_V=0$ and therefore $\partial_tV=0$ while there is anyhow no contribution to $\partial_tV$ in the disordered phase. At the quantum critical point the dimensionless combination $Vk^2$ therefore corresponds to an irrelevant coupling.

In order to judge the relevative importance of the Wilson-Fisher (WF) and the quantum critical (QC) fixed points for arbitrary microscopic couplings $V(\Lambda)$ one should consider  the critical hypersurface on which both fixed points lie. (Note that $\rho_0(\Lambda)$ varies on this hypersurface, with $\rho_0(\Lambda)=0$ for QC and $\rho_0(\Lambda)>0$ for WF. We use a common name (QC) for fixed point $(A)~(d>2)$ or fixed point $(B)~(d<2)$). The first question concerns the stability of WF with respect to the coupling $S$. Taking into account the scaling dimensions at WF one finds that WF is stable for $\eta_S<-1$ and unstable for $\eta_S>-1$. Here $\eta_S$ has to be evaluated for WF, which we have not done so far. For $\eta_S>-1$ one would observe a crossover from WF to QC on the critical hyperface. In contrast, for $\eta_S<-1$ both WF and QC are stable on the critical hypersurface. The topology of the flow would then imply the existence of a new fixed point with finite nonzero value of $S(\Lambda)/\sqrt{V(\Lambda)}$. 

For $d=1$ (and $M=1)$ the role of the Wilson-Fisher fixed point is replaced by the Kosterlitz-Thouless fixed point for the two dimensional classical $O(2)$ model. A key new ingredient is the existence of a whole line of fixed points for $S=0$. They can be parameterized by the renormalized order parameter $\rho_0$ (corresponding to $\kappa$ in Ref. \cite{GKT} at $k=0$. These fixed points govern the Goldstone regime of our model with $V(\Lambda)=0$. Thus the IR attractive fixed point $(C)$ in the truncation with $V=0$ transforms into one of the fixed points on the critical line. Now $w$ is no longer an irrelevant coupling - it can be used to parameterize the line of fixed points instead of $\rho_0$. (Indeed, $w=2(\lambda/k^2)\rho_0$ and $(\lambda/k^2)$ approaches a fixed point value depending on $\rho_0$ \cite{CWFE}.) It seems natural that $\rho_0$ depends on $\sigma$. On the other hand, $\rho_0(k=0)$ cannot take arbitrary small values, corresponding to the jump in the renormalized superfluid density of the Kosterlitz-Thouless transition \cite{Krahl}. This raises interesting questions of how the chemical potential $\sigma$ is mapped into an allowed range of $\rho_0$ or $w$. It is likely that the answer is linked to the ``initial flow'' for small $V$ with a possible influence of an approximate fixed point of type $(C)$ for which $V$ is a small perturbation. 

We observe that on the line of fixed points the anomalous dimension $\eta$ depends on $\rho_0$. The maximal value $\eta=1/4$ is reached at the Kosterlitz-Thouless transition. Our truncation therefore predicts $\eta\leq 1/4$ for eq. (\ref{XXBB}), and we infer for $d=1~,~M=1$
\begin{equation}\label{AA118}
\bar G\sim (r^2+\tau'^2)^{-\frac\eta2}.
\end{equation}
However, it is known for the Tomonaga-Luttinger liquid that the maximal value for $\eta$ is $1/2$. This brings us to the question of reliability of a truncation that only includes a potential, second order gradient terms and first and second order time derivatives with $SO(d+1)$ symmetry for $S=0$. For $V>0$ such a truncation is not compatible with Galilei invariance for $T=0$ and $\rho\to 0$. We conclude that the description in terms of a $d+1$-dimensional $O(2)$ model becomes problematic for $w\lesssim 1$. On the other hand we have presented strong arguments for its validity for $w\gg 1$. We conclude that the identification of the Tomonaga-Luttinger liquid with the Kosterlitz-Thouless phase of the two dimensional model may only be valid for large $w$. For $w\approx 1$ certain terms neglected in our tuncation and breaking of the $SO(d+1)$ symmetry can play a role. The range of $\eta$ near $1/4$ in the two-dimensional $O(2)$ model corresponds to $w\approx 1$ - this is the region of the phase transition. We conclude that our approximation is expected to hold as long as $\eta$ is small enough (large $w$), but may break down for $\eta$ near $1/4$, such that the upper bound $\eta\leq 1/4$ is not robust. 

We finally comment on the case $d=3$ where the linear $\tau$-derivative remains important, as far as naive scaling is concerned. However, in the Goldstone regime the linear $\tau$ derivative involves the radial mode $\varphi_1,\int\varphi^*\partial_\tau\varphi\sim\int \varphi_1\partial_\tau\varphi_2$. If we study the long range physics the radial mode effectively decouples for distances larger than the correlation length and we should question the criterion $s\lesssim 1$ that we have used for an estimate of the importance of the term $\sim V$. This criterion is valid for physics involving the radial mode, i.e. $k_{ph}\gtrsim \xi^{-1}_R$. We will discuss in the next section that for $k_{ph}\ll\xi^{-1}_R$ the relevant dimensionless ratio is not given by $s$ but rather by $S/\sqrt{2\lambda\rho_0 V}$. This can be seen by ``integrating out'' the radial mode and investigating the effect of $S$ in a nonlinear model for the Goldstone boson.

\section{Goldstone regime and non-linear $\sigma$-models}
\label{goldstoneregime}
In the Goldstone regime the influence of the radial mode is subdominant. We may therefore aim for a description only in terms of Goldstone bosons. This leads to the non-linear $\sigma$-models. The effective action in the nonlinear formulation can directly be computed from our results. One solves the field equation for the radial fluctuations as a functional of the Goldstone fluctuations and reinserts this solution into the effective action. For this purpose we parameterize
\begin{equation}\label{G1}
\phi(x)=\big(\phi_0+R(x)\big)e^{i\theta(x)}
\end{equation}
and eliminate the radial mode $R(x)$ as a functional of the periodic phase variable $\theta(x)$. 

Within our truncation 
\begin{equation}\label{G2}
{\cal S}=\int_x{\cal L}~,~{\cal L}=u(\phi^*\phi)+S\phi^*\partial_\tau\phi-\phi^*(\Delta+V\partial^2_\tau)\phi
\end{equation}
we find (for constant $S,V$ and removing total derivatives and constants)
\begin{eqnarray}\label{G3}
{\cal L}&=&{\cal L}_0+{\cal L}_R,\nonumber\\
{\cal L}_0&=&\phi^2_0\{\vec\nabla\theta\vec\nabla\theta+V\partial_\tau\theta\partial_\tau\theta\},\nonumber\\
{\cal L}_R&=&u(\phi^2_0+2\phi_0R+R^2)-u(\phi^2_0)\nonumber\\
&&+(2\phi_0R+R^2)
\{\vec\nabla\theta\vec\nabla\theta+V\partial_\tau\theta\partial_\tau\theta+iS\partial_\tau\theta\}
\nonumber\\
&&+\vec\nabla R\vec\nabla R+V\partial_\tau R\partial_\tau R.
\end{eqnarray}
The field equation $\delta{\cal L}_R/\delta R=0$ can be solved iteratively by expanding in 
$\{\vec\nabla\theta\vec\nabla\theta+V\partial_\tau\theta\partial_\tau\theta+iS\partial_\tau\theta\}$. In lowest order one obtains, with $\lambda=u''(\phi^2_0)$, 
\begin{equation}\label{G4}
R=-\phi_0(2\lambda\phi^2_0-\Delta-V^2\partial^2_\tau)^{-1}
\{\vec\nabla\theta\vec\nabla\theta+V\partial_\tau\theta\partial_\tau\theta+iS\partial_\tau\theta\},
\end{equation}
and reinserting this solution yields $(\rho_0=\phi^2_0)$
\begin{eqnarray}\label{G5}
&&{\cal L}_R=-\rho_0
\{iS\partial_\tau\theta-V\partial_\tau\theta\partial_\tau\theta-\vec\nabla\theta\vec\nabla\theta\}\\
&&(2\lambda\rho_0-\Delta-V\partial^2_\tau)^{-1}
\{iS\partial_\tau\theta-V\partial_\tau\theta\partial_\tau\theta-
\vec\nabla\theta\vec\nabla\theta\}.\nonumber
\end{eqnarray}
Expanding in powers of $\theta$
\begin{eqnarray}\label{G6}
{\cal L}_R&=&{\cal L}_{R,2}+{\cal L}_{int},\nonumber\\
{\cal L}_{R,2}&=&\rho_0
S^2\partial_\tau\theta(2\lambda\rho_0-\Delta-V\partial^2_\tau)^{-1}\partial_\tau\theta
\end{eqnarray}
and combining with ${\cal L}_0$ we obtain the effective inverse propagator for the Goldstone boson $(\sim\phi_0\theta)$
\begin{equation}\label{G7}
G^{-1}_g=\vec q\ ^2+Vq^2_0+
\frac{S^2 q^2_0}{2\lambda\rho_0+\vec q\ ^2+V q^2_0}.
\end{equation}
On the other hand, ${\cal L}_{int}$ contains interaction terms. These are purely derivative interactions, as appropriate for Goldstone bosons. Higher orders in the iterative expansion will only yield further interaction terms.

The inverse Goldstone propagator shows no term linear in $q_0$, even for $V=0~,~S=1$. This is a simple consequence of the observation that a linear $\tau$-derivative for a single real field is always a total derivative and can therefore be eliminated from the field equations. In lowest order in a derivative expansion (for small momenta) one has
\begin{equation}\label{G8}
G_g=\left[\vec q\ ^2+\left(V+\frac{S^2}{2\lambda\rho_0}\right)q^2_0\right]^{-1}
\end{equation}
such that both $V$ and $S$ contribute to an effectively relativistic kinetic term. (In the limit of large $q_0~,~Vq^2_0\gg2\lambda\rho_0+\vec q\ ^2$, the correction from ${\cal L}_R$ results in an effective masslike term $S^2/V$, i.e. $G_g=[\vec q\ ^2+V q^2_0+S^2/V]^{-1}$.) As mentioned at the end of the preceeding section the relevant ratio for the importance of the linear dynamic term $\sim S$ is given for the Goldstone regime by $S/\sqrt{2\lambda\rho_0 V}$.

We conclude that in the Goldstone regime the renormalized propagator always takes a relativistic form
\begin{equation}\label{G9}
G_g=\big[\vec q\ ^2+(q_0/v)^2\big]^{-1}.
\end{equation}
The quantity
\begin{equation}\label{G10}
v^2=\left(V+\frac{S^2}{2\lambda\rho_0}\right)^{-1}
\end{equation}
can be associated with the microscopic sound velocity of the Goldstone mode. Comparing with eq. \eqref{G.21} in appendix G we see that it equals the macroscopic sound velocity, $c_s$ in accordance with Ref. \cite{1A}. For $\rho_0\to~const,~S/\lambda\to~const,~S\to 0$, the second term in eq. \eqref{G10} can be neglected. This yields a simple relation between the sound velocity and the coupling $V=V(k\to 0)$
\be\label{98A}
c^2_s=V^{-1}.
\ee

The bare propagator reads correspondingly
\begin{equation}\label{G11}
\bar G_g=\big[\vec q\ ^2+(q_0/v)^2\big]^{-\left(1-\frac\eta2\right)}
\end{equation}
where $\eta$ depends on momenta according to the replacement $k^2\to \vec q\ ^2+q^2_0/v^2$. For $d>1$ the anomalous dimension vanishes asymptotically, $\eta(k\to 0)=0$. The generalization to $M$ component fields is straightforward and will briefly be addressed in the next section. We emphasize that for the nonlinear models the propagator shows no qualitative difference between a linear dynamic term $\sim S$ or a relativistic dynamic term $\sim V$. Nevertheless, the form of the dynamic term becomes important for the linear model, in particular for the flow of the order parameter $\rho_0(k)$.

\section{M-component models}
\label{mcomponentmodels}
For models with a relativistic kinetic term the critical exponents depend sensitively on the number of components of the field. For $N$ real components and scalar models with $O(N)$ symmetry the universality classes can be fully characterized by $N$ and $d$. We may investigate this question also for the nonrelativistic kinetic term and investigate models with $M$ complex scalar fields $\phi_m~,~m=1\dots M$. In terms of renormalized fields we consider models with $U(M)$-symmetry and neglect first the coupling $V$
\begin{equation}\label{Z6a}
\Gamma_k=\int_x
\big\{\sum_m(S\phi^*_m\partial_\tau\phi_m-\phi^*_m\Delta\phi_m)+u(\rho)\big\}.
\end{equation}
Due to the $U(M)$ symmetry the potential can only depend on the invariant $\rho=\sum\limits_m\phi^*_m\phi_m$. Since $\rho$ is invariant with respect to the symmetry $O(2M)$ acting on the $2M$ real components of the scalar field,  the potential exhibits an enhanced $O(2M)$ symmetry. This enhanced symmetry is shared by the term involving spatial gradients but violated by the linear  $\tau$-derivative.

Within our truncation of pointlike interactions and with a linear $\tau$-derivative, we find a very simple result: the flow equations do not depend on $M$, except for the flow of a field independent constant in $u$. Neglecting the momentum dependence of interactions all our discussion therefore holds without modifications for the $M$-component case as well.  As long as the propagator is specified by eq. (\ref{Z6a}) a dependence of the critical exponents on $M$ can therefore only arise from the momentum dependence of the interactions. This situation is quite different in presence of a relativistic kinetic term involving two $\tau$-derivatives.

In order to show this result we first note that for a constant background field the inverse propagator is block diagonal,  $G^{-1}$$=diag$$(G^{-1}_1$ , $G^{-1}_2\dots)$ where $G^{-1}_m$ are $2\times 2$ matrices corresponding to a decomposition of the complex field $\phi_m$ into real and imaginary parts. Without loss of generality we may take a real constant background field in the $m=1$-direction. Then $G^{-2}_1$ is given by eq. (\ref{aa5}) whereas for $m>1$ one has 
\begin{equation}\label{Z1}
G^{-1}_{m\neq 1}=\left(
\begin{array}{rll}
\tilde q^2+u'&,&-Sq_0\\
Sq_0&,&\tilde q^2+u'
\end{array}\right).
\end{equation}
The resulting correction to $\partial_t u$ from the $(M-1)$ additional components does not depend on $\rho$
\begin{equation}\label{Z2}
\Delta\partial_t u_{|\bar\phi}=\frac{4(M-1)v_d}{dS}
k^{d+2}\left(1-\frac{\eta}{d+2}\right).
\end{equation}
A field independent additive constant does not influence our discussion and we recover the same flow equations for $w$ and $\tilde \lambda$ as for $M=1$.

We next turn to the computation of $\eta$ and $\eta_S$. We define these quantities by the flow of the $\vec q\ ^2$ and $q_0$ dependence of the inverse propagator for the $m=1$ component, according to eqs. (\ref{x5}), (\ref{x6}). We can use eqs. (\ref{x1}), (\ref{x2}) with indices $a,b\dots$, running now from $1$ to $2M$, e.g. the $m=2$ component corresponds to $a =3,4$ etc.. In presence of a background field in the $a=1$ direction the interaction term in the effective action for the fields $\phi_b,b\neq 1$ has a discrete symmetry $\phi_b\to -\phi_b$. This implies that for nonvanishing cubic couplings $\gamma_{acd}$ the values of all indices must appear in pairs, except for the value one. For the anomalous dimension $\eta$ the indices $a$ and $b$ in eq. (\ref{x2}) take the value two. In consequence, the index pair (d,e) can only be (1,2) or (2,1), and similar for the index pair $(f,c)$. Therefore no indices $c,d,e,f$ with values three or higher appear and the additional $(M-1)$ complex fields do not influence $\eta$. For $\eta_S$ the situation is similar if one also uses the fact that $G$ is block diagonal, i.e. $G_{1,b>2}=G_{2,b>2}=0$. (The situation can be easily depicted in terms of one-loop Feynman graphs with two external legs.) This closes our argument that the flow equations are independent of the number of components $M$ in the approximation of pointlike vertices and for $V=0$. 

We conclude that the symmetric phase and the quantum phase transition are not influenced by the presence of additional components of the field. At the quantum critical point one finds for all $M$ the critical exponents $\nu=1/2~,~\eta=0~,~z=2$. In the ordered phase the number of components will influence the flow as soon as a sizeable strength $V$ for the second $\tau$-derivative is generated. In particular, the flow for $d=1$ and $d=2$ will depend on $M$. The flow for $k\to 0$ is expected to reproduce the well known behavior of the Goldstone boson physics in classical statistical systems \cite{CWAV,CWFE,BTW}. Typically, the  asymptotic behavior can be described by nonlinear $\sigma$-models on a manifold given by the coset space $U(M)/U(M-1)$. For $d+1=2$ the behavior of the abelian model for $M=1$ differs substantially from the non-abelian $\sigma$-models for $M>1$.

In the microscopic action only the coupling $S$ violates the $O(2M)$ symmetry. In analogy, we expect that for small $O(2M)$ violating couplings, i.e. small $S$ we can choose a description in terms of $O(2M)/O(2M-1)$-models. This will contain symmetry breaking effects since the $O(2M)$ symmetry is exact only for $S=0$. Such effects are proportional to the dimensionless combination $S(k)/(\sqrt{2\lambda (k)\rho_0(k) V(k)})$. Since for $d=1,2$ the flow is attracted towards $S=0$ (if $\eta_S<0$), one may suspect that the enhanced  symmetry for $S=0$, i.e. the space-time rotations $SO(d+1)$ and the internal rotations $SO(2M)$, are approximately realized in the Goldstone regime. The degree of violation of these symmetries depends on the characteristic momentum scale of the Green's functions and on $w^{-1}$. As before, a nonzero ``external momentum'' $\vec q$ may be associated with $k^2_{ph}=\vec q\ ^2>0$. Since $S(k)$ vanishes only asymptotically for $k\to 0$ the symmetry breaking of $SO(d+1)$ and $SO(2M)$ due to $S(k_{ph})>0$ will always be present. Furthermore, the vanishing of $S$ occurs for large $w$, while $S$ remains important for $w\lesssim 1$. 

The order parameter $\rho_0$ or $\bar\rho_0$ is a quantity that involves the limit of zero momentum (or $k_{ph}\approx L^{-1}$ with $L$ the macroscopic size of the probe). We may therefore take the limit $k\to 0~,~\vec q\ ^2\to 0$ for the issue of spontaneous symmetry breaking. For small momenta we may consider a derivative expansion of the effective action (for $k\to 0)$.  Let us consider up to two derivatives. For $1<d\leq 2$ both $w^{-1}$ and the ratio $S(k)/k$ vanish for $k\to0$ and $s$ can be neglected. The potential and the relativistic two-derivative term $-\phi^*(V\partial\tau^2+\Delta)\phi$ respect both $SO(d+1)$ and $O(2M)$. A term $\sim -Y\rho(\tilde V\partial\tau^2+\Delta)\rho$ violates $SO(d+1)$ only if $\tilde V\neq V$, while it preserves $O(2M)$. On the level of two derivatives an $O(2M)$ violation could only arise from a term 
$\int_x(\varphi^*_1\varphi_1\partial_\tau\varphi^*_2\partial_\tau\varphi_2+\varphi^*_2\varphi_2\partial_\tau\varphi^*_1\partial_\tau\varphi_1$ 
$-\varphi^*_2\varphi_1\partial_\tau\varphi^*_1\partial_\tau\varphi_2-\varphi^*_1\varphi_2\partial_\tau 
\varphi^*_2\partial_\tau\varphi_1)$ for $M=2$, or suitable generalizations for $M>2$. (On the level of two derivatives the invariants preserving $U(M)$ while violating $O(2M)$ are of the form $|\epsilon_{m_1\dots m_N}\phi_{m_1}\dots\phi_{m_{N-1}}\partial\phi_{m_N}|^2.)$

For $d=2$ we may speculate that the $SO(d+1)$ and $O(2M)$ violating terms involving two derivatives vanish for $k\to 0$. The behavior of the order parameter and the correlation function at small momentum are then well approximated by the classical $O(2M)$ model in three dimensions. This issue depends on whether the Lorentz symmetry violating operators are irrelevant for the trajectories with $w\to\infty$, or not. Even in presence of $O(2M)$ violating derivative terms one would still find $2M-1$ gapless excitations, due to the accidental $O(2M)$ symmetry of the potential. 

For $d=1$ the issue may be more complicated. Within the relativistic model $w$ is either constant (for $M=1$) or runs logarithmically towards zero $(M>1)$. We may therefore encounter a region where $w\lesssim 1$ such that the $SO(d+1)$ violating terms can no longer be neglected. For $M>1$ this region becomes always relevant for $k\to 0$. For $M=1$ we expect that $SO(d+1)$-symmetry becomes a good approximation for the fixed points with small $\eta$ (large $w$), whereas it is questionable for the larger values of $\eta$ corresponding to $w\approx 1$.

\section{Conclusions and outlook}
\label{conclusions}
The functional renormalization group yields a unified picture for quantum phase transitions of bosons. It covers both the ordered and the disordered phase, including the rather complicated long range dynamics of the second. We describe within the same simple truncation an arbitrary number of space-dimensions $d$ and an arbitrary number of components $M$ (for $M$ complex fields with symmetry $U(M)$). In a general setting, the properties of the phase transition depend on the parameter $s_\Lambda$ which characterizes the relative strength of the kinetic terms involving one or two time derivatives. More precisely, for a general microscopic (or classical) action (\ref{A1}) the dimensionless ratio $s_\Lambda$ is given by 
\begin{equation}\label{C1}
s_\Lambda=\frac{\bar S}{\Lambda}\sqrt{\frac{2M_B}{\bar V}}=
\frac{S(\Lambda)}{\Lambda\sqrt{V(\Lambda)}},
\end{equation}
with $\Lambda$ the ultraviolet cutoff. 

The characteristic features are described by two limits. The point $s_\Lambda=0$ describes models with a ``relativistic kinetic'' term involving two time derivatives and an enhanced space-time symmetry $SO(d+1)$. For those models the quantum phase transition in $d$ dimensions is strictly equivalent to the classical phase transition in $d+1$ dimensions. The universal critical properties correspond to the well studied $O(2M)$-models in $d+1$ dimensions. 

For the other limit $s_\Lambda\to \infty$ (i.e. $\bar V=0$) Galilei symmetry characterizes the zero temperature physics. Now the phase transition is influenced by the non-relativistic quantum critical point. This is the case we have mainly studied in this paper. For intermediate values of $s_\Lambda$ one expects near the phase transition a crossover from the ``classical'' or ``relativistic'' critical point to the non-relativistic quantum critical point. No phase transition is expected as $s_\Lambda$ is varied. However, the critical exponents and amplitudes will depend on $s_\Lambda$. They are given by the universality class of the $d+1$ dimensional relativistic $O(2M)$ models in the limit of small $s_\Lambda$ - for example by the well known Wilson-Fisher critical point for $d=2$.  As $s_\Lambda$ increases towards infinity the critical exponents smoothly change towards the mean field values for the non-relativistic quantum critical point. We note that for $s_\Lambda\neq 0$ the critical hypersurface for the phase transition typically occurs for nonzero density. 

The critical behavior for the nonrelativistic  ``quantum critical models''(i.e. $s_\Lambda\to\infty$) can be characterized by two regimes, depending on $l/D$. Here $l$ is a characteristic length scale of the system or experiment, and $D$ is proportional to the average distance between (quasi-) particles, $D\sim n^{-1/d}$. The ``dilute regime'' applies to the disordered  phase (where $D\to \infty$) and to the ``linear regime'' in the ordered phase, as shown in fig. \ref{functional-ren-fig1}. In the dilute regime simple scaling laws apply and the critical behavior is described by mean field critical exponents for all $d$ and $M$. The simplicity of its properties finds a simple explanation, since the disordered phase describes the vacuum with zero particles, and for the linear regime in the ordered phase the small density gives only subleading corrections. In more technical terms, this is the regime where $k^2_{ph}=l^{-2}\ll 2\lambda \rho_0$ or $w\ll 1$. 

For $l\gg D$ the particle density matters. For this ``dense regime'' the particle density $n$ sets a new relevant momentum scale $\sim D^{-1}$ or energy scale $(2M_BD^2)^{-1}$. The long distance physics is described by the interacting Goldstone bosons which arise from the spontaneous breaking of  $U(M)$-symmetry. We therefore often call the dense regime the ``Goldstone regime''. Since Goldstone bosons must be massless (or ``gapless'') the correlation functions decay with an inverse power of distance in space or time. 

The physics of interacting Goldstone bosons crucially depends on the number of space dimensions $d$. For $d=3$ we find that the ``Goldstone dynamics'' is influenced both by short and long wave length fluctuations. The running of dimensionless couplings is logarithmic. For small or moderate interaction strength $\lambda_\Lambda$ mean field theory remains a good guide. Quantum corrections induce quantitative corrections but do not change the qualitative behavior, except for the extreme infrared. We do not address in this paper the possibility that large $\lambda_\Lambda$ may lead to new phenomena, as the destruction of the condensate by a too large repulsion between the bosons. 

On the other hand, for $d=2$ and $d=1$ the Goldstone dynamics is infrared dominated, leading to qualitatively new features induced by quantum fluctuations. One expects strong deviations from mean field theory. We find that the renormalization flow describes a crossover to models with a relativistic kinetic term. Even for models with $s_\Lambda\to\infty$ (i.e. $\bar V=0)$ the value of $s(k)$ decreases fast for momentum scales $k\ll \Lambda$. The relativistic term $\sim V$ will be induced by the fluctuations and dominates for $k\to 0$. The Goldstone regime for $d=1,2$ is therefore characterized by $s=0$, both for $s_\Lambda\to \infty$ and for $s_\Lambda=0$. In other words, the flow is attracted fast towards the (partial) fixed point $s=0$. (The flow in the Goldstone regime differs from the flow on the critical hypersurface where $s$ presumably diverges.)

This implies for $d=1,2$ a close link between the Goldstone regime of the quantum model in $d$ dimensions and the corresponding Goldstone regime of the classical model in $d+1$ dimensions. The classical Goldstone regime is well studied. For $d=2$ one expects a description of the Goldstone regime by the classical three-dimensional $U(M)$-model. In this case the realization of order is straightforward, with both $\rho_0$ and $\bar\rho_0$ approaching constants for $k\to 0$ and $\eta$ tending to zero. The correlation function for large separations in space or time shows a simple powerlike behavior
\begin{eqnarray}\label{C2}
&&G(q)\sim \bar G(q)\sim (\vec q\ ^2+q^2_0/c^2_s)^{-1},\nonumber\\
&&G(x)\sim \bar G(x)\sim (\vec r\ ^2+c^2_s\tau^2)^{-1/2}.
\end{eqnarray}
Here we have restored dimensions, with sound velocity $c_s=v/(2M_B)$. 

For $d=1$ the ``ordered phase '' actually shows no long range order in a strict sense. The influence of  the Goldstone bosons is so strong that they destroy any long range order. This agrees with the Mermin-Wagner theorem for the classical two-dimensional models with continuous global symmetry, as applicable here. The way how this theorem is realized depends crucially on $M$. For $M=1$ the non-linear $\sigma$-model is abelian. In the ordered phase the relevant excitations include vortices. Indeed, the Kosterlitz-Thouless phase transition finds within the functional renormalization group a simple description in terms of a linear $O(2)$ model in two dimensions \cite{GKT}. As a result, the renormalized order parameter $\rho_0$ reaches a constant for $k\to 0$, while the bare order parameter $\bar \rho_0$ vanishes $\sim k^\eta$ due to a nonvanishing anomalous dimension. The correlation function decays as
\begin{eqnarray}\label{C3}
&&\bar G(q)\sim (\vec q\ ^2+q^2_0/c^2_s)^{-\left(1-\frac\eta 2\right)},\nonumber\\
&&\bar G(x)\sim (\vec r\ ^2+c^2_s\tau^2)^{-\frac\eta 2}.
\end{eqnarray}
These findings carry over to the non-relativistic model for $d=1$, the Tomonaga-Luttinger liquid. In our approach the key features of the Tomonaga-Luttinger liquid result essentially from the nonzero value of the renormalized order parameter $\rho_0$. This model simply describes the Goldstone boson associated to the ``effective $U(1)$ symmetry breaking''. The anomalous dimension depends on the density of quasi-particles, $n=\rho_0$. For large $\kappa=\sqrt{V}n=n/v=n/(2M_Bc_s)$ one finds from eq. \eqref{R6}
\be\label{106}
\eta=\frac{M_Bc_s}{2\pi n}.
\ee
After restoring dimensions the sound velocity $c_s$ is related to the coupling $V$ by eq. \eqref{G.27} 
\be\label{107}
c^2_s=\frac{1}{4M^2_BV}.
\ee
For $M>1$, as for example for the ferromagnetic to paramagnetic transition with $M=3$, the nonlinear $\sigma$-models are asymptotically free and induce a nonperturbative scale $k_{SR}$. Within a description by linear $U(M)$ models one finds that $\rho_0(k)$ and $\bar \rho_0(k)$ both vanish for $k=k_{SR}$, while being finite for $k>k_{SR}$. For $k<k_{SR}$ the flow follows the symmetric regime with a minimum of the potential at the origin. The situation with nonzero $\bar\rho_0(k)$ describes local order, with a maximal size of domains up to $k^{-1}_{SR}$. The order in larger domains is destroyed by the Goldstone boson fluctuations. Since $k_{SR}$ can be exponentially small for large enough density (the running of $\rho_0(k)$ is only logarithmic), there is always a critical density  $n_c$ beyond which the size of ordered domains $k^{-1}_{SR}$ grows beyond the macroscopic size of the experimental probe $L$. Thus for $n>n_c$ one observes effectively spontaneous symmetry breaking, despite the Mermin-Wagner theorem.

The dynamical behavior of the quantum critical models can be characterized by a dynamical critical exponent $z$. In general terms, it describes how a characteristic time scale $\hat \tau$ scales with a characteristic length $l$, i.e. $\hat\tau\sim l^z$. We have concentrated on the non-relativistic quantum critical models $(s_\Lambda\to\infty)$. We find for all $d$ and $M$ that the value of $z$ depends on the density of (quasi)-particles. For the dilute regime one has the mean field scaling $z=2$, while the dense regime shows the relativistic value $z=1$. Inbetween, there is an effective crossover. As far as time scales are concerned the Goldstone regime (dense regime) applies for $\hat\tau>D/c_s$, with $D$ the interparticle distance $\sim n^{-1/d}$ and $c_s$ the sound velocity.

This paper has demonstrated that the flow of a small set couplings, namely $\rho_0$ (or $m^2$), $\lambda,S$ and $V$ yields all qualitative features for the quantum phase transition for arbitrary $d$ and $M$. The quantitative precision can be improved by extending the truncation. Straightforward steps include the incorporation of a  nonvanishing $V$ in the flow equations for the effective potential $u$ and for $\eta_S$, as well as a differentiation between the effective renormalization constants for the radial and Goldstone modes via the inclusion of a term $\sim Y\partial\rho\partial\rho$. Together with $\gamma=u^{(3)}(\rho_0)$ we expect that the set of couplings $(\rho_0,\lambda,\gamma,S,V,Y)$ will yield already a very satisfactory quantitative accuracy. Extensions to include fermionic degrees of freedom are possible \cite{DGPW}.
Furthermore, the effects of nonvanishing temperature can easily be incorporated in our framework - one simply has to replace the $q_0$-integration by a Matsubara sum. Close to the critical temperature of the phase transition and away from the quantum critical point at $T=0$ the flow will experience an effective dimensional reduction to the classical $d$-dimensional $U(M)$ models. High accuracy for the functional flow equations has already been reached for the classical $O(N)$ models in arbitrary dimension. Including the temperature effects the flow equations should provide a rather complete picture for the thermodynamics of bosonic quantum gases. 

In this paper we have concentrated on the structural aspects. A numerical solution of the flow equations for $k\to 0$ will directly yield important thermodynamic quantities in the zero temperature limit. The density is given by $n=\rho_0$ and the condensate fraction by $\Omega_c=\bar A^{-1}$. The sound velocity obeys $c_s=1/(2M_B\sqrt{V})$ for $d=1,2$, with logarithmic corrections for $d=3$ due to $S\sim 1/\ln(k_0/k)$, cf. app. G, 
\be\label{108}
c^2_s=\frac{1}{4M^2_BV}\left[1-S\left(1+\frac{\partial\ln\Omega_c}{\partial\ln n}\right)\right].
\ee
The change of the condensate fraction with the density follows (cf. app. G)
\be\label{109}
\frac{\partial \ln \Omega_c}{\partial\ln n}=
\frac{2M^2_Bc^2_sS}{\lambda n}-1\to 
\frac{S}{2\lambda\rho_0 V}-1,
\ee
with $\lambda$ the renormalized quartic coupling. 

It may be possible to measure all these quantities by investigating the Bose-Einstein condensate for ultracold bosonic atoms. With suitable traps one may prepare essentially homogeneous systems for $d=1,2$ or $3$. For a quantitative computation one will further need to determine $\lambda_\Lambda$ in dependence on external parameters, as a homogeneous magnetic field. For this purpose one relates it to measurable properties in the vacuum, like the scattering length for $d=3$. The necessary computation for $n=0$ corresponds precisely to the phase transition discussed in sect. \ref{scaling} and can be performed in the comparatively simple disordered phase. The prospects for experimental tests of some of the features described in this paper look promising.

\medskip\noindent
{\bf Note added:}\\
After the first version of the paper interesting extensions and numerical results have been obtained in Ref. \cite{DS}.

\bigskip
\noindent
{\bf Acknowledgment:}\\
The author would  like to thank S. Diehl, S. Floerchinger, H. Gies, J. Pawlowski and M. Scherrer for  fruitful discussion and collaboration.

\section*{APPENDIX A: Flow equation for the effective potential}
\renewcommand{\theequation}{A.\arabic{equation}}
\setcounter{equation}{0}
We can write the flow at fixed renormalized field $\phi$
\begin{eqnarray}\label{aa7}
\partial_t u&=&\eta\rho u'\\
&&+\int\limits_q\Big\{\big[k^2-\frac\eta 2(k^2-\vec q\ ^2)\big]\theta(k^2-\vec q\ ^2)tr G\Big\}\nonumber\\
&=&\eta\rho u'+8v_d\int\limits^k_0 d\bar q\bar q^{d-1}\big[k^2-\frac\eta2(k^2-\bar q^2)\big]\tilde g,\nonumber
\end{eqnarray}
where $v^{-1}_1=4\pi,v^{-1}_2=8\pi,v^{-1}_3=8\pi^2$ and $\bar q=(\vec q\ ^2)^{1/2}$. The integration over  $q_0$ can be carried out easily
\begin{eqnarray}\label{aa8}
\tilde g&=&\frac12\int\frac{dq_0}{2\pi}tr G\nonumber\\
&=&\int\frac{dq_0}{(2\pi)}
\frac{\tilde q^2+u'+\rho u''}{S^2q^2_0+(\tilde q^2+u')(\tilde q^2+u'+2\rho u'')}\nonumber\\
&=&\frac{1}{2S}\frac{\tilde q^2+u'+\rho u''}{\sqrt{\tilde q^2+u'}\sqrt{\tilde q^2+u'+2\rho u''}}.
\end{eqnarray}
In the integrand we can use $\tilde q^2=k^2$ such that the $\bar q$ integration is trivial and yields eq. (\ref{aa9}).

Differentiation of eq. (\ref{aa9}) with respect to $\rho$ yields
\begin{eqnarray}\label{aa10}
\partial_t u'&=&\eta(u'+\rho u'')\nonumber\\
&&-\frac{2v_d}{dS}k^{d+2}\left(1-\frac{\eta}{d+2}\right)\nonumber\\
&&\frac{1}{\sqrt{k^2+u'}}
\frac{1}{\sqrt{k^2+u'+2\rho u''}}\nonumber\\
&&\left\{\frac{\rho u^{''2}}{k^2+u'}-
\frac{3\rho u^{''2}+2\rho^2 u'' u^{(3)}}{k^2+u'+2\rho u''}\right\}.
\end{eqnarray}
The second derivative reads
\begin{eqnarray}\label{aa11}
&&\partial_t u''=\eta(2u''+\rho u^{(3)})\nonumber\\
&&-\frac{2v_d}{dS}k^{d+2}\left(1-\frac{\eta}{d+2}\right)\nonumber\\
&&\frac{1}{\sqrt{k^2+u'}}
\frac{1}{\sqrt{k^2+u'+2\rho u''}}\nonumber\\
&&\left\{\frac{u^{''2}+2\rho u'' u^{(3)}}{k^2+u'}\right.\\
&&-\frac{3u^{''2}+10\rho u'' u^{(3)}+2\rho^2(u^{(3)2}+u''u^{(4)})}{k^2+u'+2\rho u''}\nonumber\\
&&\left.-\frac32\rho u''
\left[\frac{u^{''2}}{(k^2+u')^2}-
\frac{(3u''+2\rho u^{(3)})^2}{(k^2+u'+2\rho u'')^2}
\right]
\right\}\nonumber
\end{eqnarray}
while the third derivative becomes already quite lengthy
\begin{eqnarray}\label{A.XX}
\partial_t u^{(3)}&=&\eta(3u^{(3)}+\rho u^{(4)})\nonumber\\
&&-\frac{2v_d}{dS}\left(1-\frac{\eta}{d+2}\right)k^{d+2}\nonumber\\
&&\frac{1}{\sqrt{k^2+u'}}\frac{1}{\sqrt{k^2+u'+2\rho u''}}R
\end{eqnarray}
with 
\begin{eqnarray}\label{A7a}
R&=&\frac{X_1}{k^2+u'}-\frac{X_2}{k+u'+2\rho u''}\nonumber\\
&&-\frac{Y_1}{(k^2+u')^2}+\frac{Y_2}{(k^2+u')(k^2+u'+2\rho u'')}\nonumber\\
&&+\frac{Y_3}{(k^2+u'+2\rho u'')^2}\nonumber\\
&&+\frac{Z_1}{(k^2+u')^3}+\frac{Z_2}{(k^2+u')^2(k^2+u'+2\rho u'')}\nonumber\\
&&-\frac{Z_3}{(k^2+u')(k^2+u'+2\rho u'')^2}\nonumber\\
&&-\frac{Z_4}{(k^2+u'+2\rho u'')^3}
\end{eqnarray}
and
\begin{eqnarray}\label{A7b}
X_1&=&4u'' u^{(3)}+2\rho(u^{(3)2}+u'' u^{(4)}),\nonumber\\
X_2&=&16 u'' u^{(3)}+14\rho(u^{(3)2}+u'' u^{(4)})\nonumber\\
&&+2\rho^2(3u^{(3)}u^{(4)}+u''u^{(5)}),\nonumber\\
Y_1&=&3u''^2(u''+\frac52\rho u^{(3)}),\nonumber\\
Y_2&=&\rho u''(u'' u^{(3)}-\rho u^{(3)2}+\rho u'' u^{(4)})\nonumber\\
Y_3&=&(3u''+2\rho u^{(3)})
(9u''^2+\frac{75}{2}\rho u'' u^{(3)}\nonumber\\
&&+6\rho^2 u^{(3)2}+9\rho^2 u'' u^{(4)}),\nonumber\\
Z_1&=&\frac{15}{4}\rho u''^4~,~Z_2=\frac34\rho u''^3(3u''+2\rho u^{(3)}),\nonumber\\
Z_3&=&\frac34\rho u''^2(3u''+2\rho u^{(3)})^2~,~\nonumber\\
Z_4&=&\frac{15}{4}\rho u''(3u''+2\rho u^{(3)})^3.
\end{eqnarray}

%\newpage
\section*{APPENDIX B: Computation of anomalous dimensions}
\renewcommand{\theequation}{B.\arabic{equation}}
\setcounter{equation}{0}
For the calculation of $\eta$ and $\partial_t\ln S$ we infer from eq. (\ref{x2}) 
\begin{eqnarray}\label{x4}
\partial_t\bar P_{22}(q)&=&
\rho(u'')^2
\bar A^3\int_{q'}\partial_tR_k(q')\nonumber\\
&&\big\{(\bar G^2)_{11}(q')\bar G_{22}(q'+q)\nonumber\\
&&+(\bar G^2)_{22}(q')\bar G_{11}(q'+q)\nonumber\\
&&+(\bar{G}^2)_{12}(q')\bar{G}_{12}(q'+q)\\
&&+(\bar{G}^2)_{21}(q')\bar{G}_{21}(q'+q)+(q\to-q)\big\},\nonumber
\end{eqnarray}
and
\begin{eqnarray}
\partial_t\bar P_{11}(q)&=&
\rho(u'')^2
\bar A^3\int_{q'}\partial_tR_k(q')\nonumber\\
&&\big\{
\left(3+\frac{2\rho u^{(3)}}{u''}\right)^2
(\bar{G}^2)_{11}(q')\bar{G}_{11}(q'+q)\nonumber\\
&&+(\bar{G}^2)_{22}(q')\bar{G}_{22}(q'+q)\nonumber\\
&&+\left(3+\frac{2\rho u^{(3)}}{u''}\right)
(\bar{G}^2)_{12}(q')\bar{G}_{21}(q'+q)\nonumber\\
&&+\left(3+\frac{2\rho u^{(3)}}{u''}\right)
(\bar{G}^2)_{21}(q')\bar{G}_{12}(q'+q)\nonumber\\
&&+(q\to-q)\big\},\\
\partial_t\bar P_{12}(q)&=&\rho(u'')^2\bar A^3\int\limits_{q'}\partial_t R_k(q')\nonumber\\
&&\hspace{-1.5cm}\big\{\left(3+\frac{2\rho u^{(3)}}{u''}\right)(\bar{G}^2)_{11}(q')
\big[\bar{G}_{12}(q'+q)+\bar{G}_{21}(q'-q)\big]\nonumber\\
&&+(\bar{G}^2)_{22}(q')
\big[\bar{G}_{21}(q'+q)
+\bar{G}_{12}(q'-q)\big]\nonumber\\
&& \hspace{-1.3cm}+(\bar{G}^2)_{12}(q')\bar{G}_{22}(q'+q)
+(\bar{G}^2)_{21}(q')\bar{G}_{22}(q'-q)\nonumber\\
&&+\left(3+\frac{2\rho u^{(3)}}{u''}\right)
\big[(\bar{G}^2)_{21}(q')\bar{G}_{11}(q'+q)\nonumber\\
&&+(\bar{G}^2)_{12}(q')\bar{G}_{11}(q'-q)\big]\big\},\\
\partial_t\bar P_{21}(q)&=&\rho(u'')^2\bar A^3\int\limits_{q'}\partial_t R_k(q')\nonumber\\
&&  \hspace{-1.5cm}  \Big\{\left(3+\frac{2\rho u^{(3)}}{u''}\right)(\bar{G}^2)_{11}(q')
\big[\bar{G}_{21}(q'+q)+\bar{G}_{12}(q'-q)\big]\nonumber\\
&&+(\bar{G}^2)_{22}(q')
\big[\bar{G}_{12}(q'+q)+\bar{G}_{21}(q'-q)\big]\nonumber\\
&&\hspace{-1.3cm}+(\bar{G}^2)_{21}(q')\bar{G}_{22}(q'+q)
+(\bar{G}^2)_{12}(q)\bar{G}_{22}(q'-q)\nonumber\\
&&+\left(3+\frac{2\rho u^{(3)}}{u''}\right)\big[(\bar{G}^2)_{12}(q')\bar{G}_{11}(q'+q)\nonumber\\
&&+(\bar{G}^2)_{21}(q')\bar{G}_{11}(q'-q)\big]\big\}.\label{A.23}
\end{eqnarray}

The propagator matrix reads explicitely
\begin{equation}\label{x7}
\bar{G} =\det\nolimits^{-1}\bar A^{-1}
\left(
\begin{array}{ccc}
\tilde q^2+u'&,&Sq_0\\-Sq_0&,&\tilde q^2+u'+2\rho u''
\end{array}\right),
\end{equation}
with
\begin{eqnarray}\label{x7a}
\bar{G}^2&=&\det\nolimits^{-2}\bar A^{-2}\\
&&\hspace{-1.3cm}\left(\begin{array}{ccc}
(\tilde q^2+u')^2-S^2q^2_0&,&2Sq_0(\tilde q^2+u'+\rho u'')\\
-2Sq_0(\tilde q^2+u'+\rho u'')&,&(\tilde q^2+u'+2\rho u'')^2-S^2 q^2_0)
\end{array}\right),\nonumber
\end{eqnarray}
and
\begin{equation}\label{A.26A}
\det=(\tilde q^2+u')(\tilde q^2+u'+2\rho u'')+S^2q^2_0.
\end{equation}
For the computation of $\partial_t\bar P_{22}$ we employ the fact that only $\vec q\ '^2\leq k^2$ contributes in the integrands (\ref{A.23}) and replace
\begin{equation}\label{A.26B}
\tilde q\ '^2\to k^2~,~(q'\tilde\pm q)^2\to k^2+z_\pm.
\end{equation}

Including terms up to second order in $\vec q$ we can expand in $z_\pm$
\begin{eqnarray}\label{x8}
&&\hspace{-1.0cm}\partial_t\bar P_{22}(\vec q,0)=\lambda^2\rho_0\int_{q'_0}\int_{\vec q\ '}\theta(k^2-\vec q\ '\ ^2)
\partial_t\big[\bar A(k^2-\vec q\ '\ ^2)\big]\nonumber\\
&&\det\nolimits^{-3}_k\left\{\left(
1-\frac{2z_+(k^2+u'+\rho u'')+z^2_+}{\det_k}\right.\right.\nonumber\\
&&\left.+\frac{4z^2_+(k^2+u'+\rho u'')^2}{\det\nolimits^2_k}\right)\nonumber\\
&&\Big[(k^2+u')^2(k^2+u'+2\rho u''+z_+)\\
&&+(k^2+u'+2\rho u'')^2(k^2+u'+z_+)\nonumber\\
&&+2S^2q'^2_0(k^2+u'+\rho u''-z_+)\Big]
+(z_+\to z_-)\Big\}\nonumber
\end{eqnarray}
with
\begin{equation}\label{A.27A}
z_\pm=\big[(\vec q\ '\pm\vec q)^2-k^2\big]\theta\big[(\vec q\ '\pm \vec q)^2-k^2\big]
\end{equation}
and
\begin{equation}\label{x9}
\det\nolimits_k=(k^2+u')(k^2+u'+2\rho u'')+S^2 q'^2_0.
\end{equation}
Here we use the fact that $z_\pm$ is effectively linear in $\vec q$ for $\vec{q}\ ^2\to 0$ since only momenta $\vec q\ '^2\approx k^2$ contribute for terms involving powers of $z_+$ or $z_-$. Without loss of generality we choose $\vec q=(q,0,\dots),~q>0$ and decompose $\vec q\ '^2=q'^2_1+x_t$ (for $d>1$). The combination of the $\theta$-functions from $\partial_t R_k$ and from $z_+$ restricts the integration range for $\vec q\ '$ to 
\begin{eqnarray}\label{x9a1}
&&0<x_t<k^2,\\
&&{\rm max}\{-\sqrt{k^2-x_t},~\sqrt{k^2-x_t}-q\}<q'_1<\sqrt{k^2-x_t}.\nonumber
\end{eqnarray}
In consequence, the term $\vec q\ '^2-k^2$ in $z_+$ is effectively of the order $q$. It will be convenient to split the $\vec q\ '$-integration into two ranges
\begin{eqnarray}\label{x9B}
I:&& 0<x_t<k^2-\frac14q^2,\nonumber\\
&&\sqrt{k^2-x_t}-q<q'_1<\sqrt{k^2-x_t},\nonumber\\
II:&& k^2-\frac14 q^2<x_t<k^2,\nonumber\\
&&-\sqrt{k^2-x_t}<q'_1<\sqrt{k^2-x_t}.
\end{eqnarray}
For $d=1$ there is no $x_t$-integration and the $q'_1$ integration covers the range $k-q<q'_1<k$. Restricting the $x_t$ and $q'_1$-integrations to this range we write
\begin{eqnarray}\label{x9a}
&&\partial_t\frac{\partial}{\partial q^2}\bar P_{22|q=0}=
-2v_{d-1}\frac{\rho(u'')^2\bar A}{2\pi^2}\nonumber\\
&&\int\limits^\infty_{-\infty}dq'_0\frac{\partial}{\partial q^2}\Big\{\int dx_t
x_t^{\frac{d-3}{2}}\int dq'_1
\nonumber\\
&&\big[(2-\eta)k^2+\eta q'^2_1+\eta x_t\big]\det\nolimits^{-3}_k
(A_1\hat z_++A_2\hat z^2_+)\nonumber\\
&&+(q\to-q)\Big\}_{|q=0}
\end{eqnarray}
with
\begin{eqnarray}\label{x10}
A_1&=&\det\nolimits_k,\\
A_2&=&-(k'+u'+\rho u'')
\label{x11}\nonumber\end{eqnarray}
and
\begin{equation}\label{x12}
\hat z_+=q'^2_1+x_t-k^2+2qq'_1+q^2.
\end{equation}
For $d=1$ the $x_t$-integration and the factor $2v_{d-1}$ are absent. Using $\alpha=\sqrt{k^2-x_t}$ and 
\begin{eqnarray}\label{z1}
&&\int\limits^\alpha_{\alpha-q}dq'_1\big[(2-\eta)k^2+\eta x_t+\eta q'^2_1\big](A_1z_++A_2 z^2_+)\nonumber\\
&&=2k^2q^2\sqrt{k^2-x_t}A_1
\end{eqnarray}
we obtain for $d=1$ at the minimum $\rho=\rho_0,u''(\rho_0)=\lambda$:
\begin{equation}\label{z2}
\partial_t\frac{\partial}{\partial q^2}\bar P_{22|q=0}
=-\frac{2\lambda^2\rho_0\bar A k^3}{\pi^2}
\int\limits^\infty_{-\infty}dq'_0\det\nolimits^{-2}_k.
\end{equation}
For $d>1$ we still need to perform the $x_t$-integration and to include the region II. We employ
\begin{equation}\label{z3}
\int\limits^{k^2}_{k^2-\frac14 q^2}
dx_t F(x_t)=\frac14q^2F(k^2)
\end{equation}
and observe that the integration region II does not contribute in order $q^2$. This yields for $d>1$
\begin{equation}\label{z4}
\partial_t\frac{\partial}{\partial q^2}\bar P_{22|q=0}=-\frac{4v_{d-1}\sigma_d\lambda^2\rho_0\bar A}{\pi^2}
k^{d+2}\int^\infty_{-\infty}dq'_o\det\nolimits^{-2}_k
\end{equation}
with
\begin{equation}\label{z5}
\sigma_d=\int\limits^1_0dxx^{\frac{d-3}{2}}\sqrt{1-x}=\frac{2\pi}{d}\frac{v_d}{v_{d-1}}
\end{equation}
We therefore find for the anomalous dimension
\begin{equation}\label{z6}
\eta=\frac{8v_d}{d\pi}\lambda^2\rho_0k^{d+2}\int^\infty_{-\infty}dq'_0\det\nolimits^{-2}_k.
\end{equation}
We collect the identities $(n\geq 1)$
\begin{eqnarray}\label{z7}
\int\limits^\infty_{-\infty}&&\hspace{-0.5cm}dq'_0\det\nolimits^{-n}_k=\frac{1}{(n-1)!}2^{1-n}\nonumber\\
&&(1\cdot 3 \cdot 5\dots 2n-3)\frac{\pi}{S}B^{-\frac{2n-1}{2}},\\
S^2\int\limits^\infty_{-\infty}&&\hspace{-0.5cm}dq'_0q^{'2}_0\det\nolimits^{-(n+1)}_k=\frac{1}{2n}
\int\limits^{\infty}_{-\infty}dq'_0\det\nolimits^{-n}_k,\nonumber
\end{eqnarray}
with
\begin{equation}\label{z8}
B=(k^2+u')(k^2+u'+2\rho u'')
\end{equation}
such that
\begin{equation}\label{Z9}
\eta=\frac{4v_d}{dS}\lambda^2\rho_0 k^{d+2}B^{-\frac32}.
\end{equation}
In terms of $w$ and $\tilde\lambda$ we obtain our final result eq. (\ref{z10}). 

For the computation of $\partial_tS$ we expand in linear oder in $q_0$
\begin{eqnarray}\label{y1}
&&\partial_t\bar P_{21}(\vec  q=0,q_0)=2\rho(u'')^2 Sq_0\nonumber\\
&&\int_{\vec q\ '}\theta(k^2-\vec q\ '^2)\partial_t\big[\bar A(k^2-\vec q\ '^2)\big]\nonumber\\
&&\int\limits_{q'_0}\det\nolimits^{-3}_k\Big\{(k^2+u'+2\rho u'')^2-
\left(3+\frac{2\rho u^{(3)}}{u''}\right)(k^2+u')^2\nonumber\\
&&\qquad -2S^2q'^2_0\left(1+\frac{\rho u^{(3)}}{u''}\right)\Big\}
\end{eqnarray}
We evaluate eq. (\ref{y1}) at the minimum $u'=0,u''=\lambda,u^{(3)}=\gamma$. The $\vec q\ '$-integration is trivial and the $q'_0$-integration follows from (\ref{z7}). One finds
\begin{eqnarray}\label{y2}
\partial_t S&=&\eta S-\frac{v_d}{2d}\left(1-\frac{\eta}{d+2}\right)\lambda k^{d-2}w(1+w)^{-5/2}\nonumber\\
&&\big[8-4w-3w^2+(8+w)\rho_0\gamma/\lambda\big].
\end{eqnarray}

\section*{APPENDIX C: Quadratic frequency dependence}
\renewcommand{\theequation}{C.\arabic{equation}}
\setcounter{equation}{0}

In this appendix we extend our truncation by adding to eq. (\ref{aa1}) a term quadratic in the $\tau$-derivatives
\begin{equation}\label{F1}
\Delta\Gamma_k=- V \int_x\phi^*\partial^2_\tau\phi.
\end{equation}
The inverse propagator matrix involves now
\begin{equation}\label{F2}
\bar P=\bar A\left(\begin{array}{ccc}
\vec q\ ^2+ V q^2_0+u'+2\rho u''&,&-Sq_0\\
Sq_0&,&\vec q\ ^2+ V  q^2_0+u'\end{array}\right)
\end{equation}
and the flow equation for $ V  $ is defined by
\begin{equation}\label{F3}
\partial_t V  =\eta  V  +\frac{1}{2\bar A}\frac{\partial^2}{\partial q^2_0}\partial_t\bar P_{22|q=0}.
\end{equation}
In order to evaluate eq. (\ref{x4}) we take into account the modification of the propagator
\begin{eqnarray}\label{F4}
&&\bar G=\bar A^{-1}\det\nolimits^{-1}\\
&&\left(\begin{array}{ccc}
\tilde q^2+ V  q^2_0+u'&,&Sq_0\\
-Sq_0&,&\tilde q^2+ V  q^2_0+u'+2\rho u''\end{array}\right)\nonumber
\end{eqnarray}
with 
\begin{equation}\label{F5}
\det=(\tilde q^2+ V  q^2_0+u')
(\tilde q^2+ V  q^2_0+u'+2\rho u'')+S^2q^2_0
\end{equation}
and
\begin{eqnarray}\label{F6}
(\bar G^2)_{11}&=&\bar A^{-2}\det\nolimits^{-2}\big[(\tilde q^2+Vq^2_0+u')^2-S^2q^2_0\big],\\
(\bar G^2)_{22}&=&\bar A^{-2}\det\nolimits^{-2}
\big[(\tilde q^2+Vq^2_0+u'+2\rho u'')^2-S^2q^2_0\big],\nonumber\\
(\bar G^2)_{12}&=&-(\bar G^2)_{21}=2\bar A^2\det\nolimits^{-2}
Sq_0(\tilde q^2+Vq^2_0+u'+\rho u'').\nonumber
\end{eqnarray}

In the definition (\ref{F3}) the spacelike external momentum is taken at a vanishing value, $\vec q=0$. We can therefore replace in all propagators in eq. (\ref{x4}) $\tilde q^2\to k^2$ and perform the $\vec q\ '$ integration
\begin{eqnarray}\label{F7}
\int_{\vec q \ '}\partial_tR_k(\vec q\ ')&=&\int_{\vec q\ '}\theta(k^2-\vec q\ '^2)\partial_t
\big[\bar A(k^2-\vec q\ '^2)\big]\nonumber\\
&=&\frac{8v_d}{d}\left(1-\frac{\eta}{d+2}\right)\bar A k^{d+2}.
\end{eqnarray}
Expanding eq. (\ref{x4}) to second order in $q_0$ we find the flow equation for $ V  $
\begin{eqnarray}\label{F8}
\partial_t V  &=&\eta  V  +\frac{32v_d}{d}\left(1-\frac{\eta}{d+2}\right)\rho u''^2 k^{d+2}\int_{q'_0}\det\nolimits^{-3}_k\nonumber\\
&&\Big\{- V  \det\nolimits_k+4 V  ^2q'^2_0(k^2+ V  q'^2_0+u'+\rho u'')\nonumber\\
&&-S^2[k^2+u'+\rho u''-3 V  q'^2_0]\Big\}.\nonumber\\
\end{eqnarray}
Here $\det\nolimits_k$ replaces in eq. (\ref{F5}) $\tilde q^2\to k^2$. In the disordered phase one finds $\partial_tV=0$ in agreement with eq. (\ref{39A}). 

In the ordered phase we evaluate eq. (\ref{F8}) at the minimum $(u'=0,u''=\lambda)$ and use the integrals
\begin{equation}\label{F9}
\int_{q'_0}\det\nolimits^{-n}_k( V   q'^2_0)^m= V  ^{-1/2}k^{2m-4n+1}A_{n,m}(w,s)
\end{equation}
with
\begin{equation}\label{F10}
s=\frac{S}{k\sqrt{ V  }}
\end{equation}
and
\begin{equation}\label{F11}
A_{n,m}(w,s)=\frac{1}{2\pi}\int\limits^\infty_{-\infty}dx~x^{2m}
\big[(1+x^2)(1+w+x^2)+s^2x^2\big]^{-n}.
\end{equation}
This yields
\begin{eqnarray}\label{F12}
&&\partial_t\ln  V  =\eta+\frac{16v_d}{d}\left(1-\frac{\eta}{d+2}\right)w
\lambda k^{d-2}( V  k^2)^{-1/2}\nonumber\\
&&\Big\{-A_{2,0}+4\left(\left(1+\frac w2\right)A_{3,1}+A_{3,2}\right)\nonumber\\
&&-s^2[\left(1+\frac w2\right)A_{3,0}-3 A_{3,1}]\Big\}.
\end{eqnarray}

The computation of $\eta$ in appendix B remains essentially unchanged and one finds from eq. (\ref{z6}) 
\begin{equation}\label{F13}
\eta=\frac{8v_d}{d}w\lambda k^{d-2}( V  k^2)^{-1/2}A_{2,0}(w,s). 
\end{equation}

It is instructive to investigate the limiting cases $s\to 0$ and $s\to\infty$. For $s=0$ the propagator (\ref{F4}) becomes diagonal. The action has now a ``relativistic'' dynamic term involving two $\partial_\tau$-derivatives. This model is well understood and corresponds to the classical $O(2)$-model in $d+1$ dimensions. The euclidean space-time symmetry $SO(d+1)$ obtains by a simple rescaling $\tau= V ^{1/2}\tau', V  \partial^2_\tau=\partial^2_{\tau'}$. The effective action $\Gamma_{k\to0}$ should respect this enhanced $SO(d+1)$ symmetry. Also the flow equations would automatically respect the $SO(d+1)$ symmetry if we had chosen a cutoff function $R_k$ consistent with this symmetry. However, our cutoff (\ref{aa3}) violates $SO(d+1)$ since it only acts on $d$ momenta $\vec q$. The flow will therefore not respect $SO(d+1)$ and the full symmetry should only appear for $k\to 0$ where the cutoff effects are absent. In particular, this implies that $V(k)$ should approach a constant for $k\to 0$. We may discuss this issue in some more detail.

Combining (\ref{F19}) with (\ref{F12}) yields for $s=0$
\begin{eqnarray}\label{F19}
\partial_t\ln  V  &=&\frac{8v_d}{d}w\tilde\lambda\Big\{-\left(1+\frac{2\eta}{d+2}\right)A_{2,0}\\
&&+8\left(1-\frac{\eta}{d+2}\right)\left(\left(1+\frac w2\right)A_{3,1}+A_{3,2}\right)\Big\}\nonumber
\end{eqnarray}
where we define
\begin{equation}\label{F20}
\tilde\lambda=\lambda k^{d-2}( V  k^2)^{-1/2}.
\end{equation}

We note that the integrals $A_{n,m}$ obey relations
\begin{eqnarray}\label{F14}
&&\frac{\partial}{\partial w}A_{n,m}=-n(A_{n+1,m}+A_{n+1,m+1}),\nonumber\\
&&\frac{\partial}{\partial s^2}A_{n,m}=-nA_{n+1,m+1}.
\end{eqnarray}
We also employ the integral
\begin{equation}\label{F15}
\frac{1}{2\pi}\int\limits^\infty_{-\infty}dx(x^2+\alpha)^{-1}(x^2+\beta)^{-1}=
\frac{1}{2(\alpha\sqrt{\beta}+\beta\sqrt{\alpha})}
\end{equation}
in order to compute
\begin{eqnarray}\label{F16}
A_{1,0}(w,0)&=&\frac12(1+w+\sqrt{1+w})^{-1},\\
A_{2,0}(w,0)&=&\frac14(1+w+\sqrt{1+w})^{-3}
(1+w+3\sqrt{1+w}),\nonumber\\
A_{3,0}(w,0)&=&\frac{3}{16}(1+w+\sqrt{1+w})^{-5}\nonumber\\
&&\big\{5(2+w)\sqrt{1+w}+11+11w+w^2\big\}.\nonumber
\end{eqnarray}

Using eq. (\ref{F14}) one then obtains
\begin{eqnarray}\label{F17}
&&A_{2,1}(w,0)=\frac14(1+w+\sqrt{1+w})^{-3}(2+w),\nonumber\\
&&A_{3,1}(w,0)+A_{3,2}(w,0)=\frac{1}{8}
(1+w+\sqrt{1+w})^{-4}\nonumber\\
&&\qquad\qquad(5+2w+\frac12(1+w)^{1/2}+\frac32(1+w)^{-1/2}),\nonumber\\
&&A_{3,1}(w,0)=\frac{1}{16}(1+w+\sqrt{1+w})^{-5}\nonumber\\
&&\big\{w^2-3w-7+5(1+w)^{3/2}-9(1+w)^{1/2}\big\},
\end{eqnarray}
and 
\begin{eqnarray}\label{F18}
&&8\left(\left(1+\frac w2\right)A_{3,1}+A_{3,2}\right)=\frac14
(1+w+\sqrt{1+w})^{-5}\nonumber\\
&&\big\{w^3+5w^2+23w+28+5(1+w)^{5/2}\nonumber\\
&&-4(1+w)^{3/2}+27(1+w)^{1/2}\big\}.
\end{eqnarray}

Let us consider large $w$ where the terms $\sim \eta/(d+2)$ can be neglected. We note that the leading term in the combination (\ref{F18}) $\sim 1/(4w^2)$ cancels precisely the same term in $A_{2,0}$. While $\eta$ decreases for large $w$ as $\tilde\lambda w^{-1}$, the leading term in $\partial_t\ln  V  \sim \tilde\lambda w^{-2}$ is suppressed by an additional factor $w^{-1}$. This feature is consistent with the requirement $ V  (k\to 0)\to  V _0$. Establishing for our cutoff the asymptotic constancy for $ V (k\to 0)$ for arbitrary initial conditions with $S=0, V \neq 0$ has not yet been done. We simply recall that any valid truncation must obey this property due to the $SO(d+1)$ symmetry. 

For large $w$ we find for the anomalous dimension
\begin{equation}\label{A21a}
\eta=\frac{2v_d\lambda k^{d-3}}{d\sqrt{V}w}.
\end{equation}
For $d=1$ this yields
\begin{equation}\label{A21a1}
\eta=\frac{1}{4\pi\sqrt{V}\rho_0}
\end{equation}
and we recover the well known formula for the classical two dimensional $O(N)$ models \cite{CWFE,GKT}, with $\kappa=\sqrt{V}\rho_0$. This is closely linked to the flow of the coupling $g^2=1/(2\kappa)$ in the non-abelian nonlinear $\sigma$-models \cite{CWFE,GKT}, which obeys $(N=2M)$
\begin{equation}\label{A21b}
\partial_t g^2=-\frac{N-2}{2\pi}g^4.
\end{equation}
A perturbative expansion for small $g^2$ or small $\kappa^{-1}$ becomes possible.

In the opposite limit $s\to \infty$ we can neglect in eq. (\ref{F8}) all terms involving $ V $ such that  (with eq. (\ref{z7}) and $\tilde\lambda=\lambda k^{d-2}/S)$
\begin{eqnarray}\label{F21}
&&\partial_t V =-\frac{16v_d}{d}\left(1-\frac{\eta}{d+2}\right)w\tilde\lambda k^8
S^3\left(1+\frac w2\right)\int_{q'_0}\det\nolimits^{-3}_k
\nonumber\\
&&=-\frac{3v_d}{d}\left(1-\frac{\eta}{d+2}\right)w\left(1+\frac w2\right)(1+w)^{-5/2}
\tilde\lambda S^2k^{-2}.\nonumber\\
\end{eqnarray}
The r.h.s. is negative such that $ V $ is driven to positive values if we start with a microscopic value $ V (\Lambda)=0$. In this regime we find
\begin{eqnarray}\label{F22}
\partial_t\left(\frac{ V k^2}{S^2}\right)&=&\partial_ts^{-2}=2(1+\eta_S)s^{-2}-\alpha_ V ,\\
\alpha_ V &=&\frac{3v_d}{d}\left(1-\frac{\eta}{d+2}\right)w\left(1+\frac w2\right)(1+w)^{-5/2}\tilde\lambda.\nonumber
\end{eqnarray}
For (approximately) constant $w$ and $\tilde\lambda$ one observes two qualitatively different behaviors. For $\eta_S>-1$ the evolution of $V\sim s^{-2}$ tends towards an infrared stable partial fixed point
\begin{equation}\label{F23}
s^{-2}_*=\frac{\alpha_ V }{2(1+\eta_S)}.
\end{equation}
In contrast, for $\eta_S<-1$ the combination $s^{-2}$ increases fast to large values. The linear dynamic term $\sim S$ becomes subdominant as $s$ approaches zero according to 
\begin{equation}\label{F24}
\partial_t s=-(1+\eta_S)s+\frac12\alpha_ V s^3.
\end{equation}
Values $\eta_S<-1$ therefore suggest a crossover from an initial evolution where the term linear in $\partial_\tau$ characterizes the effective action to a ``relativistic regime'' where the term quadratic  in $\partial_\tau$ dominates. If this happens and the flow is not stopped due to $w\to\infty$ one expects the long distance behavior to be governed by the relativistic model. For the infrared physics of the Goldstone modes the relativistic regime applies for $\eta_S\leq 0$. Our findings suggest that this is realized for $d\leq 3$. 

Of course, once $s$ is small eq. (\ref{F24}) is no longer quantitatively correct since it was obtained in the limit $s\to \infty$. Also the computation of $\eta_S$ has now to be performed in the relativistic regime. For small $s$ we can use (\ref{F12})
\begin{equation}\label{F25}
\partial_ts=-(1+\eta_S+\frac12\partial_t\ln  V )s
\end{equation}
and observe that the $s$-independent term in $\partial_t\ln  V $ should be small due to the relativistic $SO(d+1)$-symmetry, while the $s$-dependent term is negative, reproducing qualitatively eq. (\ref{F24}). One concludes that the criterion for $s(k\to 0)\to 0$ remains $\eta_S<-1$, whereas the criterion for a relativistic Goldstone regime $s\ll\sqrt{w}$ applies for $\eta_S\leq 0$.

\section*{APPENDIX D: Fixed point properties for $V=0$}
\renewcommand{\theequation}{D.\arabic{equation}}
\setcounter{equation}{0}
In this appendix we briefly address some properties of the possible fixed points for $w\neq0$ in truncations with a linear $\tau$-derivative, i.e. for $V=0$. We start with the simplest truncation (\ref{Q1}). In terms of the variable
\begin{equation}\label{M1}
\sigma=\frac{v_d}{d}\frac{\tilde\lambda w}{\sqrt{1+w}}
\end{equation}
the condition $\partial_t w=0$ reads
\begin{eqnarray}\label{M2}
&&-2+\left(\frac32+\frac{2}{1+w}-\frac{27}{2(1+w)^2}\right)\sigma\nonumber\\
&&-\frac{3}{(d+2)(1+w)}\left(1-\frac{9}{(1+w)^2}\right)\sigma^2=0.
\end{eqnarray}
Similarly, $\partial_t\tilde\lambda=0$ requires either $\tilde\lambda=0$ or
\begin{equation}\label{M3}
d-2+\frac{(2-w)^2}{w(1+w)^2}\sigma-\frac{2(4-6w-w^2)}{(d+2)w(1+w)^3}\sigma^2=0.
\end{equation}
No fixed point exists for $\tilde\lambda=0~,~w\neq 0$. One may use a linear combination of eqs. (\ref{M2}), (\ref{M3}) in order to express $\sigma$ in terms of $w$ and then solve the remaining equation for $w$ numerically. Alternatively, one may numerically solve the flow equations (\ref{Q1}) for $k\to 0$ - an infrared stable fixed point can be found easily without the need of tuning initial conditions. One finds a fixed point for all $d<2$, as shown in table 1. 

As we have discussed in sect. \ref{crossover} the properties of fixed point $(C)$ are strongly affected by the inclusion of the relativistic kinetic term $\sim V$. It disappears for $d>1$ and remains for $d=1, M=1$. Nevertheless, the truncation $V=0$ may be relevant for the initial running before a sizeable $V$ is built up by the flow. It is therefore interesting to know to what extent the properties of fixed point $(C)$ are robust with respect to extensions of the truncation which keep $V=0$. For small $\eta/(d+2)$ the terms $\sim\sigma^2$ in eqs. (\ref{M2}) (\ref{M3}) are subleading. The fixed point  value $w_*$ is strongly influenced by the relative size of the contributions linear in $\sigma$ in eq. (\ref{M2}). While the radial mode contributes with a negative sign, the Goldstone mode gives a positive contribution. This allows us to roughly estimate the effects of extended truncations. Adding the coupling $\gamma=u^{(3)}(\rho_0)$ will enhance the weight of the radial contribution, thus disfavoring very high values of $w_*$. On the other hand, a contribution $\sim Y\rho_0$ in the wave function renormalization of the radial mode will diminish its weight. We have investigated in appendix E the role of the coupling $\gamma$. While fixed point $(C)$ persists, its location becomes rather unstable with respect to the order of the truncation as $d$ approaches two. Only for $d=1$ the fixed point seems relatively robust. 

\section*{APPENDIX E: \\Extended truncation with six point vertex}
\renewcommand{\theequation}{E.\arabic{equation}}
\setcounter{equation}{0}

The neglected third derivative of $u$ with respect to $\rho$ contributes to $\eta_S$ (cf. eq. (\ref{y1})) and to the running of $\rho_0$ (cf. eq. (\ref{aa10})). In this section we will extend the truncation by including $u^{(3)}$, while still neglecting $u^{(4)}$ and higher $\rho$-derivatives. Within the approximation of pointlike interactions (and neglecting V), the formulae for $\partial_t\rho_0,\eta$ and $\eta_S$ are then complete, while the neglected term $u^{(4)}$ is missing in $\partial_t\lambda$ as well as $\partial_tu^{(3)}$.

In the symmetric phase we define $\gamma=u^{(3)}(\rho=0)$ and infer from eq. (\ref{A.XX}) the flow equation 
\begin{eqnarray}\label{E1}
\partial_t\gamma&=&3\eta\gamma+\frac{24v_d}{dS}
\left(1-\frac{\eta}{d+2}\right)\nonumber\\
&&\frac{k^{d+2}\lambda}{(k^2+m^2)^2}
\left(\gamma-\frac{2\lambda^2}{k^2+m^2}\right).
\end{eqnarray}
In terms of the dimensionless coupling 
\begin{equation}\label{E2}
\tilde\gamma=\frac{\gamma}{S^2}k^{2(d-1)}
\end{equation}
we obtain the scaling form of the flow
\begin{eqnarray}\label{E3}
\partial_t\tilde\gamma&=&\big[2(d-1)+3\eta+2\eta_S\big]\tilde\gamma\\
&&+\frac{24v_d}{d}\left(1-\frac{\eta}{d+2}\right)
\frac{\tilde\lambda}{(1+w)^2}
\left(\tilde\gamma-\frac{2\tilde\lambda^2}{1+w}\right).\nonumber
\end{eqnarray}
We note that for $\tilde\lambda\neq0$ the flow has no solution $\tilde\gamma=0$. For fixed point $(B)$ for $d<2$ with $w_*=0~,~\tilde\lambda_*\neq 0~,~\eta=\eta_S=0$ one finds an IR-attractive fixed point for $\tilde\gamma$
\begin{eqnarray}\label{E4}
\tilde\gamma_*&=&2\tilde\lambda^2_*\big[2(d-1)+\frac{24v_d}{d}\tilde\lambda_*\big]^{-1}\nonumber\\
&=&\frac{\tilde\lambda^2_*}{5-2d}.
\end{eqnarray}

In the SSB regime we define $\gamma=u^{(3)}(\rho_0)$ and use again eq. (\ref{E2}), resulting in the flow equation
\begin{eqnarray}\label{E5}
\partial_t\tilde\gamma&=&\big[2(d-1)+3\eta+2\eta_S\big]\tilde\gamma\nonumber\\
&&-\frac{2v_d}{d}\left(1-\frac{\eta}{d+2}\right)
\frac{\tilde\lambda^3}{(1+w)^{7/2}}\tilde R,
\end{eqnarray}
with 
\begin{equation}\label{E6}
\tilde R=A_1-A_2\frac{\tilde\gamma}{\tilde\lambda^2}-A_3\left(\frac{\tilde\gamma}{\tilde\lambda^2}\right)^2
+A_4\left(\frac{\tilde\gamma}{\tilde\lambda^2}\right)^3 
\end{equation}
and
\begin{eqnarray}\label{E7}
&&A_1=24-33w-\frac92 w^2+\frac{15}{4}w^3+\frac{15}{8}w^4,\nonumber\\
&&A_2=12-42w+\frac32 w^2+\frac{33}{4}w^3+\frac{27}{8}w^4,\nonumber\\
&&A_3=6w-12w^2-\frac32w^3-\frac38w^4\nonumber\\
&&A_4=\frac32w^3-\frac38w^4.
\end{eqnarray}
For $w=0$ we recover eq. (\ref{E3}). For $w\to \infty$ one finds for the leading term $\sim w^{1/2}$ 
\begin{eqnarray}\label{E8}
&&\partial_t\tilde\gamma=2\eta_S\tilde\gamma-\frac{v_d}{4d}\tilde\lambda^3 w^{1/2}\nonumber\\
&&\left[15-27\frac{\tilde\gamma}{\tilde\lambda^2}+3\left(\frac{\tilde\gamma}{\tilde\lambda^2}\right)^2-3
\left(\frac{\tilde\gamma}{\tilde\lambda^2}\right)^3\right].
\end{eqnarray}

While the anomalous dimension $\eta$ is not affected by $\gamma$ we find for $\eta_S$ a correction (\ref{y2}) 
\begin{equation}\label{E9}
\Delta\eta_S=\frac{v_d}{4d}
\left(1-\frac{\eta}{d+2}\right)
\frac{\tilde\gamma}{\tilde\lambda}w^2(8+w)(1+w)^{-5/2}.
\end{equation}
This contribution is positive and increases $\sim w^{1/2}$ for large $w$. Combining with eq. (\ref{E8}) yields for the flow of $\tilde\gamma$ at large $w$
\begin{eqnarray}\label{E10}
\partial_t\tilde\gamma&=&-\frac{v_d}{4d}\tilde\lambda^3w^{1/2}\nonumber\\
&&\left[15-15\frac{\tilde\gamma}{\tilde\lambda^2}+\left(\frac{\tilde\gamma}{\tilde\lambda^2}\right)^2-3
\left(\frac{\tilde\gamma}{\tilde\lambda^2}\right)^3\right].
\end{eqnarray}
Again, for fixed $\lambda_*$ and $w_*$ this yields an $IR$-stable fixed point for
$\tilde\gamma/\tilde\lambda^2$ and therefore for $\tilde\gamma$. In view of the fixed point behavior for $w\to0$
 and $w\to\infty$ it may not be surprising that numerical solutions of the flow for $d<2$ will show an
 infrared stable fixed point $(w_*~,~\tilde\lambda_*~,~\tilde\gamma_*)$, corresponding to $(C)$.

We finally need the corrections to the flow of $w$ and $\tilde\lambda$. In the symmetric regime they vanish, just as the correction to $\eta_S$. The six-point vertex does not influence the lower vertices. In contrast, we find in the SSB regime
\begin{equation}\label{A9a}
\Delta(\partial_t\rho_0)=-\frac{2v_d}{d}\left(1-\frac{\eta}{d+2}\right)
\frac{\tilde\gamma}{\tilde\lambda}\frac{w}{(1+w)^{3/2}}\rho_0.
\end{equation}
For the evolution of $\lambda$ we now have to include the effect of the change of the location of the minimum
\begin{equation}\label{A9b}
\partial_t\lambda=\partial_t u''(\rho_0)+\gamma\partial_t\rho_0.
\end{equation}
This yields
\begin{eqnarray}\label{A9c}
\Delta(\partial_t\tilde\lambda)=\Delta\eta_S\tilde\lambda-\frac{2v_d}{d}
\left(1-\frac{\eta}{d+2}\right)
\frac{\tilde\lambda^2}{(1+w)^{5/2}}\nonumber\\
\left\{\frac{w\tilde\gamma}{\tilde\lambda^2}(-3+2w+\frac12 w^2)
+\frac{3w}{4}\left(\frac{w\tilde\gamma}{\tilde\lambda^2}\right)^2\right\}
\end{eqnarray}
and
\begin{eqnarray}\label{A9d}
\Delta\partial_t w&=&-\frac{2v_d}{d}\left(1-\frac{\eta}{d+2}\right)
\frac{\tilde\gamma}{\tilde\lambda}
\frac{w^2}{(1+w)^{5/2}}\nonumber\\
&&\left\{-2+3w+\frac12 w^2+\frac{3w^2\tilde\gamma}{4\tilde\lambda^2}\right\}.
\end{eqnarray}

We conclude that the fixed points $(A)$ and $(B)$ have the same values $w_*$ and $\tilde\lambda_*$ as computed in the simple truncation $\gamma=0$. For fixed point $(A)$ one has $\tilde\gamma_*=0$ and the $\tilde\gamma$-direction is $IR$-stable for $d>1$. For fixed point $(B)$ the value of $\tilde\gamma_*$ is given by eq. (\ref{E4}) and the $\tilde\gamma$-direction is $IR$-stable for $d<5/2$. As before, the quantum phase transition corresponds to (B) for $d<2$ and to (A) for $d>2$. The location of fixed point $(C)$, however, depends on the truncation. The values for the extended truncation are shown in table 2. 

\medskip
\begin{center}
\begin{tabular}{|l|l|l|l|l|l|}\hline
$d$&$w_*$&$\tilde\lambda_*$&$\tilde\gamma_*/\tilde\lambda^2_*$&$\eta$&$\eta_S$\\\hline
$1$&$3.22$&$35.6$&$0.45$&$2.1$&$-3.14$\\
$1.9$&$3.69$&$123.5$&$0.48$&$2.07$&$-3.99$\\
$1.99$&$3.74$&$140.6$&$0.49$&$2.06$&$-4.07$\\\hline
\end{tabular}
\end{center}

\noindent
table II: Fixed point values for $(C)$ in $\rho^3$ truncation.

\medskip
\noindent
We observe that in this truncation the fixed point comes close to a value where $\eta=2~,~\eta_S=-(d+2)$ for which the fluctuation effects are relatively weak and the running of $\bar\rho_0~,~\bar\lambda~,~\bar\gamma$ is therefore slow. This may well be an artefact of the truncation and an investigation beyond the approximation of pointlike couplings becomes necessary. In particular, we note that in the present truncation the expansion in small powers of $\tilde\lambda$ near $d=2$ gets modified. For $\partial_t\tilde\lambda$ and large $w$ the terms $\sim\tilde\lambda^2 w^{1/2}$ do not cancel anymore if $\tilde\gamma\sim\tilde\lambda^2$, implying $\partial_t\tilde\lambda\sim\tilde\lambda$ once $w\sim\tilde\lambda^{-2}$. In any case, the large negative value of $\eta_S$ indicates that the flow around fixed point $(C)$ will strongly be affected by the ``relativistic dynamic term'' $\sim V\varphi^*\partial^2_\tau\varphi$, as discussed in sect. IX.

\section*{APPENDIX F: Scaling behavior for large $w$}
\label{scalingbehavior}
\renewcommand{\theequation}{F.\arabic{equation}}
\setcounter{equation}{0}
In this appendix we investigate the flow equations (\ref{Q1}) for large values of $w$. In particular, for $d\geq 2$ the flow for $k\to 0$ necessarily ends in this region since no fixed point with $w\neq 0$ is present even in the simplest truncation. Including the ``relativistic dynamic term'' $\sim V$ one expects that for all $d>1$ the flow in the ordered phase obeys $w(k\to 0)\to \infty$. We study here the simplest truncation and comment only briefly the extended truncations. 
A comparison of the results for the simplest truncation in this appendix with the results for a relativistic dynamic term in sect. \ref{crossover} demonstrates the importance of the relativistic dynamic term.

In the limit $w\to\infty$ one observes for the equation \eqref{Q1}
\begin{eqnarray}\label{Q4}
\partial_tw&=&w\left(-2+\frac{3v_d}{2d}\tilde\lambda
w^{1/2}-\frac{3v^2_d}{d^2(d+2)}\tilde\lambda^2\right)+\dots,\nonumber\\
\partial_t\tilde\lambda&=&\tilde\lambda\left(d-2+\frac{v_d}{d}\tilde\lambda w^{-1/2}+
\frac{2v^2_d}{d^2(d+2)}\tilde\lambda^2 w^{-1}\right)+\dots\nonumber\\
\end{eqnarray}
The evolution of $w$ depends on 
\begin{equation}\label{Q4a}
\zeta=\frac{3v_d}{2d}\tilde\lambda w^{1/2}
\end{equation}
which obeys
\begin{eqnarray}\label{Q5}
\partial_t\zeta&=&\zeta
\left\{d-3+\frac{\zeta}{2}\right.\\
&&\left.+\frac{2}{3w}\left[\zeta-\frac{\zeta^2}{d+2}+\frac{4\zeta^2}{3(d+2)w}\right]\right\}.
\end{eqnarray}
For large $w$ the last term can be neglected and we obtain a simple closed equation for the flow of $\zeta$.

For $d\geq 3$ the combination $\zeta$ vanishes in the infrared $(k\to 0)$ and we obtain the leading behavior for $d>3$
\begin{eqnarray}\label{Q6}
w\sim k^{-2}~,~\tilde\lambda \sim k^{d-2}~,~\lambda\sim S,
\rho_0\sim S^{-1}.
\end{eqnarray}
Here the case $d=3$ is special due to the slow logarithmic running of $\zeta$
\begin{equation}\label{XXAA}
\zeta(k)=
\frac{\zeta(k_0)}{1+\big(\zeta(k_0)/2\big)\ln(k_0/k)},
\end{equation}
where actually $w\sim k^{-2+\zeta}~,~\tilde\lambda\sim k~,~\rho_0\sim\zeta^2/S$. For large $w$ the leading term for $\eta_S$ is given by eq. (\ref{y3}) $\eta_S=-\zeta$ such that $S$ approaches a constant in the infrared for $d>3$. For $d=3$ one obtains $S\sim\zeta^2$ such that $S$ and $\lambda$ vanish logarithmically according to eq. (\ref{XXAA}).  
\be\label{F.6A}
S\sim \lambda\sim\ln^{-2}\left(\frac{k_0}{k}\right).
\ee
Also $\eta$ vanishes for $w\to\infty$ and the infrared behavior of the flow simply stops, with fixed $\rho_0$ and $\bar A$. For $d>3$ the flow is ultraviolet dominated such that 
$\xi^{-2}_R=2\lambda(k=0)\rho_0(k=0)\sim \lambda(\Lambda)\rho_0(\Lambda)=\tilde\sigma$, corresponding to a critical exponent $\nu=1/2$. We recall, however, that $\xi_R$ only appears in the exponential decay of $\bar G_{11}$ for $r\to\infty$, while $\bar G_{22}$ shows a powerlike decay given for $\eta=0$ by eq. (\ref{a12g}). For large $r$ the correlation function $\langle\bar\phi^*(\vec r)\bar\phi (0)\rangle =\frac12(\bar G_{11}+\bar G_{22})$ is dominated by the ``Goldstone contribution'' $\bar G_{22}$.

The situation is different for $d<3$. Now the flow of $\zeta$ exhibits a partial infrared fixed point
\begin{equation}\label{Q7}
\zeta_*=2(3-d).
\end{equation}
The flow of $w$ obeys near this fixed point $(d\neq 2)$
\begin{equation}\label{Q8}
\partial_t w=(4-2d)w
\end{equation}
and we find the asymptotic behavior
\begin{eqnarray}\label{Q9}
w&\sim&k^{4-2d}~,~\tilde\lambda\sim k^{d-2}~,~\lambda\sim S,\nonumber\\
\rho_0&\sim&S^{-1}k^{2(3-d)}.
\end{eqnarray}
For $d>2$ the asymptotic value of $w$ increases and the flow always ends in the regime where $w\to \infty$. For $d<3$ the infrared behavior of $S$ depends on $k$ and we obtain for $\zeta=\zeta_*~,~w\to\infty$
\begin{equation}\label{Q10}
\eta_S=2(d-3)~,~S\sim k^{2(3-d)}.
\end{equation}
According to eq. (\ref{FFA1}) we note the modified relative scaling of time and space in the extreme infrared, 
\begin{equation}\label{B6c}
z=2(d-2).
\end{equation}
The anomalous dimension $\eta$ vanishes in this limit. We conclude that $\rho_0$ reaches a constant value, while $\lambda$ and $S$ vanish asymptotically
\begin{equation}\label{Q11}
\lambda\sim S\sim k^{2(3-d)}~,~\rho_0{\to{\rm const.}}~,~\bar A\to{\rm const.}
\end{equation}

Eq. (\ref{Q10}) suggests that $\eta_S$ increases monotonically with $d$. We may define $d_r$ such that $\eta_S>-1$ for $d>d_r$. Eq. (\ref{Q10}) would imply $d_r=2.5$. For $d<d_r$ the relativistic  kinetic term $\sim V$ dominates and the asymptotic equations for large $w$ discussed in this appendix loose their validity. As we argue in sects. \ref{crossover}, \ref{goldstoneregime} the linear kinetic term $\sim S$ becomes subdominant for the infrared behavior of the Goldstone boson physics whenever $S$ vanishes, i.e. for $\eta_S\leq 0$. We conclude from eq. \eqref{Q10} that this happens for all dimensions $d\leq 3$. 

Even for $d<d_r$ only the infrared behavior in the dense regime is modified whereas other features continue to be reasonably represented by the simplest truncation. As an example we may consider the correlation length for the radial mode. For the definition of the correlation length we include only fluctuations with momenta $\vec q\ ^2>k^2_{ph}=\xi^{-2}_R$, resulting in $\xi^{-2}_R=2\lambda(k_1)\rho(k_1)$ with $w(k_1)=1$. The behavior for  $w\gg1$ corresponds to momenta $\vec q\ ^2\ll\xi^{-2}_R$ and does not affect the scaling $\xi_R\sim\tilde\sigma^{-1/2}$. 

The particular scaling properties for $2<d<3$ mainly affect the extreme infrared behavior of the propagator $G_{11}$. For $\vec q\ ^2\to 0$ we may replace $2\lambda\rho_0\to 2\lambda(\vec q)\rho_0$ with
\begin{equation}\label{B6a}
\lambda(\vec q)=\lambda(k_1)\left(\frac{\vec q\ ^2}{k^2_1}\right)^{-\frac{\eta_S}{2}}.
\end{equation}
We expect a similar qualitative behavior even in presence of the relativistic dynamic term. The value of $\eta_S$ may be modified, however. We argue that the infrared physics for $d\leq 3$ is actually described by the $d+1$-dimensional $O(2)$ model. This suggests $\eta_S=d-3$ instead of $2(d-3)$.  This yields the leading part of the static propagator $G_{11}$
\begin{equation}\label{B6b}
\lim_{\vec q\ ^2\to 0}G_{11}=\frac{1}{2\lambda(k_1)\rho_0}
\left(\frac{\vec q\ ^2}{k^2_1}\right)^{d-3}.
\end{equation}

Summarizing our simplest truncation for the ordered phase, we find that for $d>2$ the flow drives $w$ always towards large values and the condensate $\rho_0$ or $\bar \rho_0$ settles at a constant value. For $d> 3$ also the interaction strength $\lambda$ and the coefficient $S$ reach constant values, whereas for $2<d<3$ both $\lambda$ and $S$ vanish asymptotically. For $d<2$, in contrast, the flow drives $w$ towards a fixed point value $w_*$. Indeed, starting with very large $w$ eq. (\ref{Q8}) implies for $d<2$ a decrease of $w$. This continues until corrections $\sim w^{-1}$ begin to be important. 

For the upper critical dimension $d=2$ the situation is special. The coupling $\tilde\lambda$ becomes now a marginal coupling. For small enough $\tilde\lambda$ the running effectively stops and we can take a small $\tilde\lambda$ as a free parameter. Expanding in powers of $\tilde\lambda$ yields
\begin{eqnarray}\label{H128a}
\partial_tw&=&w\left\{-2+\frac{\tilde\lambda}{32\pi}\frac{w}{\sqrt{1+w}}\right.\nonumber\\
&&\left.\left(3+\frac{4}{1+w}-
\frac{27}{(1+w)^2}\right)\right\},\nonumber\\
\partial_t\tilde\lambda&=&(w-2)^2(1+w)^{-5/2}\frac{\tilde\lambda^2}{16\pi}.
\end{eqnarray}
One finds an infrared stable (approximate) fixed point for large $w$
\begin{equation}\label{H128b}
w_*\approx\left(\frac{64\pi}{3\tilde\lambda}\right)^2.
\end{equation}
As $\tilde\lambda$ moves slowly towards zero, $w_*$ increases to infinity. Indeed, for this fixed point the evolution of $\tilde\lambda$ follows
\begin{equation}\label{H128c}
\partial_t\left(\frac{\tilde\lambda}{16\pi}\right)\approx\frac34\left(\frac{\tilde\lambda}{16\pi}\right)^3
\end{equation}
implying a very slow running once $\tilde\lambda(k)\ll 16\pi$
\begin{equation}\label{H128d}
\tilde\lambda(k)=\tilde\lambda(k_0)
\left[1+\frac32\left(\frac{\tilde\lambda(k_0)}{16\pi}\right)^2\ln
\frac{k_0}{k}\right]^{-1/2}.
\end{equation}
Here $k_0$ denotes the scale where $w\approx w_*$ becomes valid. The effective anomalous dimension is very small
\begin{equation}\label{H128e}
\eta=\frac32\left(\frac{\tilde\lambda}{16\pi}\right)^2.
\end{equation}

On the other hand, the flow equation
\begin{equation}\label{Q14}
\partial_tS=2S~,~\eta_S=-2
\end{equation}
implies
\begin{equation}\label{Q15}
S=S_0\frac{k^2}{k^2_0},
\end{equation}
corresponding to the limit $d\to 2$ of eq. (\ref{Q10}). Similar to (\ref{Q11}) we find the asymptotic behavior
\begin{equation}\lambda=\tilde\lambda S\sim k^2.\end{equation}
The order parameter is approximately constant
\begin{eqnarray}\label{Q17}
\rho_0(k)=\frac{k^2}{2S}
\frac{(64\pi)^2}{9\tilde\lambda^3}=
\frac{(64\pi)^2k^2_0}{18S_0\tilde\lambda^3}.
\end{eqnarray}
More precisely, the tiny running of $\rho_0$ for large $w_*$ and $d=2$ may be directly inferred from eq. (\ref{aa15}) (up to corrections $\sim w^{-1}_*$)
\begin{eqnarray}\label{C6}
\partial_t\rho_0&=&\frac{\tilde\lambda\rho_0}{3\pi(1+w_*)^{-3/2}}\nonumber\\
&=&-\frac{9\tilde\lambda^4\rho_0}{(64)^3\pi^4}=-\frac{\rho_0}{\ln^2\frac{k_0}{k}}.\nonumber\\
\end{eqnarray}
The evolution of $\rho_0(k)$ stops for $k\to 0$
\begin{equation}\label{E6a}
\rho_0(k)=\rho_0(k_0)\exp \left\{\frac{1}{\ln\frac{\Lambda}{k_0}}-\frac{1}{\ln\frac{\Lambda}{k}}\right\}.
\end{equation}

The flow of the bare order parameter $\bar \rho_0=\rho_0/\bar A$
\begin{equation}\label{E6b}
\partial_t\ln\bar\rho_0=\partial_t\ln\rho_0+\eta
\end{equation}
is dominated by $\eta$. For very small $k<k_l$,
\begin{equation}\label{E6c}
\ln\frac{k_0}{k_l}\gg\frac23\left(\frac{\tilde\lambda(k_0)}{16\pi}\right)^{-2},
\end{equation}
one has
\begin{equation}\label{E6d}
\eta\approx \ln^{-1}(k_0/k).
\end{equation}
This implies that $\bar\rho_0(k)$ vanishes logarithmically for $k\to 0$
\begin{equation}\label{E6e}
\bar\rho_0(k)=\bar\rho_0(k_l)\frac{\ln(k_0/k_l)}{\ln(k_0/k)}.
\end{equation}
We conclude that for $d=2$ no long range order exists in a strict sense in this truncation. In this respect the upper critical dimension $d=2$ is similar to $d<2$. In fact, we may take the limit $d\to 2$ of eq. (\ref{A14e}) and observe that with
\begin{equation}\label{E6f}
\lim_{d\to 2}\eta_S=-d
\end{equation}
the behavior $\bar\rho_0\sim k^\eta$ becomes essentially logarithmic as $\eta$ approaches zero for 
$d\to 2$. 

Let us compare these findings with the true evolution in presence of the relativistic kinetic term $\sim V$. Without $V$ the critical dimension below which order disappears would be $d_c=2$. This is shifted to $d_c=1$ in presence of $V$. The presence of $V$ reduces the disordering power of the Goldstone fluctuations. For a relativistic kinetic term the effective power counting is different - the Goldstone regime corresponds now to the classical model in dimension $d+1$. 

We conclude this appendix by addressing the issue of the scale $k_F$ associated to the density. We expect that for $k\gg k_F$ the density plays no role. The regime where the flow is essentially independent of the density effects corresponds to $w\ll 1$. For $k\ll k_F$ one expects a transition to a new qualitative regime where the density matters. This may be roughly associated with the Goldstone regime for $w\gg 1$.

The transition between the Goldstone regime for $w\gg1$ and the linear regime for $w\ll 1$ typically occurs for $w$ near one and we may define the scale $k_G$ where $w=1$ by
\begin{equation}\label{E6g}
k^2_G\approx2\lambda(k_G)\rho_0(k_G).
\end{equation}
The detailed relation between $k_F$ and $k_G$ may depend on the dimension, but we expect that they are of a similar magnitude. For a demonstration, we discuss this issue even in our simplest truncation. Within the truncation (\ref{Q1}) we have found for $d\geq 2$  that $\rho_0(k\to0)\equiv\rho_0$ approaches a positive constant. Since $\rho_0$ scales proportional to the density we can define a dimensionless quantity 
\begin{equation}\label{E6h}
L=2\lambda\rho_0^{\frac{d-2}{d}}
\end{equation}
such that
\begin{equation}\label{E6i}
w=\frac{L\rho_0^{\frac2d}}{k^2}.
\end{equation}
Since $L$ is evaluated for $k=k_G$ we expect $L$ to be a constant that is neither extremely small nor large. This implies $k_G=\sqrt{L}\rho^{1/d}_0\sim k_F$. 

For $d<2$ the issue is more involved. Besides $w$ the crossover to a relativistic kinetic term plays a role. Within the truncation (\ref{Q1}) one may wonder where the scale set by the density appears in a situation for which the couplings $w$ and $\tilde\lambda$ flow to their fixed point values irrespective of their microphysical values. If all couplings are irrelevant the information about the density would be lost in $\Gamma_{k\to 0}$. Expressed in terms of renormalized fields the effective action should therefore contain at least one parameter that is not determined by the fixed point. Such a parameter is given by $S$ and the presence of a scale can  therefore be  encoded in $S$, despite the fact that $S$ is dimensionless. Indeed, the nontrivial scaling with $\eta_S\neq 0$ implies the generic form
\begin{equation}\label{E6j}
S=S_0\left(\frac{k}{k_0}\right)^{-\eta_S}
\end{equation}
and a momentum scale appears by dimensional transmutation in the form of $k_0$. Since the size of $S$ determines the scale where $V$ will start to dominate, it is plausible that $k_0$ can be related to the final value of $\bar\rho_0$ and therefore to $n$ for $d>1$. For $d\leq 1$ the condensate contribution to the density $\bar\rho_0$ vanishes and the properties of $\bar G(\vec q)$ for $\vec q\neq 0$ play a decisive role. 

We have avoided these subtle points by choosing a fixed definition \eqref{51C} for $k_F$. The price to pay is that the transition between the qualitatively different behavior for the dense and the dilute regime occurs not necessarily for $k_{ph}\approx k_F$, but may involve a nontrivial proportionality constant.

\section*{APPENDIX G: Symmetries and thermodynamic relations}
\renewcommand{\theequation}{G.\arabic{equation}}
\setcounter{equation}{0}

In this appendix we summarize symmetries and Ward identities \cite{1D}, \cite{1B}, \cite{KT} on the level of an appropriate truncation of the effective action \cite{FW}. Let us consider, for real time (Minkowski signature), the classical action $S_M$ in presence of local sources
\bear\label{G.1}
{\cal S}_M=\int_x \Big\{\chi^* \Big[i\partial_t +\sigma +(\vec \nabla-i\vec A)^2\Big]\chi\n\\
-V(\chi^*\chi)+j^*\chi+j\chi^*\Big\}.
\ear
Here $\sigma(x)$ and $\vec A(x)$ are real source fields, while $j(x)$ is complex. For a homogeneous setting the physical values will be $\sigma(x)=\mu, \vec A(x)=0, j(x)=0$. The action is invariant under time- and space-translations, rotations, parity reflections $x_k\to -x_k,~A_k\to -A_k$, as well as time reflection $t\to -t,~\varphi\to\varphi^*,~j\to j^*$. It is real and exhibits a local $U(1)$-gauge symmetry associated to conserved particle number
\bear\label{G.2}
&&\chi(x)\to e^{i\vartheta(x)}\chi(x),~j(x)\to e^{i\vartheta(x)}j(x),~\n\\
&&A_\mu(x)\to A_\mu(x)+\partial_\mu\vartheta(x),
\ear
where $A_\mu=(\sigma,\vec A),~\partial_\mu=(\partial_t,\vec\nabla)$. For $\vec A=0$ the action remains invariant under Galilei transformations
\bear\label{G.3}
&&\chi(t,\vec x)\to e^{if}\chi(t,\vec x - 2\vec P t),~j(t,\vec x)\to e^{if} j(t,\vec x-2\vec P t),\n\\
&&f=\vec P \vec x -\vec P^2 t,~\sigma(t,\vec x)\to \sigma (t, \vec x-2\vec P t).
\ear
In eq. \eqref{G.1} we have scaled our units such that effectively $2M_B=1$, with $M_B$ the mass of the particles, such that the velocity obeys $\vec v=\vec P/M_B\widetilde{=} 2\vec P$. Thus for a plane wave with $\omega=\vec p^2$ the Galilei transformation indeed produces the appropriate shifts in momentum and energy, $\vec p'=\vec p+\vec P,~\omega'=(\vec p+\vec P)^2$, i.e.
\be\label{G.4}
\chi= e^{i(\vec p\vec x-\omega t)}\to e^{i(\vec p'\vec x-\omega' t)}.
\ee
We note that Galilei transformations and $\vec x$-dependent local $U(1)$ transformations are not compatible. In the following we take $\vec A=0$ and restrict the $U(1)$-transformations to transformation parameters $\vartheta(t)$ depending on time but not on the space coordinates.

Including the fluctuation effects yields the effective action $\Gamma[\bar\varphi,\sigma]$, with $\bar\varphi(x)=\langle\chi(x)\rangle$ evaluated for a corresponding source $j(x)$. In absence of anomalies from the functional measure, and for $T=0$, the effective action has the same symmetries as the classical action, where $\bar\varphi$ transforms in the same way as $\chi$. This extends to the average action $\Gamma_k$ if the cutoff $R_k$ is consistent with the symmetries. For an investigation of the long distance behavior we expand the effective action in the derivatives $\partial_t$ and $\Delta$. Consistency with the symmetries requires
\bear\label{G.5}
\Gamma&=&\int_x\Big[U_0(\bar\rho)
-\frac12 Z(\bar\rho)
\Big\{\bar\varphi^*\big[i\partial_t+\sigma+\Delta\big]\bar\varphi+c.c\Big\}\n\\
&&-\frac12 \bar V(\bar\rho)\Big\{\bar\varphi^*\big[i\partial_t+\sigma+\Delta\big]^2
\bar\varphi+c.c\Big\}\n\\
&&+\frac14\bar Y_t(\bar\rho)\bar\rho
\Big\{\partial^2_t\bar\rho-2i \partial_t\vec\nabla
(\bar\varphi^*\bar\nabla
\bar\varphi-\bar\varphi\vec\nabla\bar\varphi^*)\n\\
&&+2\vec\nabla
(\vec\nabla\bar\varphi^*\Delta\bar\varphi+\vec\nabla\bar\varphi\Delta\bar\varphi^*)-\Delta(\bar\varphi^*\Delta\bar\varphi+\bar\varphi\Delta\bar\varphi^*)\Big\}\n\\
&&-\frac14\bar Y(\bar\rho)\bar\rho\Delta\bar\rho+\Delta{\cal L}\Big].
\ear
Here we use $\bar\rho=\bar\varphi^*\bar\varphi$ and $\Delta{\cal L}$ contains further gradient terms like $(\vec\nabla\bar\rho\vec\nabla\bar\rho)^2$ or $\vec\nabla\rho\vec\nabla(\bar\varphi^*(i\partial_t+\sigma+\Delta)\bar\varphi$). In particular, Galilei symmetry implies that all time derivatives appear only in the combination 
${\cal  D}_t=\partial_t-i\sigma-i\Delta$, and additional gradients act only on the invariants $\bar\rho$ and $\bar\varphi^*{\cal D}^n_t\bar\varphi$. 

We next perform an analytic continuation to the euclidean effective action by replacing $-i\partial_t\to\partial_\tau$ in eq. \eqref{G.5}. (We have introduced an overall minus sign in the transition from $S_M$ to $\Gamma$, such that the standard conventions for the classical action in Minkowski spacetime match with the standard conventions for the euclidean effective action.) This allows for an extension to thermal equilibrium with $T\neq 0$. For $T\neq 0$ additional terms can appear in the effective action. First, Galilei invariance is broken since the heat bath singles out a particular reference frame. This permits the appearance of additional terms involving gradients, as $\bar\varphi^*\Delta\bar\varphi$. Second, a continuation of the local $U(1)$ invariance with real functions $\vartheta(\tau)$ requires complex $\sigma$, transforming as $\sigma\to\sigma+i\partial_\tau\vartheta$. However, the real part of $\sigma$ is  invariant under local transformations. In consequence, an arbitrary dependence of $U,Z,\bar V$ on $\sigma+\sigma^*$ becomes possible. In practice, only the global $U(1)$ symmetry is relevant for $T\neq 0$, since the local transformations only constrain possible couplings of the imaginary part of $\sigma$ which play no physical role. The Ward identities related to the local $U(1)$ symmetry and the Galilei transformations therefore only restrict the limiting behavior of $\Gamma$ for $T\to 0$. We emphasize that all Ward identities are automatically implemented by the invariant form of $\Gamma$ in eq. \eqref{G.5}.

The effective potential in presence of a nonzero chemical potential $\sigma$ reads 
\be\label{G.6}
U(\bar\rho,\sigma)=U_0(\bar\rho)-Z(\bar\rho)\bar\rho\sigma-\bar V(\bar\rho)\bar\rho\sigma^2.
\ee
The order parameter is given by the condensate density $n_c=\bar\rho_0$, which corresponds to the minimum of $U$ at a fixed value of $\sigma$
\be\label{G.7}
\frac{\partial U}{\partial\bar\rho}_{|\sigma,\bar\rho_0}=0~,~\partial_{\bar\rho}U_{0_{|\bar\rho_0}}=\sigma
\partial_{\bar\rho}(Z\bar\rho+\sigma\bar V\bar\rho)_{|\bar\rho_0}.
\ee
The total particle density is related to the $\sigma$-derivative of $U$ at $\bar\rho_0$
\be\label{G.8}
n=-\frac{\partial U}{\partial \sigma}(\bar\rho_0)=
\big[Z(\bar\rho_0)+2\sigma\bar V(\bar\rho_0)\big]\bar\rho_0.
\ee
In terms of renormalized fields
\be\label{G.9}
\phi=\bar A^{1/2}\bar\phi,~\rho=\bar A\bar\rho,~\bar A=Z(\bar\rho_0)+2\sigma\bar V(\bar\rho_0)
\ee
one finds
\be\label{G.10}
n=\rho_0.
\ee
Thus a nonvanishing density requires for $T=0$ a nonvanishing renormalized order parameter. (We recall that there are $\bar\rho$-independent contributions to $U$ for $T\neq 0$. For example, a piece $\Delta U=-n_T\sigma$ contributes to $n$ a piece $n_T$ that does not vanish for $\rho_0=0$.) This observation is particularly interesting for $d=1$ where the consendate density $\bar\rho_0$ vanishes in the infinite volume limit. Indeed, we find a diverging $\bar A$ as the IR-cutoff $k$ runs to zero.

We may also compute thermodynamic susceptibilities like the response of $n$ to a change in the chemical potential
\bear\label{G.11}
&&\frac{\partial n}{\partial\sigma}_{|T=0}=
\frac{\partial\rho_0}{\partial\sigma}=2\bar V(\bar\rho_0)\bar\rho_0+ 
\bar S\frac{\partial\bar\rho_0}{\partial\sigma},\n\\
&&\bar S=\bar A\left(1+\frac{\partial\ln\big(Z(\bar\rho_0)+2\sigma\bar V(\bar\rho_0)\big)}{\partial\ln\bar\rho_0}\right).
\ear
Differentiating the minimum condition \eqref{G.7} yields
\be\label{G.12}
\frac{\partial\bar\rho_0}{\partial\sigma}=\frac{1}{\bar\lambda}\frac{\partial n(\bar\rho_0,\sigma)}{\partial\bar\rho_0}_{|\sigma}~,~
\bar\lambda=\frac{\partial^2U(\bar\rho,\sigma)}{\partial\bar\rho^2}_{|\sigma,\bar\rho_0},
\ee
where
\be\label{G.11A}
\frac{\partial n}{\partial\bar\rho_0}_{|\sigma}
=\frac{\partial\big(Z(\bar\rho_0)\bar\rho_0\big)}{\partial\bar\rho_0}
+2\sigma\frac{\partial\big(\bar V(\bar\rho_0)\bar\rho_0\big)}{\partial\bar\rho_0}=\bar S,
\ee
and therefore
\be\label{12A}
\frac{\partial\bar\rho_0}{\partial\sigma}=\frac{\bar S}{\bar\lambda}\quad,\quad
\frac{\partial n}{\partial\sigma}_{|T=0}=\frac{2\bar V(\bar\rho_0)n}{\bar A}+\frac{\bar S^2}{\bar\lambda}.
\ee

The pressure 
\be\label{13}
p=-U(\rho_0)
\ee
is normalized such that it vanishes for $T=n=0$, i.e. $U_0(\rho=0)=0$. Its response to a change of the chemical potential obeys
\be\label{14}
\frac{\partial p}{\partial \sigma}_{|T}=-
\frac{\partial U}{\partial \sigma}_{|\bar\rho_0}-
\frac{\partial U}{\partial\bar\rho_0}_{|\sigma,\bar\rho_0}\frac{\partial\bar\rho_0}{\partial\sigma}=n.
\ee
In our units $(2M_B=1)$ the energy density obeys $\epsilon=n/2$ and one obtains for the macroscopic sound velocity
\be\label{14A}
c^2_s=\frac{\partial p}{\partial\epsilon}_{|T}=2
\frac{\partial p}{\partial\sigma}_{|T}
\left(\frac{\partial n}{\partial\sigma}_{|T}\right)^{-1}.
\ee
For $T=0$ one finds
\be\label{15}
c^{-2}_s=\frac{\bar V(\bar\rho_0)}{\bar A}+
\frac{\bar S}{2\rho_0}\frac{\partial\bar\rho_0}{\partial\sigma}
=\frac{\bar V(\bar\rho_0)}{\bar A}
+\frac{\bar S^2}{2\rho_0\bar\lambda}.
\ee

In presence of spontaneous symmetry breaking we are interested in the propagator for small fluctuations around the expectation value $\varphi_0$ that we take here to be real. We decompose 
$\varphi(x)=\varphi_0+\frac{1}{\sqrt{2}}\big(\delta\varphi_1(x)+i\varphi_2(x)\big)$. The inverse propagator matrix is encoded in the part of $\Gamma$ that is quadratic in $\delta\varphi_1$ and $\varphi_2$. Retaining terms with up to two derivatives one may parameterize
\bear\label{16}
\Gamma_2&=&\int_x\left\{\frac{iS}{2}
(\delta\varphi_1\partial_\tau\varphi_2-\varphi_2\partial_\tau\delta\varphi_1)\right.\n\\
&&-\frac V2 (\varphi_2\partial^2_\tau\varphi_2+B_r\delta\varphi_1\partial^2_\tau\delta\varphi_1)\\
&&\left. -\frac12(\varphi_2\Delta\varphi_2+C_r\delta\varphi_1\Delta\delta\varphi_1)+
\frac{m^2_r}{2}\delta\varphi^2_1\right\}.\n
\ear
Here we have defined the renormalized field $\varphi$ such that the term $\sim\varphi_2\Delta\varphi_2$ has coefficient one. For $T=0$ we may extract the different couplings by expanding eq. \eqref{G.5} (with $\bar\lambda=\partial^2_{\bar\rho}U_{|\bar\rho_0})$
\bear\label{17}
\Gamma_2&=&\int_x\Big\{\frac i2
\Big[Z(\bar\rho_0)+2\sigma\bar V(\bar\rho_0)
+\bar\rho_0(\partial_{\bar\rho}Z+2\sigma\partial_{\bar\rho}\bar V)_{|\bar\rho_0}\Big]\n\\
&&\qquad (\delta\bar\varphi_1\partial_\tau\bar\varphi_2-\bar\varphi_2\partial_\tau\delta\bar\varphi_1)\n\\
&&-\frac12\bar V(\bar\rho_0)\bar\varphi_2\partial^2_\tau\bar\varphi_2\n\\
&&-\frac12\big[\bar V(\rho_0)+\bar\rho_0(2+\partial_{\bar\rho}V+\bar Y_t+\bar\rho\partial_{\bar\rho}\bar Y_t)_{|\bar\rho_0}\big]
\delta\bar\varphi_1\partial^2_\tau\delta\bar\varphi_1\n\\
&&-\frac12\big[Z(\bar\rho_0)+2\sigma\bar V(\bar\rho_0\big]\bar\varphi_2\Delta\bar\varphi_2\n\\
&&-\frac12\Big[Z(\bar\rho_0)+2\sigma\bar V(\bar\rho_0)\n\\
&&+\bar\rho_0(2\partial_{\bar\rho}Z+4\sigma\partial_\rho\bar V
+\bar Y+\bar\rho\partial_{\bar\rho}Y)_{|\bar\rho_0}\Big]
(\delta\bar\varphi_1\Delta\delta\bar\varphi_1)\n\\
&&+\bar \lambda\bar\rho_0\delta\bar\varphi^2_1\Big\}.
\ear
The renormalized field is the same as in eq. \eqref{G.9} and we identify
\bear\label{19}
S&=&1+\frac{\partial \ln\big(\bar Z(\bar\rho_0)+2\sigma\bar V(\bar\rho_0)\big)}{\partial\ln\bar\rho_0}=\frac{\bar S}{\bar A},\n\\
V&=&\frac{\bar V(\bar\rho_0)}{\bar A}~,\n\\
B_r&=&1+2\frac{\partial\ln\bar V(\bar\rho_0)}{\partial\ln\bar\rho_0}
+\frac{\bar Y_t(\bar\rho_0)\bar\rho_0}{(\bar V(\bar\rho_0)}
\left(1+\frac{\partial\ln\bar Y_t(\bar\rho_0)}{\partial\ln\bar\rho_0}\right),\n\\
C_r&=&2S-1+\frac{\bar Y(\bar\rho_0)\bar\rho_0}{\bar A}
\left(1+\frac{\partial\ln\bar Y(\bar\rho)}{\partial\ln\bar\rho_0}\right),\n\\
m^2_r&=&2\bar\lambda\bar\rho_0/\bar A=2\lambda\rho_0~,~\lambda=\frac{\bar\lambda}{\bar A^2}.
\ear

We will find that $S$ vanishes in the infinite volume limit for $d\leq 3$. The solution of eq. \eqref{19} implies that $\bar A(\bar\rho)$ diverges for $\bar\rho\to 0$
\be\label{20}
\bar A(\bar\rho)=\bar Z(\bar\rho)+2\sigma\bar V(\bar\rho)\to\frac{\rho_0}{\bar\rho}.
\ee
For $d>1$ one finds a nonzero $\bar\rho_0$ such that $\bar A =\bar A(\bar\rho_0)$ remains finite. For $d=1$, however, $\bar\rho_0\to 0$ and $\bar A$ diverges in the infinite volume limit. Since $V$ remains finite also $\bar V$ must diverge $\sim \bar\rho^{-1}$ in this case. 

Comparing the definitions \eqref{G.11} and \eqref{19} one has 
\be\label{G.21}
c^{-2}_s=V+\frac{S^2}{2\lambda\rho_0}.
\ee
For $S\to 0$ this implies for the macroscopic sound velocity
\be\label{22}
c^2_s=V^{-1}.
\ee

In summary, Galilei and gauge symmetry relate for $T=0$ the properties of the inverse propagator, like the microscopic sound velocity $v$, the superfluid density $n_S=\bar A\bar\rho_0$ (as defined by the stiffness with respect to phase changes), or the term linear in the frequency $\sim S$, to macroscopic thermodynamic quantities. Using 
\be\label{G.22a}
\partial n/\partial\sigma=2Vn+S^2/\lambda=2n/c^2_s
\ee
we can replace the $\sigma$-derivatives by $n$-derivatives or derivatives with respect to the volume $\Omega_d~(\partial\ln n/\partial\ln\Omega_d=-1$ for fixed particle number $N$)
\be\label{G.26}
\frac{\partial \ln n_c}{\partial\sigma}=\frac{2}{c^2_s}
\frac{\partial\ln n_c}{\partial \ln n}=-
\frac{2}{c^2_s}\frac{\partial\ln n_c}{\partial \ln\Omega_d}.
\ee
This gives a direct physical interpretation of the running renormalized couplings evaluated for $k=0$
\bear\label{G.27}
\rho_0&=&n=n_s~,~\bar\rho_0=n_c~,~
\bar A^{-1}=\frac{n_c}{n}=\Omega_c,\n\\
\frac{S}{\lambda}&=&\frac{2n}{c^2_s}\frac{\partial\ln n_c}{\partial\ln n},\n\\
V&=&\frac{1}{c^2_s}\left(1-S\frac{\partial\ln n_c}{\partial \ln n}\right)~,~c_s=v.
\ear

\end{document}